\documentclass[a4paper,10pt]{article}
\usepackage{a4wide}
\usepackage{color}
\usepackage[round, sort, numbers, authoryear]{natbib}
\usepackage{amsbsy,amssymb,amsgen,amsfonts,graphics}
\usepackage{amsmath,amsfonts,color,latexsym,gensymb}
\usepackage[english]{babel}
\usepackage{inputenc}
\usepackage[pdftex,bookmarks=true,colorlinks=true,linkcolor=blue,citecolor=blue,
filecolor=blue,urlcolor=blue]{hyperref}
\usepackage[T1]{fontenc}
\usepackage{graphicx}
\usepackage{titlesec}
 \titleformat{\section}
   {\normalfont\sffamily\Large\bfseries}
   {\thesection}{1em}{}
\usepackage[font={small}]{caption}
\usepackage{tikz}
\usetikzlibrary{shapes,arrows,positioning,calc,through,backgrounds}
\tikzstyle{every picture}+=[remember picture]
\usetikzlibrary{patterns}
\definecolor{lightgray}{rgb}{0.75,0.75,0.75}
\definecolor{green}{rgb}{0.0, 0.75, 0.0}
\definecolor{cyan}{rgb}{0.0, 0.75, 0.75}
\definecolor{magenta}{rgb}{0.75, 0.0, 0.75}
\definecolor{grey}{rgb}{0.45, 0.45, 0.45}
\def\solidthick{\protect\rule[2pt]{10.pt}{1pt}}
\def\solidshort{\protect\rule[2pt]{3.pt}{1pt}}
\def\dashed{\solidshort$\,$\solidshort$\,$\solidshort}

\usepackage{relsize}
\usepackage[percent]{overpic}

\usepackage[toc,page]{appendix}
\usepackage[normalem]{ulem}
\usepackage{mathrsfs}
\usepackage{enumitem}
\begin{document}
\reversemarginpar
\normalsize
\begin{center}
\large{\textbf{\textsf{
Heat and water vapor transfer in the wake of a falling ice sphere and its implication for secondary ice formation in clouds
}}}\\[2ex]
\normalsize
{Agathe Chouippe$^{(1)}$, Michael Krayer$^{(1)}$, Markus Uhlmann$^{(1)}$, Jan Du\v{s}ek$^{(2)}$, Alexei Kiselev$^{(3)}$, Thomas Leisner$^{(3)}$}\\[2ex]
\small{$^{(1)}$ Institute for Hydromechanics, Karlsruhe Institute of Technology (KIT), 76131 Karlsruhe, Germany}\\[0.5ex]
\small{$^{(2)}$ ICUBE, Fluid Mechanics Group, Universit\'e de Strasbourg, 67000 Strasbourg, France} \\[0.5ex]
\small{$^{(3)}$ Institute for Meteorology and Climate Research, Atmospheric Aerosol Research, Karlsruhe Institute of Technology (KIT), Germany} \\[1ex]
\footnotesize{(\today)}\\
\end{center}

\normalsize
\begin{abstract} 
We perform direct numerical simulations of the settling of an ice sphere in an ambient fluid accounting for heat and mass transfer with the aim of studying in a meteorological context the case of falling graupel in humid air. The study is motivated by the fact that falling graupels in clouds are heated by the latent heat released during the accretion of liquid water droplets. They may therefore be considerably warmer than their surrounding and evaporate water vapor, which mixes with the surrounding air in the wake of the graupel, thereby creating transient zones of supersaturation there.   
{The problem of a falling graupel is modeled as that of a
heated sphere falling in a quiescent ambient fluid under the action of gravity. The coupling between the temperature and
velocity fields  is accounted for by the Boussinesq approximation. This problem can be parameterized by four parameters: the particle/fluid density ratio $\rho_p/\rho_\infty$, the Galileo number $Ga=u_g D/\nu$ (where $D$ is the diameter of the sphere, $\nu$ the viscosity of the fluid,  $u_g=\sqrt{\lvert( \rho_p/\rho_\infty-1)g\rvert D}$, and $g$ the gravitational acceleration)%
, the Prandtl number $Pr=\nu/D_T$ (where $D_T$ stands for the thermal diffusivity), and the Richardson number
 $Ri_T= \beta (T_p-T_\infty)/(\frac{\rho_p}{\rho_\infty}-1)$, where $T_p-T_\infty$ is the temperature difference between the sphere and the ambient fluid and $\beta$ the thermal expansion coefficient of the fluid. A separate scalar transport equation accounts for the vapor transport.}
Typical cloud conditions involve small temperature differences between the sphere and the surrounding, yielding relatively small Richardson numbers for both heat and mass transport.
{We give} a special emphasis to the Galileo numbers 150, 170, 200 and 300 in
order to analyze the specificities of each settling regime. The questions addressed in this study are mainly methodological and concern the influence of the settling regime and the mobility of the sphere on the structure of the scalar fields, the possible influence of modest Richardson numbers on the structure of the wake, and the possible application of this simulation framework to the investigation of the saturation in the wake of a falling graupel.
{We observe that the body behaves similar to a body with infinitely large density. Buoyancy effects upon the wake at the values of the Richardson number corresponding to the atmospheric context are found to be negligible.}
We discuss the necessity to distinguish between the diffusivity of temperature and vapor content and for this the requirement to solve both scalar transport equations separately.
The simulations reveal the structure of the saturation field which features zones of supersaturation that might indeed be the sites of secondary ice nucleation {(formation of additional ice crystals)}.
{The potential error in not solving both fields separately is relatively low but affects the regions of the flow that feature the largest supersaturation, such that it could be preferable to separate both transport equations depending on the future questions addressed}.
\end{abstract}

\section{Introduction}  \label{sec:introduction}
Convection associated with mass transfer around a settling sphere is of
relevance for many industrial and environmental systems such as clouds or
combustion chambers. In the case of a graupel falling in a cloud an additional
effect becomes important. The ice particle can be considerably warmer than the
surrounding due to the latent heat released by the riming process, i.e. the
freezing of liquid cloud droplets colliding with the falling ice particle. The
diffusion of heat and vapor in the wake of the ice ball can lead to local vapor
supersaturation that may even lead to secondary ice nucleation in its wake
{through heterogeneous ice nucleation mechanisms}, if ice nuclei are present.
{Heterogeneous ice nucleation, in contrast to homogeneous ice
nucleation, requires the presence of another substance usually called ice
nucleating particle (INP) and is more likely to occur in the present context since homogeneous ice nucleation requires very low temperatures (below -38$\degree C$) and large relative humidity (above 145$\%$)}. The conditions up to which this {heterogeneous }nucleation can occur depend on both temperature and water content as well as the type of ice nuclei \citep{hoose:12}.
{Heterogeneous ice nucleation can take place under various modes depending on the atmospheric condition and the type of INP \citep{vali15}: ice can directly form from water vapor (deposition) or from supercooled liquid (freezing) with different possible freezing mechanisms (immersion, contact, condensation freezing).}
A description of the advection-diffusion of both temperature and water content in the wake of a particle is therefore to be seen as a first step for a better understanding of the fundamental processes involved in {secondary ice nucleation induced by a falling graupel}.
\begin{figure}[t]
\begin{minipage}{\linewidth}
\centering
\includegraphics[width=1.0\linewidth]
   {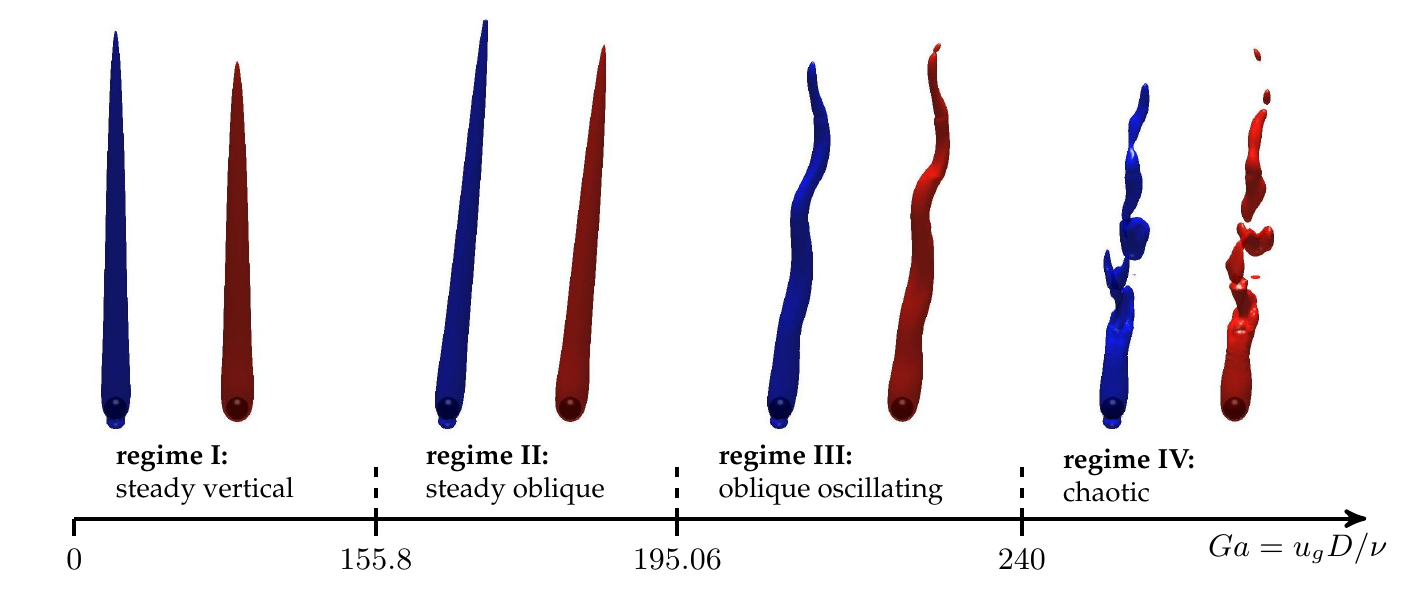}
\end{minipage}
\caption{{Settling regimes observed for a density ratio $\rho_p/\rho_\infty=10$, for different Galileo numbers and in the absence of buoyancy effects, with illustration of the wake structure for one Galileo number in each regime (i.e. from left to right $Ga=\lbrace 150,\;170,\;200,\;300 \rbrace$). 
Blue surfaces represent isocontour of the velocity $\left|\mathbf{\tilde{u}-\tilde{u}_p}\right|=0.80\left|\mathbf{\tilde{u}_\infty}-\mathbf{\tilde{u}_p}\right|$ (where $\mathbf{\tilde{u}}$ is the non-dimensional fluid velocity, $\mathbf{\tilde{u}}_\infty$ the velocity of the incoming fluid and $\mathbf{\tilde{u}}_p$ the sphere's velocity) and red surfaces isocontour of the temperature $\tilde{T}=0.1$ (where $\tilde{T}=(T-T_\infty)/(T_p-T_\infty)$, with $T$ the fluid temperature, $T_\infty$ the temperature of the incoming fluid and $T_p$ the temperature at the surface of the sphere).}}
\label{fig:sketch_regimes}
\end{figure}

The case of a single heavy sphere falling in a quiescent environment, without accounting for heat or mass transfer, could seem to be simple but is known to feature complex properties \citep{Ern_AnnuRev2012,Jenny_JFM2004}. It can be parametrized by two dimensionless quantities, as described in \cite{Jenny_JFM2004}, namely the ratio between particle and fluid density {$\rho_p/\rho_\infty$} and the Galileo number $Ga=u_gD/\nu$, where {$u_g=\sqrt{\lvert (\rho_p/\rho_\infty-1 )g\rvert D }$} is the gravitational velocity, $D$ the diameter of the particle, $g$ the gravitational acceleration and $\nu$ the kinematic viscosity of the fluid.
{Figure \ref{fig:sketch_regimes} summarizes the different settling regimes to be expected for a density ratio $\rho_p/\rho_\infty=10$: for} the lowest Galileo number the particle falls on a straight vertical path and the wake is steady axisymmetric. When $Ga$ increases, above a critical value $Ga_{c1}$, the particle follows an oblique path and the wake loses its axisymmetry to become planar reflectional symmetric.
In this steady oblique regime
the value of $Ga_{c1}$ is independent of the density ratio \citep{fabre:12}.
Steadiness is lost when $Ga$ becomes larger than $Ga_{c2}$ where a Hopf bifurcation occurs: the wake oscillates in time while the particle continues to fall in a plane. This is the oblique oscillating regime \citep{Jenny_JFM2004}. The flow becomes then chaotic when $Ga$ further increases above $Ga_{c3}$ and the particle motion is fully three-dimensional. 
The density ratio has been shown to have negligible effect on the transition scenario for values above 2.5 \citep{Zhou_IJMF2015}.
In the case of the flow around a fixed sphere, which can be thought of as a case for which $\rho_p/\rho_{\infty}$ tends to infinity, the transition scenario of the wake structure is parametrized by the Reynolds number $Re=u_\infty D/\nu$, where $u_\infty$ is the relative velocity of the unperturbed fluid.
This scenario is similar to that of free spheres of large density ratio, and it has been investigated in many experimental and numerical works \citep{ormieres:99,johnson:99, Ghidersa_JFM2000}.

Turning now to the case of flows with heat or mass transfer, the literature mostly focused on configurations where the ratios between scalar diffusivity and fluid viscosity (namely the Prandtl number $Pr=\nu/\mathcal{D}_T$ for the temperature and the Schmidt number $Sc=\nu/\mathcal{D}_m$ for the mass) {is equal to $0.7$, $1$ or $7$}.  
The numerical work of \cite{Bagchi_ASME2001} for passive scalar transport featuring $Pr=0.72$ and a fixed sphere with Reynolds number ranging from 50 to 500, showed that the structure of the scalar field is highly dependent on the structure of the flow field with a clear influence of three-dimensionality and vortex shedding. Ignoring 3D effects indeed leads to an underprediction of drag and an overprediction of the Nusselt number.
{Other studies from the literature explored the influence of density variation due to temperature inhomogeneities, with more attention given to the configurations of assisting and opposing flow (i.e. where gravity points in the same or in the opposite direction of the unperturbed flow respectively). This influence of temperature variations on the buoyancy term is }%
usually parametrized by the Grashof number $Gr=\beta g (T_p-T_\infty)D^3/\nu^2$ (where $\beta$ is the thermal expansion coefficient, $T_p$ the temperature of the sphere and $T_\infty$ the temperature of the incoming flow) or Richardson number $Ri$, that are linked by the relation $Ri=Gr/Re^2$.
It has been observed in the literature that this can induce modifications of the structure of the flow and of the heat transfer coefficient \citep{Kotouc_IJHMT2008,Kotouc_JFM2009,Bhattacharyya_IJHMT2008}.
 \cite{Bhattacharyya_IJHMT2008} studied the influence of buoyancy for a Reynolds number ranging from $1$ to $200$ and a
Richardson number ranging from $0$ and $1.5$.
They observed the development of a recirculating eddy in the downstream of the sphere which further collapses when the temperature of the sphere increases, consequently leading to the development of a buoyant plume above the sphere. 
A main effect is the delay of the flow separation with $Ri$. \cite{Kotouc_IJHMT2008} also observed that convection tends to stabilize the flow in the assisting flow configuration by preventing detachment of the boundary layer and the formation of a recirculation zone, while in the case of opposing flow a destabilizing effect is observed with a decrease of the critical Reynolds number for the onset of recirculation (\cite{Kotouc_JFM2009}). The transition scenario remains unchanged for weakly heated spheres, while new regimes appear for larger values of the Richardson number \citep[of the order of 0.3][]{Kotouc_IJHMT2008,Kotouc_JFM2009}. One can cite for example steady flow regimes with four or six vorticity threads and respectively two and three symmetry planes.
{Even a rather weak level of buoyancy with $Ri=0.1$ appeared to have an influence on the critical Reynolds number at which the transition between the different regimes appears \citep{Kotouc_JFM2009}.}
As expected, buoyancy can also lead to a modification of the drag, lift and transfer coefficients.
An increase of the drag coefficient with $Ri$ in the assisting configuration has been observed by \cite{Bhattacharyya_IJHMT2008} , \cite{Kotouc_JFM2009}, while the lift coefficient seems to decrease with $\left| Ri \right|$ \citep{Kotouc_JFM2009}. Concerning the heat transfer, an influence is only visible for high values of the Grashof number.

{The references mentioned above focused on the influence of buoyancy on fixed spheres and very little attention has been given to the case of freely falling objects.
\cite{Dan_IJHMT2010} compared the terminal velocity of a sphere for Reynolds numbers ranging between $10$ and $130$ and Grashof numbers
$Gr=\{-100,\;0,\;100\}$ and observed that a cold sphere settles faster than a warm one, with larger modifications induced at lower Reynolds numbers.}
Such modifications on the wake structure can further influence particle trajectories as well as the way that they might interact. Indeed, the numerical work of \cite{Gan_JFM2003}  showed  that the equilibrium position and the trajectory of a particle settling in a vertical channel might change with the Grashof number.
They further observed that two particles settling in a channel tend to separate if they are colder than the fluid while they would aggregate if hotter.\\
Numerical simulations have been employed more recently, in a meteorological context, to investigate the flow around falling graupels of different shapes \citep{wang:13} and the ventilation coefficient of a spherical hailstone \citep{cheng:14}. {Both of them neglected the influence of buoyancy, and set the graupel to be fixed.}
{\newline Based on this several questions can be raised, which will be the object of the current study:}
\begin{itemize}[leftmargin=*]%
\setlength\itemsep{0em}
\item {As will be seen in the next section, the typical Galileo numbers for falling graupels are relatively large (from $\mathcal{O}(10^2)$ to $\mathcal{O}(10^3)$), corresponding to the chaotic regime.
Therefore the question arises: is it necessary to consider the chaotic regime when studying secondary ice formation? For this purpose we will detail the characteristics of the thermal wake, and will base our analysis on the description of the wake in a tilted coordinate system similar to \cite{Uhlmann_IJMF2014}. This is, to our knowledge, the first time that this has been proposed for a thermal wake.}
\item {The density ratio representative for falling graupel is typically large (several hundreds), meaning that the density might be large enough for the system to be equivalent to a fixed sphere (i.e. $\rho_p/\rho_\infty \rightarrow \infty$). The second question which will be addressed concerns thus the influence of the density of the sphere on the structure of the thermal wake.}
\item {The third point concerns the influence of buoyancy effects: The typical thermal Richardson number of the system is expected to be small ($\mathcal{O}(10^{-4})$), but the large Galileo number implies that the corresponding Grashof number is not necessarily small.
Since we consider lower Galileo numbers than the largest values observed in clouds the Grashof numbers investigated here are smaller than for clouds. We suggest therefore to investigate configurations featuring larger Richardson numbers ($\mathcal{O}(10^{-3})$) and above to test if buoyancy can modify the wake.
It is not clear whether buoyancy would potentially affect the system, since the literature showed that weak buoyancy ($Ri=0.1$) can modify the flow structure \citep{Kotouc_JFM2009}. This issue will also be addressed in the current study.}
\item {We will be then concerned more specifically by secondary
ice nucleation and will focus the discussion on the structure of the saturation field. The first question which we will address is purely methodological and concerns
the necessity to separate the transport equation of heat from the transport
equation of mass. As it will be described in the next sections, it is important
to take into account the variations of both temperature and water vapor in the
flow induced by the presence of the graupel. The transport equations of both
scalars can be formulated in non-dimensional form by the same type of
advection-diffusion equation, with the same set of boundary conditions, the only
difference lying on the difference of diffusivities. As those diffusivities are
relatively similar, it might be unnecessary to solve two separate transport equations for the two scalars, which will be discussed in the current study.}
\item {The last point concerns a qualitative description of the saturation that can be reached in the wake of the sphere: does the graupel increase the level of saturation in its vicinity such that it could have an effect on the ice content \citep[according to the values indicated by][]{hoose:12}?}
\end{itemize}
Let us finally emphasize that the goal of the current study is not to investigate in details the question of secondary ice nucleation that might be induced, but to discuss the methodological framework that could be used to address this question {and provide a new database which can be further used to study the possibility of condensation/deposition in the wake \citep{prabhakaran:17}.}

The paper is organized as follows: the methodology followed in the current paper is presented in section \ref{sec:methodology}, then we will describe the structure of the velocity and temperature fields for different settling regimes first by neglecting the buoyancy effect (in section \ref{sec:passive}) and then for Richardson numbers, which we will define here as $Ri_T=Gr/Ga^2$, ranging from $0.001$ to $0.1$ (in section \ref{sec:active}). Both sections \ref{sec:passive} and \ref{sec:active} concern purely non-dimensional scalar and velocity fields, while section \ref{sec:discussion} will be devoted to {the description of the saturation field and} the necessity to separate both heat and mass transport equations for the computation of the saturation with respect to ice in the wake of a graupel, for given sets of temperature conditions. We will finally draw conclusions with respect to the questions addressed in the paper.

%
\section{Methodology}  \label{sec:methodology}
\begin{figure}[t]
\begin{minipage}{\linewidth}
\centering
\includegraphics[width=0.45\linewidth]
   {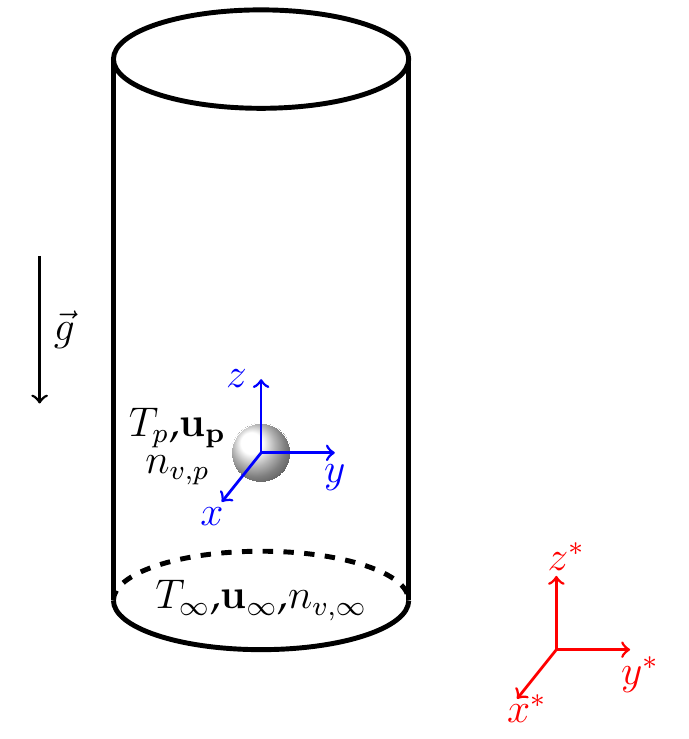}
\end{minipage}
\caption{Sketch of the system studied here: a sphere of diameter $D$ and temperature $T_p$ is falling at a velocity $\mathbf{u_p}$ in a mix of dry air and water vapor. The inflow is set at constant velocity $\mathbf{u_\infty}$, temperature $T_\infty$ and concentration $n_{v,\infty}$. The concentration at the surface of the sphere is assumed to be constant equal to $n_{v,p}$. Velocities are expressed with respect to the fixed coordinate system {$(x^{*},y^{*},z^{*})$} and projected onto the cartesian mesh $(x,y,z)$ translated with the sphere.}
\label{fig:system_sketch}
\end{figure}

We propose to consider a system representing a sphere of diameter $D$ falling in quiescent moist air as sketched in figure \ref{fig:system_sketch}. We follow the Boussinesq hypothesis and model the fluid with constant viscosity $\nu$, and a density {$\rho$} which depends linearly on the local variations of temperature $T$.
The density variations due to variable vapor concentration are negligible in the present context since they are about two orders of magnitude smaller.

The sphere is supposed to have a uniform and constant temperature $T_p$, the incoming flow has a constant velocity $\mathbf{u_\infty}$, a constant temperature $T_\infty$ and, consequently, a constant density {$\rho_\infty$}.
The system is represented by the Navier-Stokes equations (eq. \ref{eq-continuity} and \ref{eq-momentum}) coupled to one advection-diffusion equation for the transport of heat (eq. \ref{eq-heat}), under the Boussinesq approximation for the representation of the buoyancy effects.
A second (passive) scalar field representing the water vapor concentration is transported according to the advection-diffusion (eq. \ref{eq-mass}).
The system is described in a dimensionless form with reference quantities chosen as in \cite{Kotouc_JFM2009} and \cite{Jenny_JCP2004}, where $\mathbf{u_p}$ represents the velocity of the sphere and $\mathbf{u}$  corresponds to the fluid velocity with respect to a fixed frame {$(0,x^{*},y^{*},z^{*})$} and {evaluated on a grid $(x,y,z)$} moving with the sphere center as represented in figure \ref{fig:system_sketch}. 
Here lengths are made non-dimensional with respect to the particle diameter, velocity components with the gravitational velocity {$u_g$}, timescale with the gravitational timescale $D/u_g$, pressure with {$\rho_\infty u_g^2$}.
As a convention, henceforth we denote non-dimensional quantities with a tilde. The dimensionless temperature and vapor concentration are defined as {in equations \ref{eq-T_NonDim}-\ref{eq-nv_NonDim}.}
Under these assumptions the governing equations read as follows {(for more details on the formulation the reader is referred to the appendix \ref{sec:governing_equations}):}
\begin{eqnarray}
   \mathbf{\tilde{\nabla}}\cdot\mathbf{\tilde{u}} 
      &=&  0 \;,\label{eq-continuity}\\
   \frac{\partial \mathbf{\tilde{u}}}{\partial \tilde{t}}
      + \left[(\mathbf{\tilde{u}}-\mathbf{\tilde{u}_p})\cdot\mathbf{\tilde{\nabla} 
      \tilde{u}} \right]
      &=&  -\mathbf{\tilde{\nabla}}\tilde{P}
        +\frac{1}{Ga}(\mathbf{\tilde{\nabla}})^2\mathbf{\tilde{u}}
        +(Ri_T \tilde{T})\mathbf{k}
        \;,\label{eq-momentum} \\
   \frac{\partial \tilde{T}}{\partial \tilde{t}}
      + \left[(\mathbf{\tilde{u}}-\mathbf{\tilde{u}_p})\cdot\mathbf{\tilde{\nabla}} \tilde{T} \right] 
      &=&  \frac{1}{PrGa} (\tilde{\nabla})^2 \tilde{T} 
        \;,\label{eq-heat}\\
   \frac{\partial \tilde{n}_v}{\partial \tilde{t}}
      + \left[(\mathbf{\tilde{u}}-\mathbf{\tilde{u}_p})\cdot\mathbf{\tilde{\nabla}} \tilde{n}_v \right] 
      &=&
        \frac{1}{ScGa} (\tilde{\nabla})^2 \tilde{n}_v \;,
        \label{eq-mass}\\
   \tilde{T} &=& (T-T_\infty)/(T_p-T_\infty)\;, \label{eq-T_NonDim}\\
   \tilde{n}_v &=& (n_v-n_{v,\infty})/(n_{v,p}-n_{v,\infty})\; \label{eq-nv_NonDim},
\end{eqnarray}%
where $\mathbf{k}$ is the normalized unit vertical vector such that {$\mathbf{g}=-\lvert \mathbf{g} \rvert\mathbf{k}$}.
We introduce the dimensionless Richardson number $Ri_T$ defined as
\begin{eqnarray}
   Ri_T &=& \frac{1}{\left(\frac{\rho_p}{\rho_\infty} -1\right)}\frac{T_p-T_\infty}{T_\infty}.
      \label{eq:Ri_heat}
\end{eqnarray}
The sphere is free to move and its motion and rotation are governed by the following equations (with $\boldsymbol{\omega_p}$ the angular velocity of the sphere, $\tilde{P}$ the hydrodynamic pressure without the hydrostatic part and $\mathlarger{\boldsymbol{\tilde{\tau}}}$ the dimensionless viscous stress tensor whose components are $\tilde{\tau}_{ij}=\left[ {\partial\tilde{u}_{i}}/{\partial \tilde{x}_j} +  {\partial\tilde{u}_{j}}/{\partial \tilde{x}_i}\right]/Ga$):
\begin{eqnarray}
   \frac{\rho_p}{\rho_\infty}\frac{\mathrm{d} 
      \mathbf{\tilde{u}_p}}{\mathrm{d}\tilde{t}}
      &=& \frac{6}{\pi} \oint ( \mathlarger{\boldsymbol{\tilde{\tau}}}\mathbf{n}-\tilde{P}\mathbf{n})\mathrm{d}\tilde{S}
      - \mathbf{k} \label{eq-part-translation}\;,\\
   \frac{\rho_p}{\rho_\infty}\frac{\mathrm{d} \boldsymbol{\tilde{\omega}_p}}{\mathrm{d}\tilde{t}}
      &=& \frac{60}{\pi}\oint (\mathbf{\tilde{r}_s} \times \mathlarger{\boldsymbol{\tilde{\tau}}}\mathbf{n})\mathrm{d}\tilde{S} \label{eq-part-rotation}\;.
\end{eqnarray}
%
\begin{figure}[t]
\centering
   \begin{minipage}{2ex}
      \rotatebox{90}{\centerline{$r$}}
   \end{minipage}
   \begin{minipage}{0.7\linewidth}
      \includegraphics[width=1\linewidth]
         {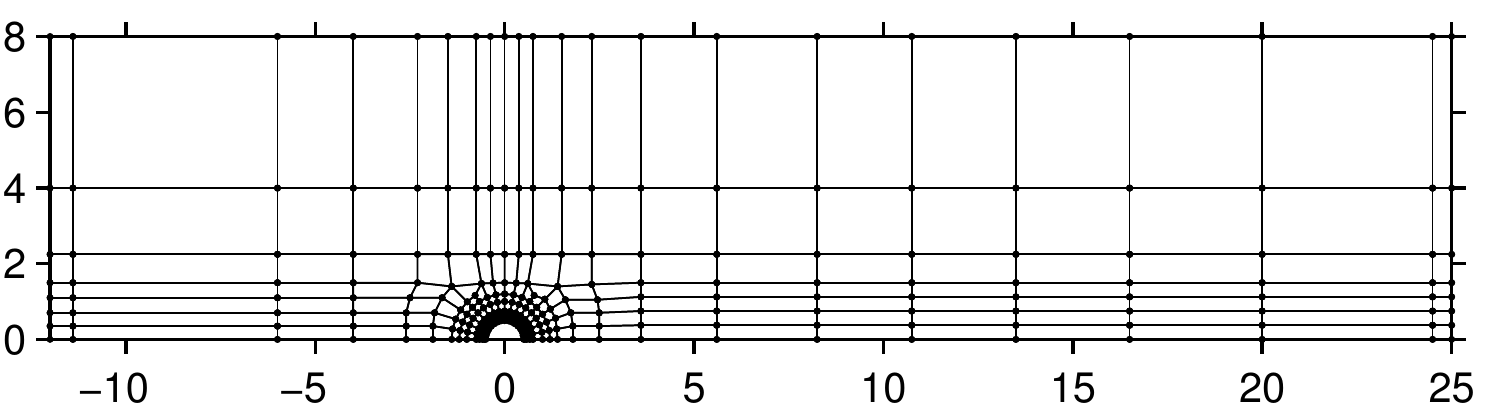}\\
      \centerline{$z$}
   \end{minipage}
   \caption{Visualization of the elements mesh in the $(z,r)$ plane. The polynomial expansion within each element is equal to 6 and we truncate above the 6th or 10th azymuthal Fourier mode.}
   \label{fig:spectral_mesh}
\end{figure}
\newline
Equations \ref{eq-continuity}-\ref{eq-part-rotation}  are solved in a strongly coupled fashion with the method of \cite{Kotouc_IJHMT2008} and \cite{Jenny_JFM2004}, namely a spectral/spectral-element method on a cylindrical domain with axis parallel to the gravity vector, and the sphere center placed at the domain axis. We use a spectral element discretization {in} the axial/radial $(z,r)$ plane and a Fourier decomposition in the azimuthal direction which was shown to be an efficient choice for the simulation of the different transitions to be observed \citep{Ghidersa_JFM2000}. 
We have used the same computational domain as \cite{Zhou_IJMF2015} with an
extension of $37D$ and $8D$ in the axial and radial directions, respectively,  and a discretization into 245 elements with 6 collocation points per element in each direction. Figure \ref{fig:spectral_mesh} shows the structure of this mesh in a $(z,r)$ plane. We used a discretization in the azimuthal direction with a truncation at the 7th Fourier mode for steady cases and 10th for unsteady ones. Tests have been performed with 8 collocation points in each direction and a truncation up to the 16th mode and no significant difference has been observed.

The Prandtl number is taken equal to $Pr=0.72$ throughout this work, meaning that the thermal and mass boundary layers should be resolved with the same quality as the background flow and no extra refinement is required here.

%
A no-slip boundary condition is used at the particle's surface and the incoming flow is set constant. A zero-gradient boundary condition is applied at the outflow face for the velocity field as well as for the temperature. At the lateral boundaries of the domain a no-stress, adiabatic and hermetic Neumann condition is imposed with zero pressure. 

In a meteorological context, the transport of {water vapor} has to be taken into account. It is indeed important to consider the local saturation $S$ which is defined as the ratio between the local vapor partial pressure $e$ and the saturation pressure $e_{sat}$ {(which in our case we assume to be only function of temperature)}. The local partial pressure and vapor concentration are linked according to the relation
\begin{eqnarray}
   n_v &=& e/k_bT 
   \label{eq:nv_definition}\;,
\end{eqnarray}
with $k_b$ the Boltzmann constant.
The saturation will therefore depend on both variation of temperature and vapor concentration. The transport of $n_v$ is governed by an advection-diffusion (equation \ref{eq-mass}) with a diffusion coefficient yielding $Sc=\nu/\mathcal{D}_{m} \sim 0.63$ \citep{young:93} and a non-dimensional number concentration number defined according to equation \ref{eq-nv_NonDim}. The boundary conditions $n_{v,\infty}$ and $n_{v,p}$ are also set constant and uniform both at the inflow and on the sphere surface. The corresponding Richardson number is of the order $\mathcal{O}(10^{-6})$ such that density variations due to vapor transport can be neglected.
Microdroplets will often collide with the falling ice particle, and freeze when they come in contact with the ice particle. This leads to an increase of the sphere's surface temperature due to the release of latent heat during freezing. For this reason we propose to focus here on configurations for which $T_p > T_\infty$, i.e.\ a sphere which is warmer than the surroundings.
{Concerning the boundary conditions on the partial pressure (at infinity and on the particle) we assume that the saturation is equal to unity with respect to the liquid phase for the incoming flow and of unity with respect to ice at the sphere's surface, yielding $n_{v,p}=e_{sat,i}(T_p)/k_bT_p$ and $n_{v,\infty}=e_{sat,w}(T_\infty)/k_bT_\infty$ (with $e_{sat,w}$ the saturation vapor pressure with respect to liquid water and $e_{sat,i}$ the saturation vapor with respect to ice).}
The boundary condition at infinity is chosen to reflect the presence of supercooled droplets in mixed phase clouds. 
Please note that the boundary condition at the particle is capable of describing dry growth as well as wet growth, since the latter occurs at $T_p \approx 0\degree C$, for which $e_{sat,i}=e_{sat,w}$.
We used the correlations of \cite{murphy:05} to set the evolution of the saturation partial pressure $e_{sat,j}$ for the phase $j$ as a function of the temperature.
For more details on this step the reader is referred to appendix (\ref{sec:appendix_MurphyKoop}).

The results in section \ref{sec:passive} and \ref{sec:active} are presented in a dimensionless form and do not include the evolution of the dimensionless concentration number. The difference between the Schmidt and Prandtl numbers are indeed too small to induce significant difference between the non-dimensional temperature and concentration fields.
Contrarily, the results of section \ref{sec:discussion} dealing with the description of the evolution of the saturation field include the solution for the transport equation for $\tilde{n}_v$.\\
%
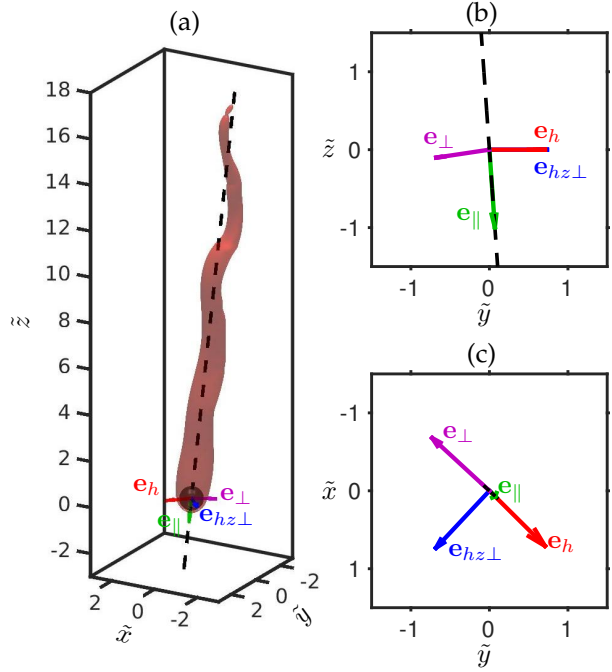
\begin{figure}[t] 
\centering
   \begin{minipage}{2ex}
      \rotatebox{90}{\centerline{$\tilde{z}$}}
   \end{minipage}
   \begin{minipage}{0.25\linewidth}
      \centerline{(a)}
      \begin{overpic}[width=1\linewidth]
         {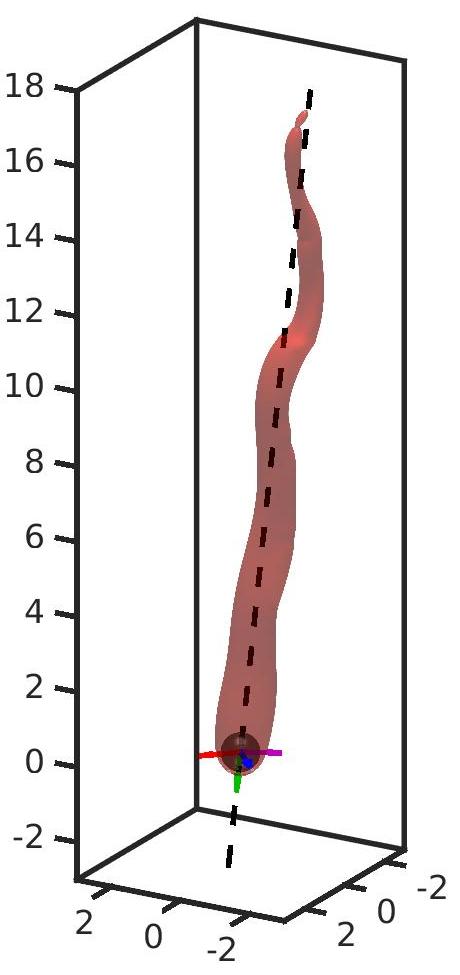}
      \put (12,-1) {\rotatebox{-15}{$\tilde{x}$}}
      \put (41,1.5) {\rotatebox{50}{$\tilde{y}$}}
      \put (19,18) {\color{green}{$\mathbf{e}_\parallel$}}
      \put (15,24) {\color{red}{$\mathbf{e}_h$}}
      \put (29.5,22.5) {\color{magenta}{$\mathbf{e}_\bot$}}
      \put (26,19) {\color{blue}{$\mathbf{e}_{hz\bot}$}}
      \end{overpic}
   \end{minipage}
   \begin{minipage}{4ex}
   \end{minipage}
   \begin{minipage}{0.25\linewidth}
      \centerline{(b)}
      \begin{overpic}[width=0.95\linewidth]
         {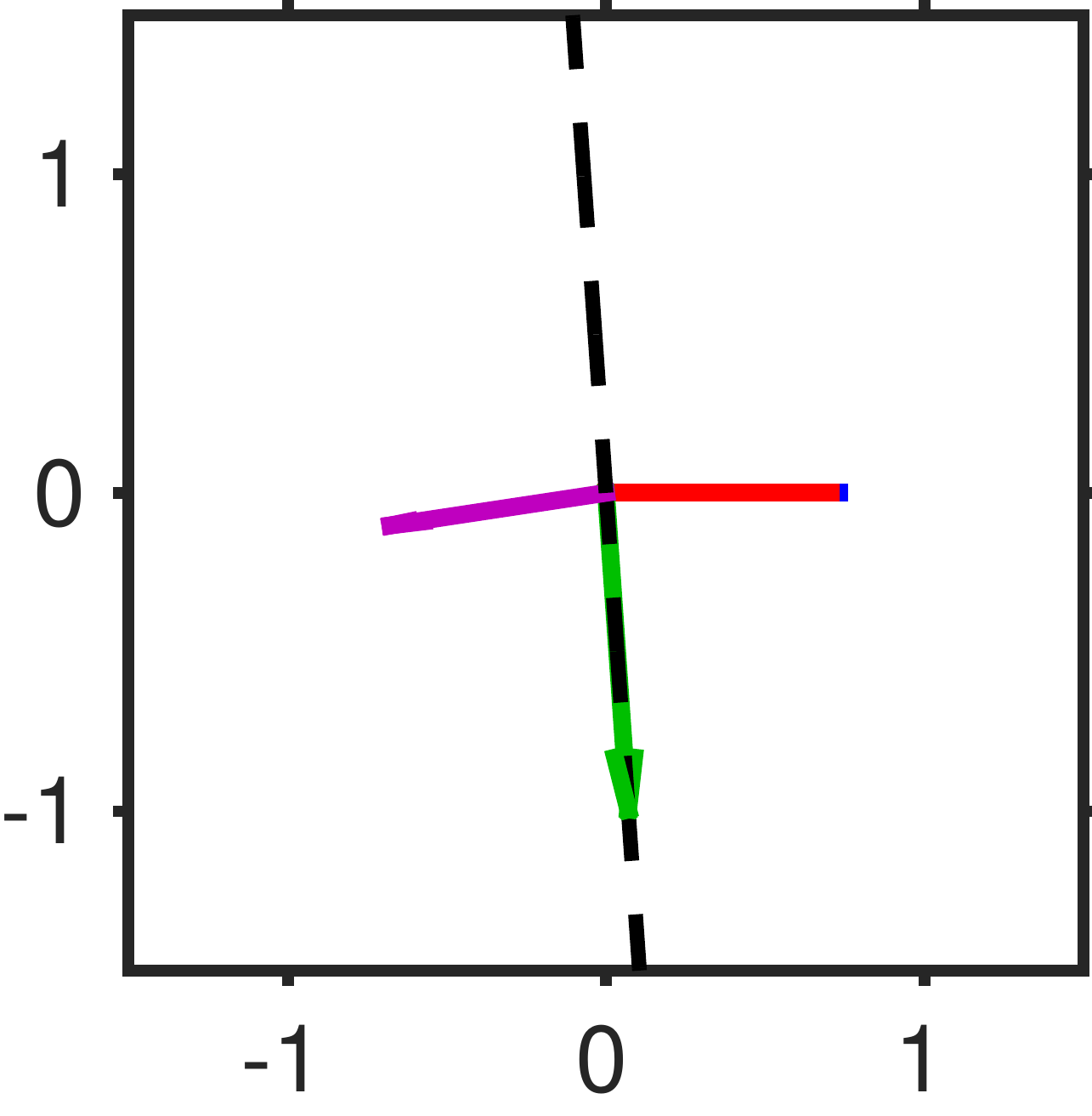} \\
         \put (40,30) {\color{green}{$\mathbf{e}_\parallel$}}
         \put (28,58) {\color{magenta}{$\mathbf{e}_\bot$}}
         \put (68,60) {\color{red}{$\mathbf{e}_h$}}
         \put (68,47) {\color{blue}{$\mathbf{e}_{hz\bot}$}}
         \put (47.5,-7) {$\tilde{y}$}
         \put (-9,53) {$\tilde{z}$}
      \end{overpic}
      \mbox{}\\[2.0ex]
      \centerline{(c)}
      \begin{overpic}[width=0.95\linewidth]
         {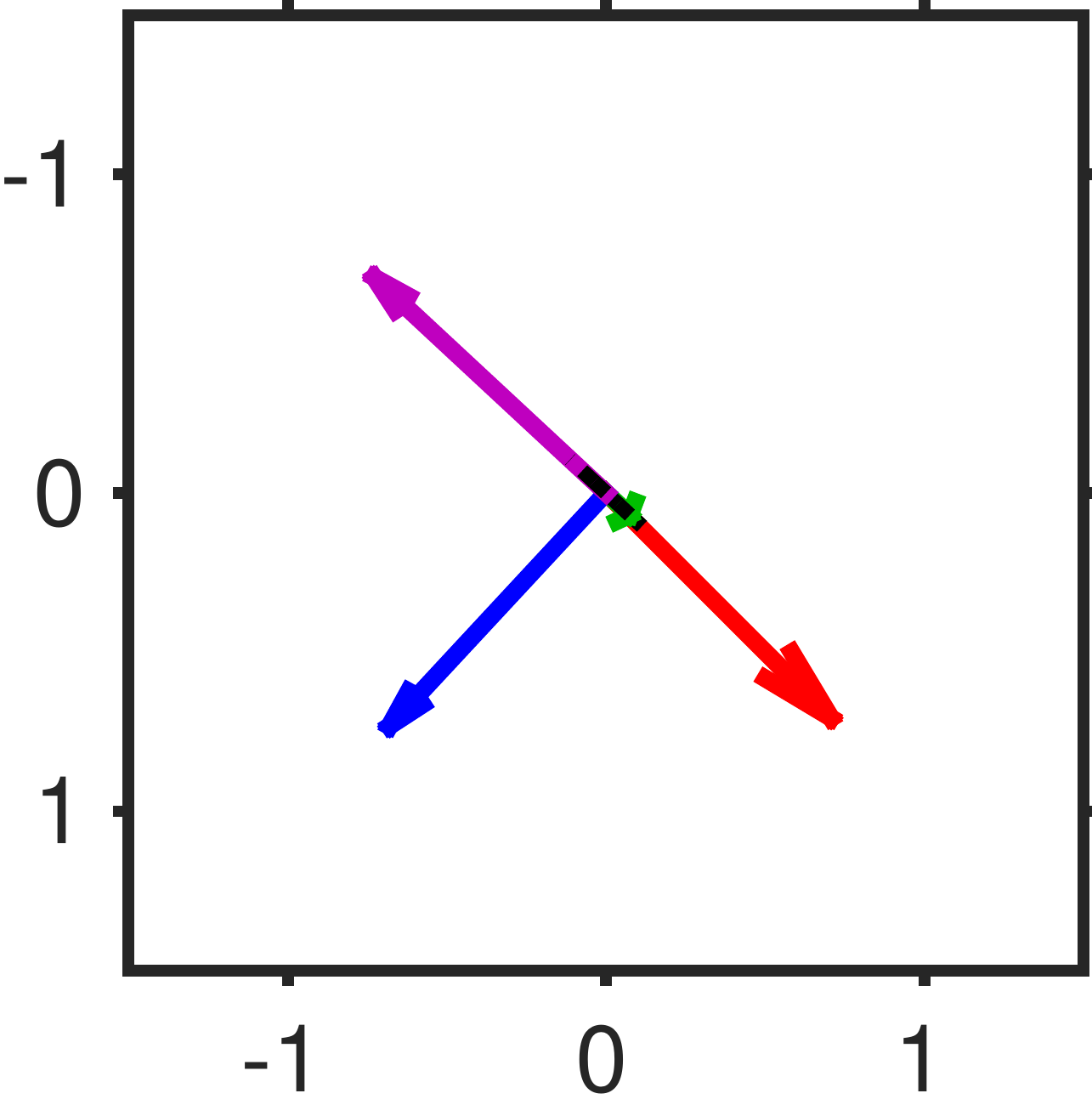} \\
         \put (37,32) {\color{blue}{$\mathbf{e}_{hz\bot}$}}
         \put (35,75) {\color{magenta}{$\mathbf{e}_\bot$}}
         \put (73,35) {\color{red}{$\mathbf{e}_h$}}
         \put (55,55) {\color{green}{$\mathbf{e}_\parallel$}}
         \put (47.5,-7) {$\tilde{y}$}
         \put (-9,53) {$\tilde{x}$}
      \end{overpic}
   \end{minipage}
   \caption{Sketches of the orientation of the unit vectors introduced for the analysis in the oblique coordinate system: (a) Position of the sphere and isocontour of the temperature for an oscillating oblique case and orientation of the vectors $\mathbf{e}_\parallel,\mathbf{e}_h,\mathbf{e}_{hz\bot},\mathbf{e}_\bot$ for this specific snapshot, with a projection of the vectors in a vertical plane (b) and a horizontal plane (c). The dashed lines pass through the center of the sphere and follow the orientation of $\mathbf{e}_\parallel$.} 
   \label{fig:sketch_wake_base}
\end{figure}
A part of the analysis in the wake of the sphere is made with the help of a coordinate system that is not only centered at position of the particle, but also follows the orientation of its settling velocity, as previously used by \cite{Uhlmann_IJMF2014}. For this purpose we first introduce the velocity of the particle with the Cartesian coordinates $\mathbf{u}_{p}=(u_{px},u_{py},u_{pz})^T$ and the corresponding unit vector 
\begin{eqnarray} 
   \mathbf{e}_\parallel &=& \mathbf{u}_{p}/\lVert \mathbf{u}_{p} \rVert
   \;.
\end{eqnarray}
We also introduce the projection of the relative velocity onto the $(x,y)$ plane $\mathbf{u}_h$ and its associated unit vector $\mathbf{e}_h$, namely
\begin{eqnarray}
   \mathbf{u}_h &=& (u_{px},u_{py},0)^T \;, \\
   \mathbf{e}_h &=& \mathbf{u}_{h}/\lVert \mathbf{u}_{h} \rVert \;,
\end{eqnarray}
then we define another horizontal vector $\mathbf{e}_{hz\bot}=\mathbf{e}_z \times \mathbf{e}_h$, and finally a unit vector which is now perpendicular to both $\mathbf{e}_\parallel$ and $\mathbf{e}_{hz\bot}$ and which is defined by
\begin{eqnarray}
   \mathbf{e}_\bot=\mathbf{e}_{hz\bot} \times \mathbf{e}_\parallel \;.
\end{eqnarray}
An example of the structure of this new Cartesian coordinate system is shown in figure \ref{fig:sketch_wake_base} for a configuration in the oscillating oblique regime. 
{We will preferentially consider this coordinate system to describe the structure of the wake.
We} denote the position along the $\mathbf{e}_\parallel$ axis as $\tilde{z}_\parallel$ and along the $\mathbf{e}_\bot$ axis as {$\tilde{x}_\bot$. A tilted cylindrical coordinate system with $ \tilde{z}_\parallel$ as a main axis will also be used, and we denote the corresponding radial position by $\tilde{r}_\bot$,} as will be used in section \ref{sec:passive}.
Statistics which imply time averaging of the temperature or velocity field, presented in this coordinate system, are performed as follows: the instantaneous velocity and temperature fields are first projected onto this instantaneous inclined frame for each snapshot and then averaged over time. 
%

%
We focus here on a density ratio $\rho_p/{\rho_\infty}=10$ and investigate Galileo numbers ranging from 10 to 300 such that all different regimes to be observed with such density ratio will be covered. A special emphasis is given on $Ga=150$, $170$, $200$ and $300$ in order to describe the specificities of each regime. In the case of a falling graupel the density ratio would typically range between 330 and 650 with a diameter between 0.5 and 6mm, leading then to Galileo numbers ranging from 60 and 2800 \citep{pruppacher:10}. In the present study we propose to focus on intermediate Galileo numbers in order to test the influence of the wake regime on the structure of the scalar field.
{It would be necessary in the future to get also more insight on the influence of turbulence or collective effects but this would go beyond the scope of the current paper and we neglect those effects in the current study.}
{We} set {first} $\rho_p/{\rho_\infty}=10$ and not {$\rho_p/\rho_\infty=600$} as it has the advantage of reducing the characteristic timescale of the particle and with it the duration of the simulations, but is large enough to behave in a similar way as $\rho_p/{\rho_\infty} \simeq 600$. A real density ratio of 600 can be seen as an intermediate point between $\rho_p/{\rho_\infty}=10$ and a fixed sphere {(i.e.\ $\rho_p/\rho_\infty=\infty$)}.
{In order to test the necessity to reproduce the influence of the mobility in a meteorological context we investigated the density ratios $\rho_p/\rho_\infty=\lbrace 1.5;\;10;\;\infty\rbrace$}.

We focus on ambient temperature $T_\infty$ ranging between -15$\degree C$ and
-5$\degree C$ and sphere temperature ranging between -10$\degree C$ and
0$\degree C$, the resulting thermal Richardson number for this system is
therefore estimated to vary between $0.5 \times 10^{-4}$ and $1.75 \times
10^{-4}$. The corresponding Grashof number $Gr=RiGa^2$ would range between $Gr=\mathcal{O}(1)$ and $Gr=\mathcal{O}(10^3)$.
The highest Galileo number considered in the current study (300) combined with the highest expected Richardson number value yield $Gr \sim 15$. Considering the real values of $Ri_T$, with this Galileo number range, would not yield any visible effect on the flow. For this reason we propose to explore larger ranges and consider $Ri_T=(0;\;0.001;\;0.05;\;0.1)$, in order to cover larger values of the Grashof number.
%
%

%
\section{The case of passive scalar transport} \label{sec:passive}

\begin{figure}[h]
\centering
   \begin{minipage}{1.0\linewidth}
   \centering
      \includegraphics[width=0.75\linewidth]
         {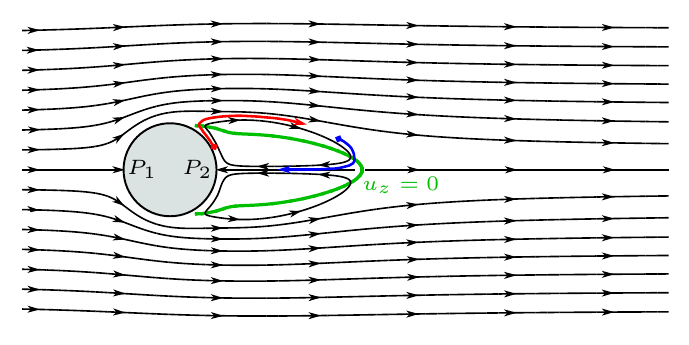}\\
   \end{minipage}
   \caption{{Sketch of the typical streamlines and flow regions for a steady axisymmetric configuration. The points $P1$ and $P2$ refer to the stagnation points, the limit of the recirculation region is represented in green by the isocontour $u_z=0$. The red line highlights the transport of warm fluid from the rear of the sphere to the shear region, and the blue line the transport of cold fluid from the shear region to the rear of the sphere.}}
   \label{fig:sketch_recirculation}
\end{figure}

We propose in this section to focus on the structure of the scalar field in the configuration where no buoyancy effects are accounted for ($Ri_T=0$), for a Prandtl number $Pr=0.72$, with the aim of providing a first description on the influence of the settling regime as well as the density ratio on the structure of velocity and the scalar field.
Figure {\ref{fig:sketch_regimes}} gives a visual impression of the velocity and temperature field for different Galileo numbers and for $\rho_p/\rho_\infty=10$. The main characteristics of the flow regime are easily recognizable on both velocity and temperature field: $Ga=150$ features an axisymmetric wake while for $Ga=170$ and $Ga=200$ the wake is oblique, with oscillations in space only visible for $Ga=200$. The structure of the velocity and thermal wakes are chaotic for $Ga=300$. 
{Figure \ref{fig:sketch_recirculation} gives also a visual impression of the different zones of the flow.}
%
\subsection{Recirculation length}
%
\begin{figure}[p]
   \centering
   \begin{minipage}{2ex}
      \rotatebox{90}{\centerline{$\tilde{z}_\parallel$}}
   \end{minipage}
   \begin{minipage}{0.2\linewidth}
      \centerline{(a)}
      \begin{overpic}[width=1\linewidth]
         {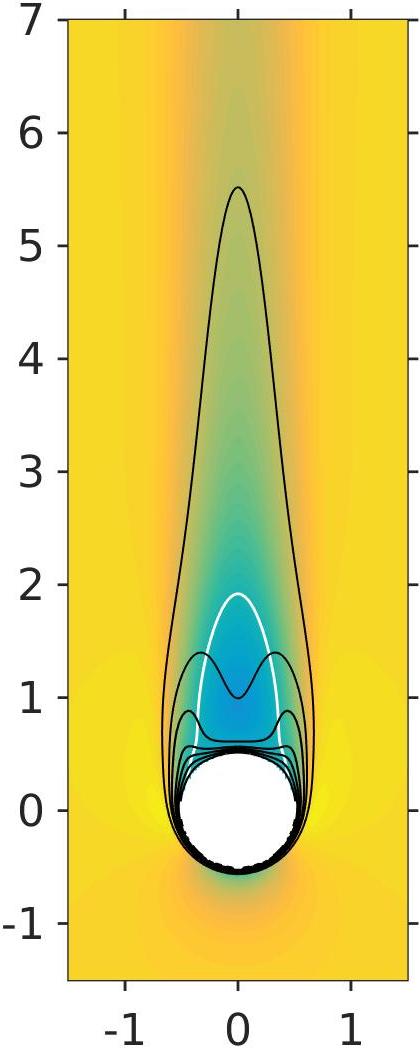}
         \put(13,93){$Ga=150$}
      \end{overpic}
      \centerline{$\tilde{x}_{\perp}$}\\
   \end{minipage}
   \begin{minipage}{0.2\linewidth}
      \centerline{(b)}
      \begin{overpic}[width=1\linewidth]
         {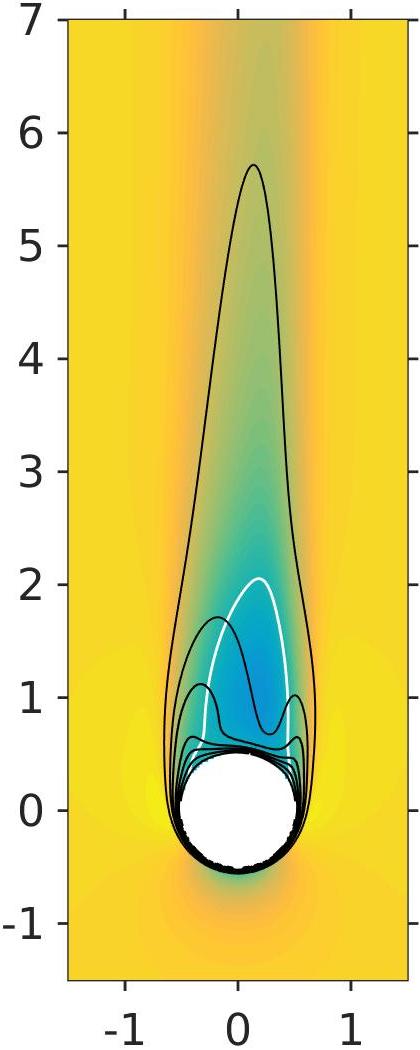}
         \put(13,93){$Ga=170$}
      \end{overpic}
      \centerline{$\tilde{x}_{\perp}$}\\
   \end{minipage}
   \begin{minipage}{0.2\linewidth}
      \centerline{(c)}
      \begin{overpic}[width=1\linewidth]
         {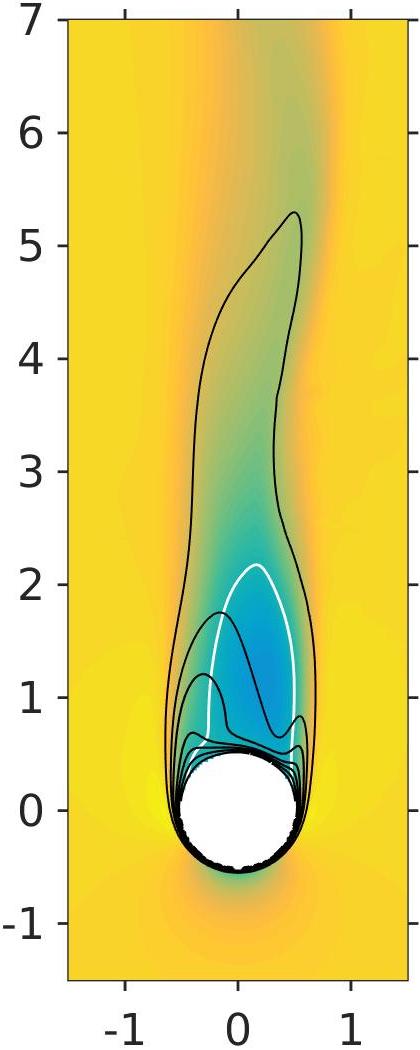}
         \put(13,93){$Ga=200$}
      \end{overpic}
      \centerline{$\tilde{x}_{\perp}$}\\
   \end{minipage}
   \begin{minipage}{0.2825\linewidth}
      \centerline{(d)}
      \begin{overpic}[width=1\linewidth]
         {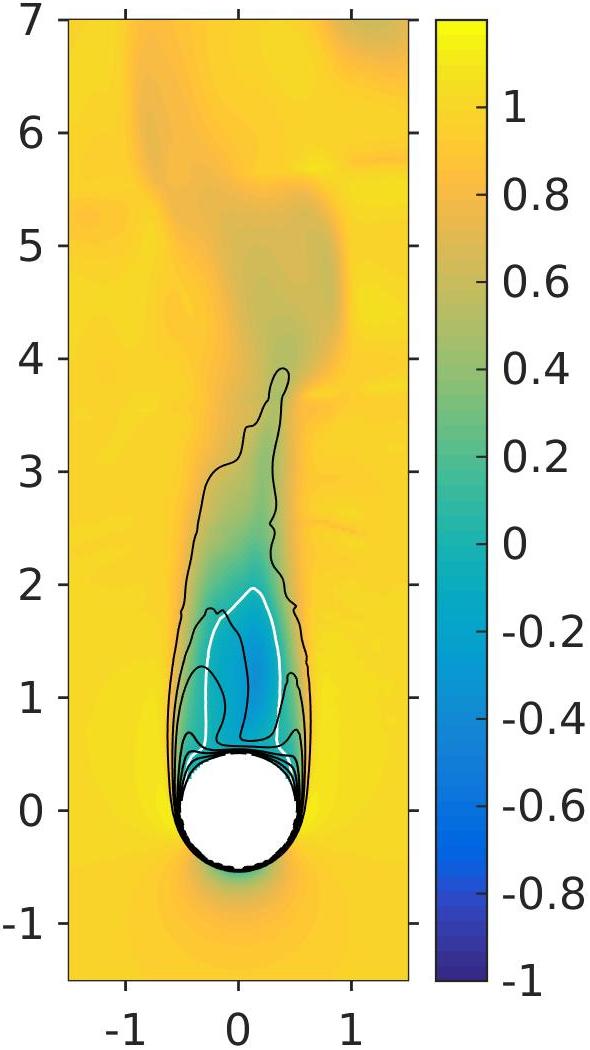}
         \put(13,93){$Ga=300$}
      \end{overpic}
      \centerline{$\tilde{x}_{\perp}$}\\
   \end{minipage}
   \mbox{}\\
%
   \begin{minipage}{2ex}
      \rotatebox{90}{\centerline{$\tilde{z}_\parallel$}}
   \end{minipage}
   \begin{minipage}{0.2\linewidth}
      \centerline{(e)}
      \begin{overpic}[width=1\linewidth]
         {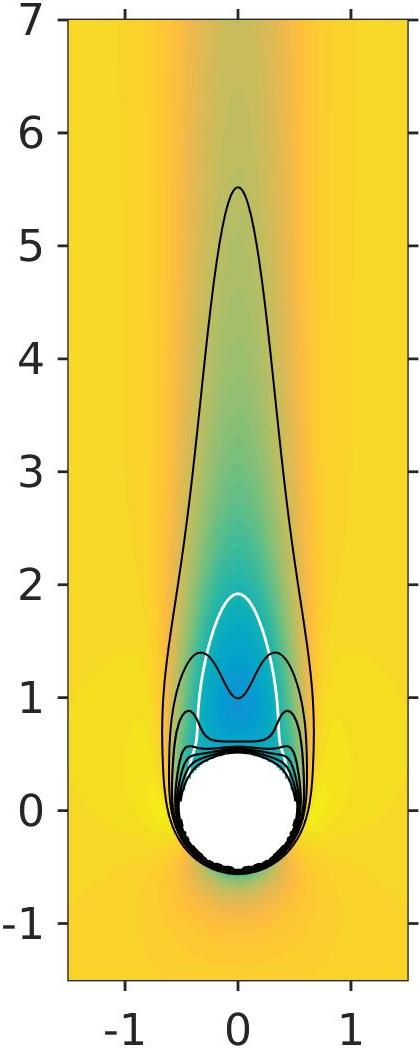}
         \put(13,93){$Ga=150$}
      \end{overpic}
      \centerline{$\tilde{x}_{hz\perp}$}\\
   \end{minipage}
   \begin{minipage}{0.2\linewidth}
      \centerline{(f)}
      \begin{overpic}[width=1\linewidth]
         {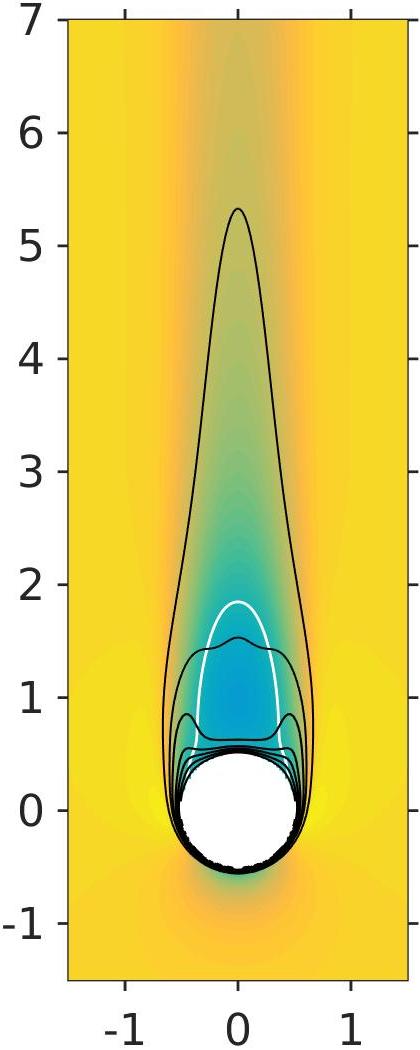}
         \put(13,93){$Ga=170$}
      \end{overpic}
      \centerline{$\tilde{x}_{hz\perp}$}\\
   \end{minipage}
   \begin{minipage}{0.2\linewidth}
      \centerline{(g)}
      \begin{overpic}[width=1\linewidth]
         {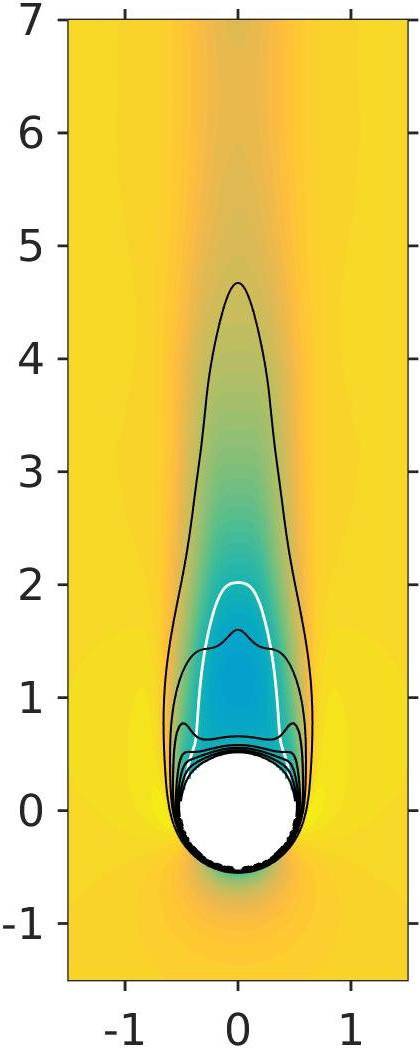}
         \put(13,93){$Ga=200$}
      \end{overpic}
      \centerline{$\tilde{x}_{hz\perp}$}\\
   \end{minipage}
   \begin{minipage}{0.2825\linewidth}
      \centerline{(h)}
      \begin{overpic}[width=1\linewidth]
         {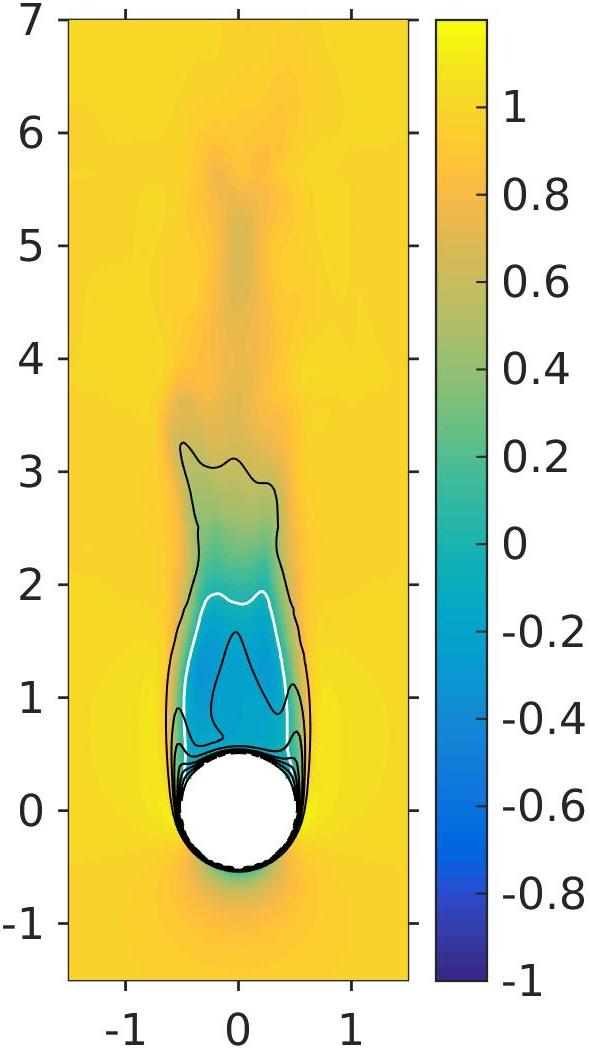}
         \put(13,93){$Ga=300$}
      \end{overpic}
      \centerline{$\tilde{x}_{hz\perp}$}\\
   \end{minipage}
   \caption{{Instantaneous vertical velocity (in colors) and temperature isocontours (in black) projected upon a tilted plane for different Galileo numbers (a-e) $Ga=150$, (b-f) $Ga=170$, (c-g) $Ga=200$, (d-h) $Ga=300$. For the velocity the isocontour $\tilde{u}_z^{*}=0$ is indicated in white. For the temperature the contours represent the temperature levels $\lbrace 0.2,\; 0.35,\; 0.45,\; 0.6,\; 0.7,\; 0.8,\; 0.9\rbrace$.}}
   \label{fig:passive_2Dvisus}
\end{figure}
%
We start our description with the evolution of the extent of the recirculation region behind the sphere as well as the evolution of the mean transfer coefficients as they provide important information for the following description.
{Figure \ref{fig:passive_2Dvisus} shows the evolution of the vertical component of the instantaneous velocity as well as the instantaneous temperature field in planes containing the tilted axis $x_\perp$ and $x_{hz\perp}$ for different Galileo numbers.
It highlights the influence of the recirculation region on the temperature distribution in the wake. This region is indeed characterized by a toroidal vortex whose projection onto two perpendicular planes represents counter-rotating vortices that affect the temperature as sketched on figure \ref{fig:sketch_recirculation}: cold fluid is transported from the outer shear region towards the rear stagnation point. Conversely warm fluid is transported from the back of the sphere to the shear region, leading to local maximum of temperature in the core of the vortices and local minimum between them.
The influence of the symmetry breaking is also visible. If one considers for
instance the steady oblique case, projected into the
$(x_\perp,z_\parallel)$ plane ($Ga=170$ in figure \ref{fig:passive_2Dvisus}b,f),
the recirculation is mainly located on the side $x_\perp>0$ and therefore cold ambient fluid is primarily entrained on this side, making the temperature
globally lower in this region of the wake than on the side $x_\perp<0$.
}
\begin{figure}[t]
\centering
   \begin{minipage}{2ex}
      \rotatebox{90}{\centerline{$L_r$}}
   \end{minipage}
   \begin{minipage}{0.95\linewidth}
      \includegraphics[width=0.95\linewidth]
         {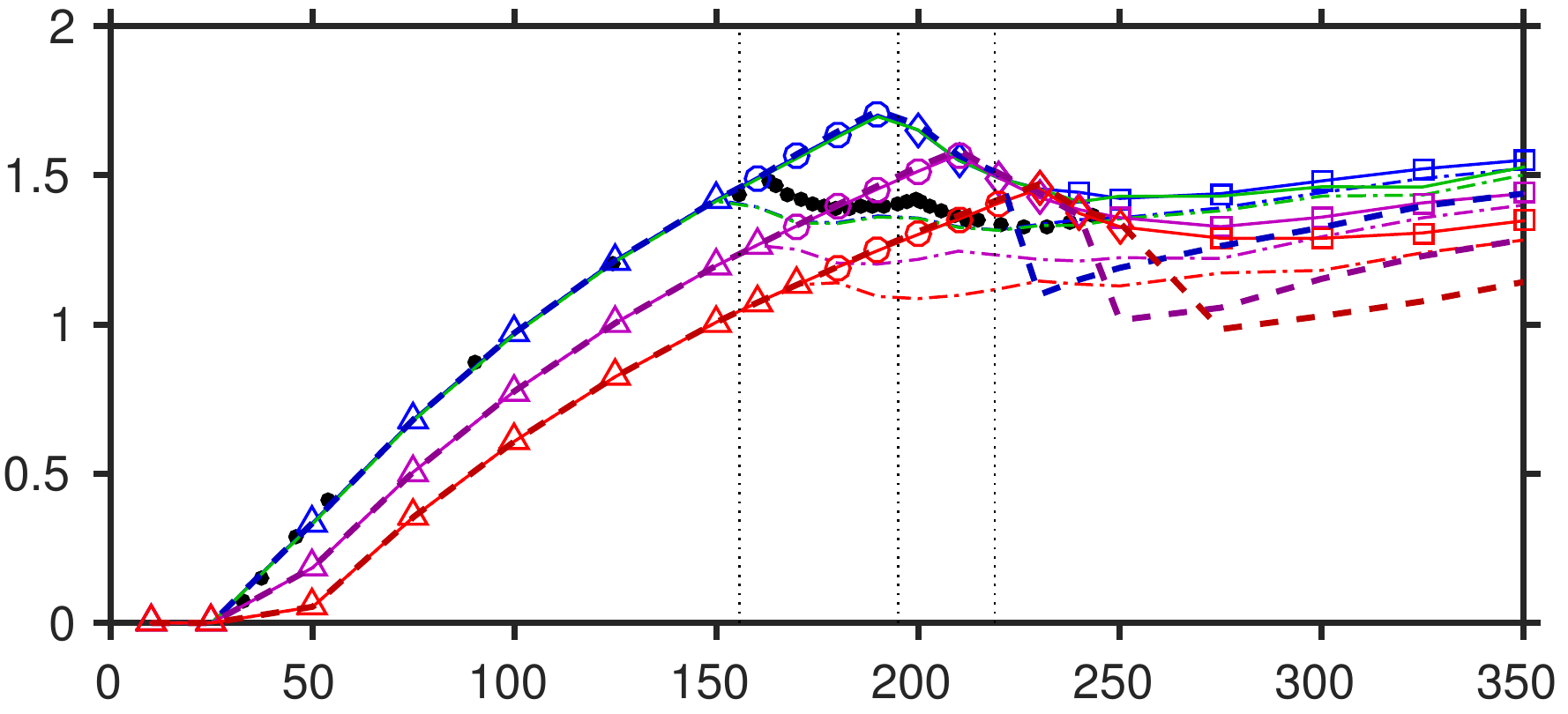}\\
      \centerline{$Ga$ , $\tilde{Ga}$}
   \end{minipage}
   \caption{Evolution of the recirculation length with the Galileo number, for different Richardson numbers $Ri_T$ and for $\rho_p/{\rho_\infty}=10$. Markers indicate the regime for each Richardson number ($\mathlarger{\mathlarger{\vartriangle}}$ steady axisymmetric, $\mathlarger{\mathlarger{\mathlarger{\circ}}}$ steady oblique, $\lozenge$ oscillating oblique, $\mathsmaller{\square}$ chaotic) {and the vertical dotted lines give the limits between the regimes at $\rho_p/\rho_\infty=10$}.
   Colorstyle: 
   $\color{blue}{\solidthick}$ $Ri_T=0$,
   $\color{green}{\solidthick}$ $Ri_T=0.001$,
   $\color{magenta}{\solidthick}$ $Ri_T=0.05$,
   $\color{red}{\solidthick}$ $Ri_T=0.1$, dashed lines correspond to the original coordinates and solid lines to the oblique coordinate system (in this case the time averaging operator is applied after projection on the oblique coordinate system). {Dash-dotted lines correspond to the recirculation length obtained if defined as the distance from the particle to the point on the axis for which the velocity crosses zero (computed on the tilted mesh). The black dots represents the data from \cite{Bouchet_EJM2006} for a fixed sphere, and for which we used a surrogate Galileo number $\tilde{Ga}$ defined by $\tilde{Ga}=\sqrt{3C_D/4}Re$.}
}
   \label{fig:recirculation_length}
\end{figure}
{Figure \ref{fig:recirculation_length} shows} the recirculation length which is defined as the largest distance to the sphere surface for which the time averaged streamwise velocity is equal to zero, estimated on the three dimensional contours of the mean velocity field (with this definition the position of the corresponding point can thus be located away from the wake axis {as can be clearly seen in figure \ref{fig:passive_2Dvisus}}).
We make the distinction here between two time averaged flow fields: The velocity field is either averaged in the original coordinate system $(x,y,z)$ or after rotation to the oblique coordinate system $(\mathbf{e}_\parallel,\mathbf{e}_{hz\bot},\mathbf{e}_\bot)$.
It shows that the recirculation length first increases with the Galileo number until unsteadiness is reached. It then decreases in the {oblique oscillating} regime for further increasing in the chaotic regime.
The increase of $L_r$ with $Ga$ is consistent with the observation, for a fixed sphere, of \cite{magnaudet:95} and \cite{Bouchet_EJM2006} for the steady axisymmetric regime, but they observe a first decay once the steady non-axisymmetric flow develops while here the decay starts with the onset of unsteadiness. This discrepancy is most probably due to estimation of the recirculation length not on the wake axis but as the largest distance of the isocontour $\tilde{u}_z=0$ to the sphere surface.{ We indeed computed the length on the axis and found similar trend as in \cite{Bouchet_EJM2006}.}
{The decrease of $L_r$ of the time-averaged flow for the oblique oscillating case is a signature of the vortex shedding and the oscillation of the counter-rotating vortices observed in the plane $(x_\perp,z_\parallel)$.}
For more clarity we omitted the configurations featuring $\rho_p/\rho_\infty=1.5$ or a fixed sphere on the figure, for which we observed either a negligible influence on $L_r$ (fixed sphere), or a small decrease of $L_r$ ($\rho_p/\rho_\infty=1.5$).
%
%
%
\subsection{Transfer coefficient}
We use the same definition of the local Nusselt number as \cite{Kotouc_IJHMT2008}, namely
\begin{eqnarray}
   Nu_{loc}&=&-2\lambda \left( \frac{\partial T}{\partial n} \right)\frac{\pi D^2}{\dot{Q}_{cond}}\;,
\end{eqnarray}
where $\dot{Q}_{cond}$ is the theoretical purely conductive heat flux:
\begin{eqnarray}
   \dot{Q}_{cond} &=& 2\pi\lambda(T_p-T_\infty)D\;,
\end{eqnarray}
with $\lambda$ the thermal conductivity of the fluid.
The mean Nusselt number is then defined as the integral of $Nu_{loc}$ over the surface of the sphere, which is time-averaged in unsteady flows:
\begin{eqnarray}
   Nu=\frac{4}{\pi D^2}\left<\int_{\mathcal{S}}Nu_{loc}\mathrm{d}S\right>_t,
\end{eqnarray}
where the operator $\left<\;.\;\right>_t$ refers to time averaging.
\begin{figure}[t]
   \begin{minipage}{2ex}
      \rotatebox{90}{\centerline{$Nu$}}
   \end{minipage}
   \begin{minipage}{0.45\linewidth}
      \centerline{(a)}
      \includegraphics[width=0.95\linewidth]
         {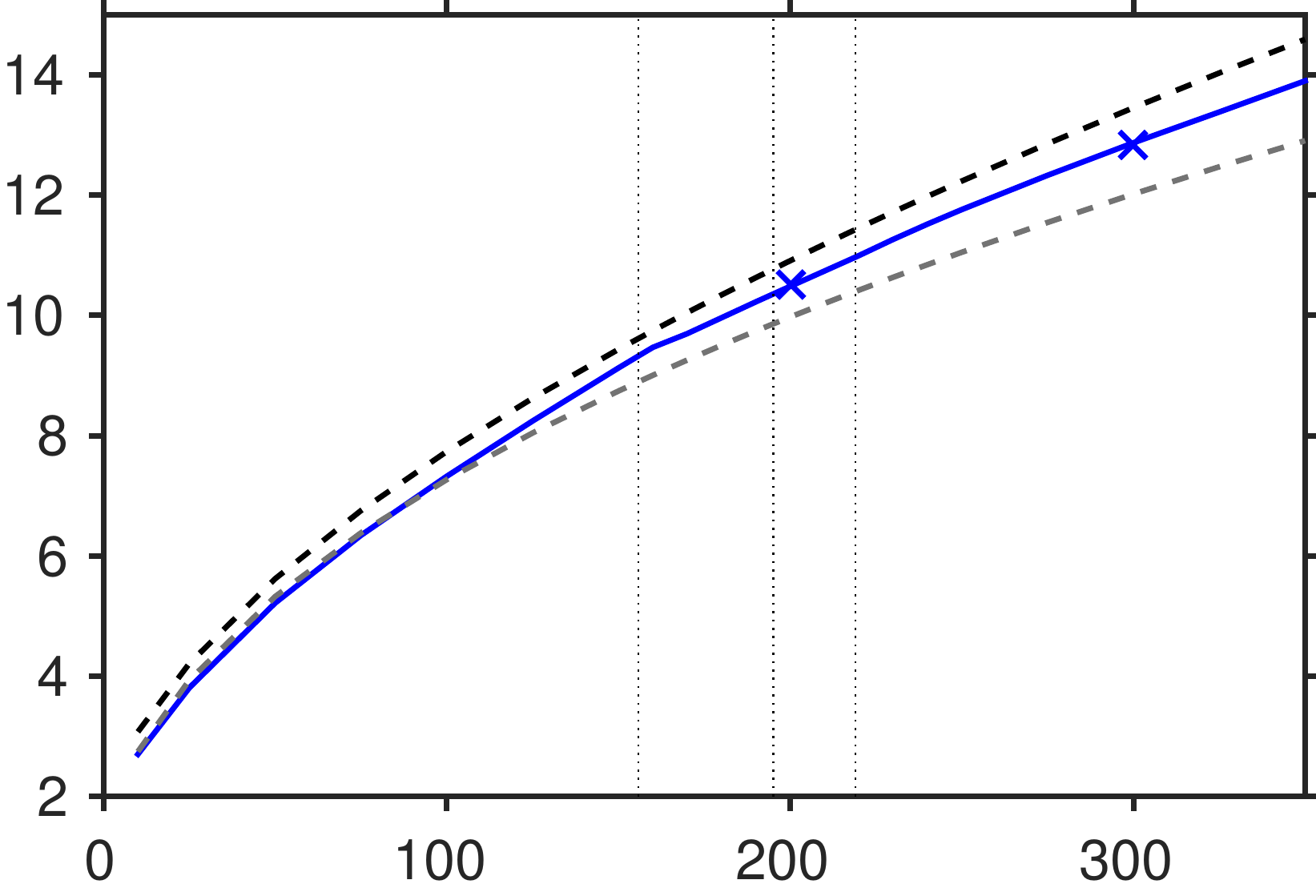}
      \centerline{$Ga$}
   \end{minipage}
      \begin{minipage}{2ex}
      \rotatebox{90}{\centerline{{$Re(t)/\langle Re \rangle_t$ , $Nu(t)/\langle Nu \rangle_t$}}}
   \end{minipage}
   \begin{minipage}{0.45\linewidth}
      \centerline{(b)}
      \includegraphics[width=0.95\linewidth]
         {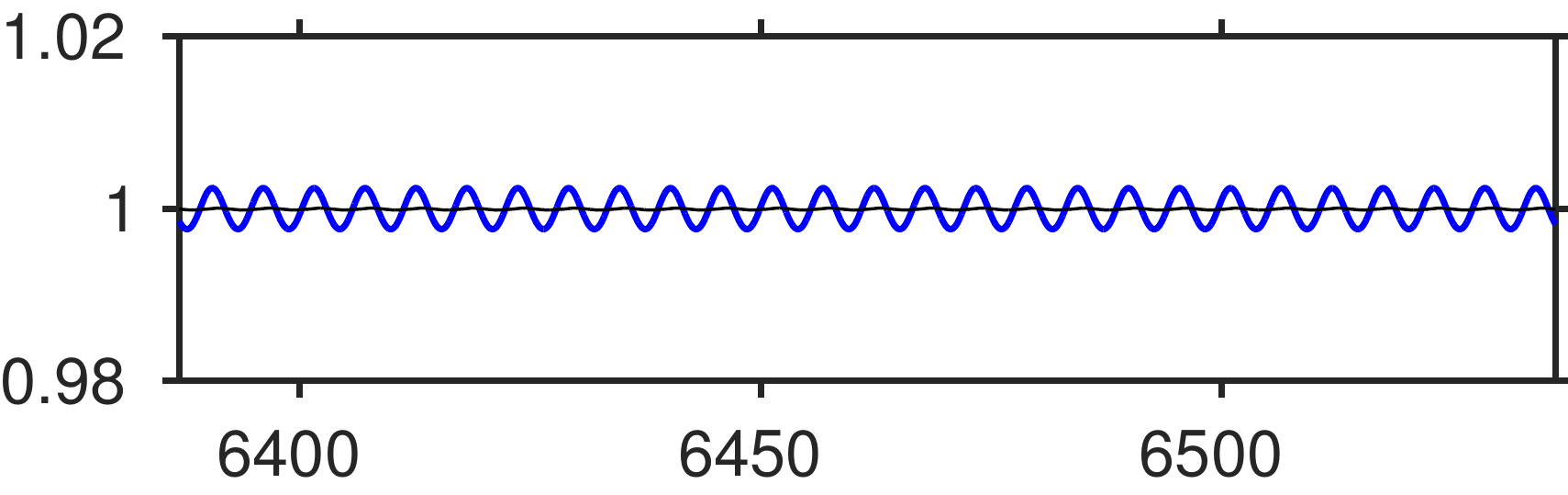}\\
      \includegraphics[width=0.95\linewidth]
         {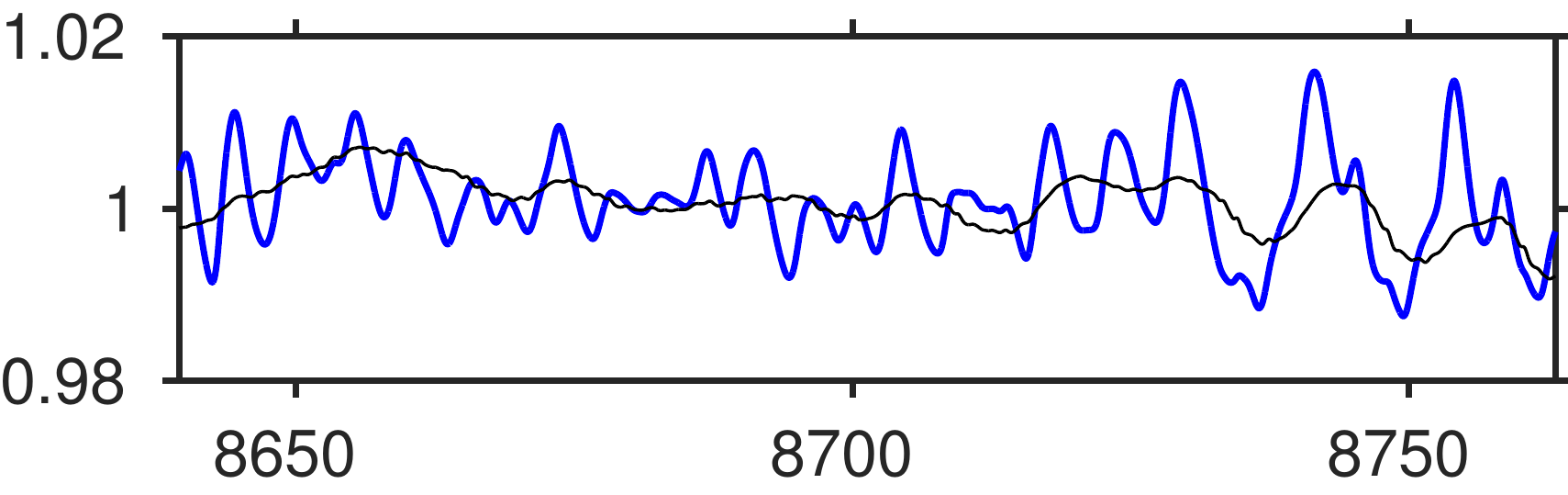}\\
      \centerline{$\tilde{t}$}
   \end{minipage}\\[2ex]
   \caption{(a) Evolution of the time averaged Nusselt number as a function of the Galileo number for $Pr=0.72$. Solid line and symbols depict results from the simulations and dashed lines correspond to correlations $Nu_{c1}$ and $Nu_{c2}$ where the dependence with the Reynolds number has been {taken for $\langle Re_T \rangle_t$ and} replaced by a dependence with $Ga$ according to the relation {$\sqrt{3C_D/4}\langle Re_T \rangle_t=Ga$} (as proposed in \cite{Clift_1978}).
Linestyle: 
$\color{blue}{\solidthick}$ $Ri_T=0$, $\rho_p/{\rho_\infty}=10$, 
$\color{blue}{\boldsymbol{\times}}$ $Ri_T=0$, $\rho_p/{\rho_\infty}=1.5$,  
$\solidshort\;\solidshort$ $Nu_{c1}$,
$\color{grey}{\solidshort\;\solidshort}$ $Nu_{c2}$.
(b) Time evolution of the Nusselt number (colored lines) and Reynolds number (black lines) for $Ga=200$ (top) and $Ga=300$ (bottom) for $Pr=0.72$ and $\rho_p/\rho_\infty=10$.
}
   \label{fig:transfer_coefficients}
\end{figure}
Figure \ref{fig:transfer_coefficients} gives the evolution of the mean heat transfer coefficient as a function of the Galileo number and a comparison with the correlation of Ranz and Marshall \citep{ranz:52a}, namely:
\begin{eqnarray}
   Nu_{c1} &=& 2+0.6(Re)^{1/2}(Pr)^{1/3} \label{eq:corr_RanzMarshall}\;,
\end{eqnarray}
which can be used for Reynolds number up to $Re=5\times10^4$, as well as another correlation which has been validated against direct numerical simulation for $1<Re<400$ and $0.25<Pr<100$ \citep{Clift_1978}:
\begin{eqnarray}
   Nu_{c2} &=& 1+\left( 1+\frac{1}{RePr}\right)^{1/3}Re^{0.41}Pr^{1/3}
   \;.
       \label{eq:Nu_corr2}
\end{eqnarray}
Applied to this study, the correlations would refer to a time-averaged Reynolds number, defined in two ways. The first one accounts for the three components of the sphere's velocity $Re(t)={\left| \mathbf{u}_p\right|D}/{\nu}$, while the other accounts only for its vertical component $Re_T(t)={\left| u_{p,z}\right|D}/{\nu}$.
Using the balance of external and hydrodynamical forces one can show that the mean Reynolds number $\left<Re_T\right>_t$ is linked to the Galileo number according to the relation $\sqrt{3C_D/4}\left<Re_T\right>_t\sim Ga$ \citep{Jenny_JFM2004}, so that the mean Reynolds number can be estimated as a function of the Galileo number once assuming that the drag coefficient follows a known correlation of \cite{Schiller_1935} {with a difference of at most one percent}. 
The combination of the correlations provides a good estimate of the Nusselt number as a function of the Galileo number,
with almost no difference observed between the values computed with the actual time averaged Reynolds number or the with the one obtained from the drag correlation (figure omitted).
The standard deviation of those coefficients are larger for the chaotic regimes than the oblique oscillating one with a standard deviation of approximately one percent for the chaotic regimes as can be for example observed on the time variations of $Nu$ and $Re$ at $\rho_p/\rho_\infty=10$ on figure \ref{fig:transfer_coefficients}.
Interestingly, the time averaged value of the Nusselt coefficient shows little dependence on the density ratio, which is in accordance with the formulation of the correlations and the relation between the drag coefficient and the Reynolds number.
%
\subsection{Structure of the scalar field and influence of the density ratio}
\begin{figure}[h!]
   \begin{minipage}{4ex}
      \rotatebox{90}{\centerline{$u_d(\tilde{r}_\bot=0,\tilde{z}_\parallel)$}}
   \end{minipage}
   \begin{minipage}{0.45\linewidth}
      \centerline{(a)}
      \begin{overpic}[width=1\linewidth]
         {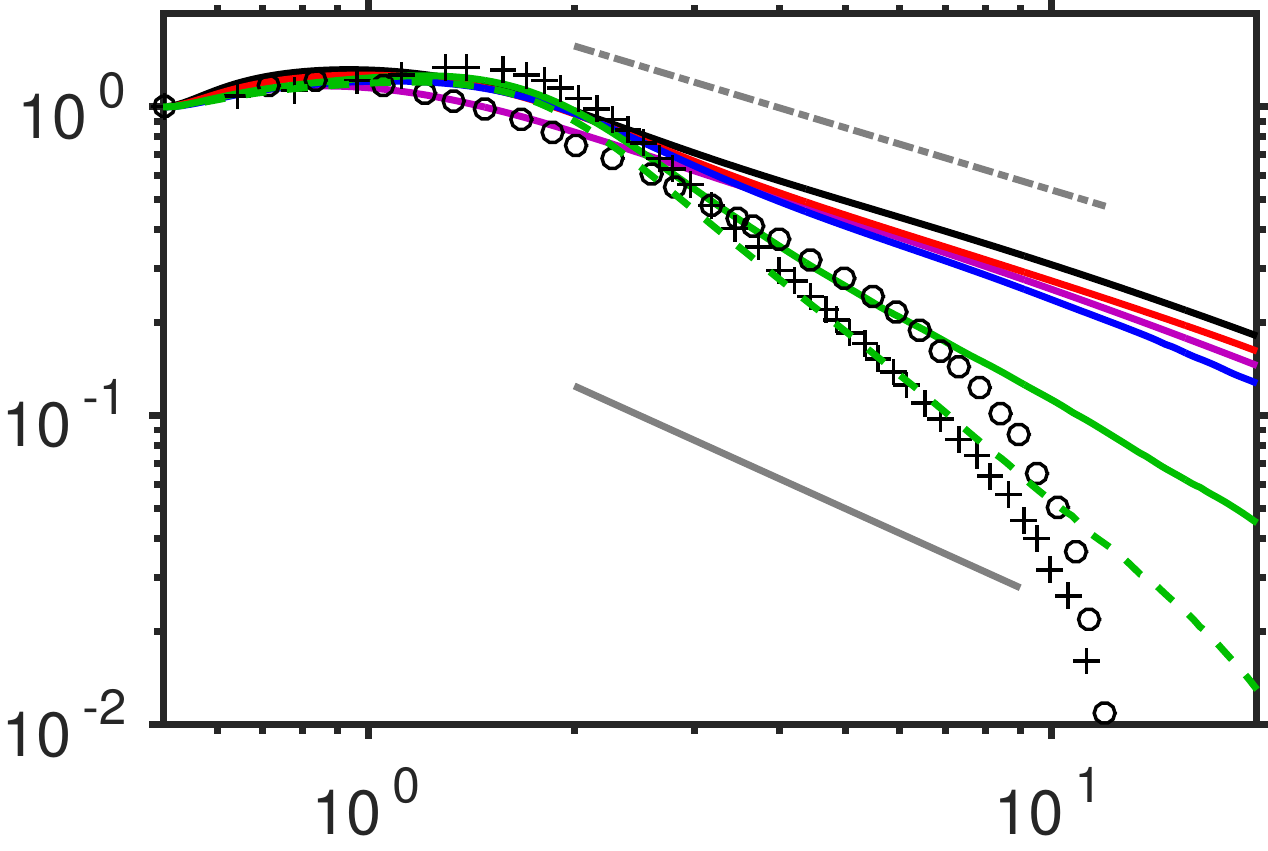}\\[-1.5ex]
         \put(70,58) {$z^{-2/3}$}
         \put(58,22) {$z^{-1}$}
      \end{overpic}
     \centerline{$\tilde{z}_\parallel$}
   \end{minipage}
   \begin{minipage}{2ex}
      \mbox{}
   \end{minipage}
   \begin{minipage}{4ex}
      \rotatebox{90}{\centerline{$\left<\tilde{T}\right>_t(\tilde{r}_\bot=0,\tilde{z}_\parallel)$}}
   \end{minipage}
   \begin{minipage}{0.45\linewidth}
      \centerline{(b)}
      \begin{overpic}[width=1\linewidth]
         {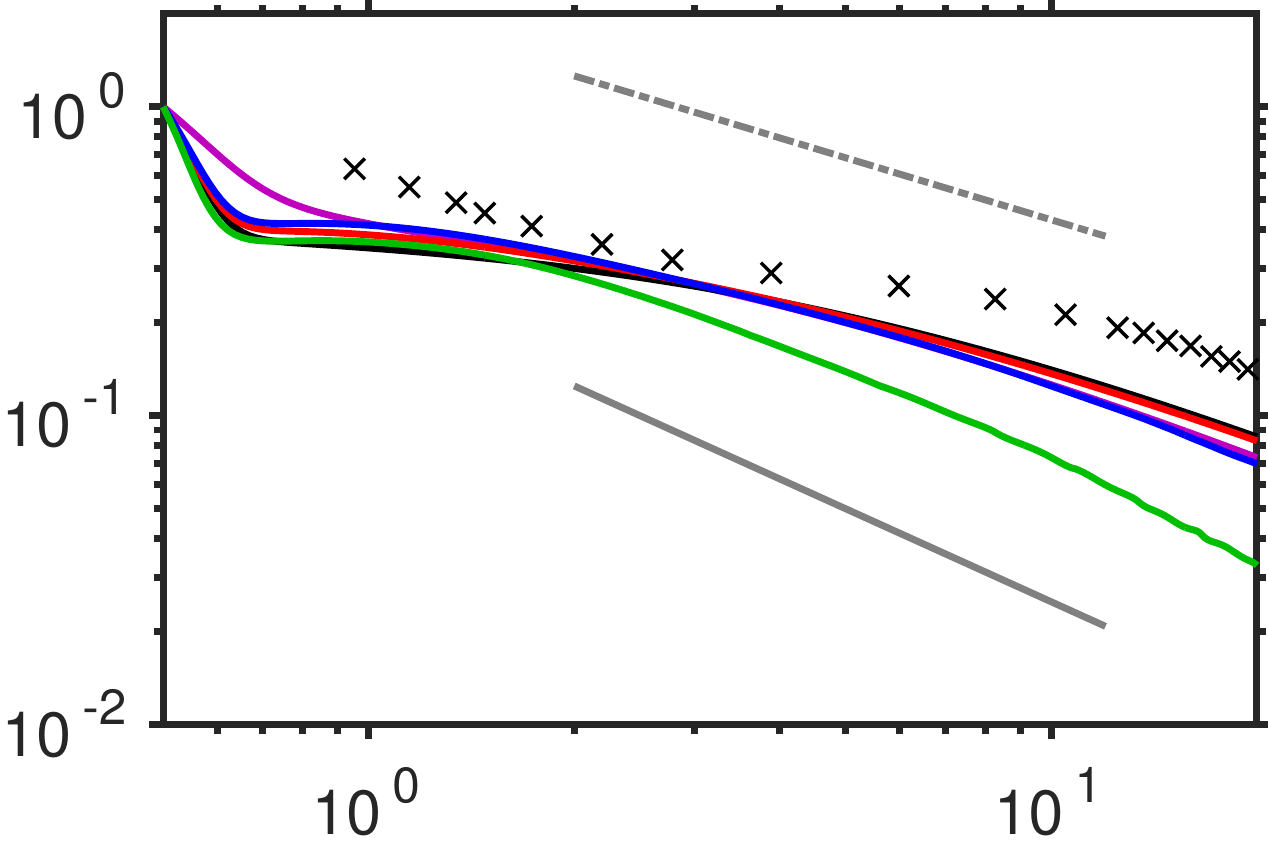}\\[-1.5ex]
         \put(70,55) {$z^{-2/3}$}
         \put(58,22) {$z^{-1}$}
      \end{overpic}
      \centerline{$\tilde{z}_\parallel$}
   \end{minipage}
   %
   \caption{Evolution of the velocity deficit (a) and mean temperature (b) on the centerline of the wake for different Galileo numbers, $Ri_T=0$, $Pr=0.72$ and $\rho_p/{\rho_\infty}=10$, with comparison with the numerical work of \cite{Bagchi_JFM2004} and \cite{Bagchi_PoF2008}.
   \newline
   Linestyle:  $\color{magenta}{\solidthick}$ $Ga=100$, ${\solidthick}$ $Ga=150$, $\color{red}{\solidthick}$ $Ga=170$, $\color{blue}{\solidthick}$ $Ga=200$, $\color{green}{\solidthick}$ $Ga=300$. 
   Grey lines indicate the trends for a laminar (solid lines) and a turbulent wake (dash-dotted lines). The green dashed line on figure (a) gives the streamwise evolution of the velocity deficit computed on the original coordinate system.
   \newline
   Data for the velocity field taken from \cite{Bagchi_JFM2004} for a fixed sphere in uniform flow at 
   $+$ $\left<Re\right>_t=610$, $\circ$ $Re=107$, and data from the temperature field taken from \cite{Bagchi_PoF2008} for a fixed sphere in uniform flow at $\left<Re\right>_t=250$, and $Pr=1.0$ $\times$. 
   }
   \label{fig:decay_centerline}
\end{figure}

We now move on to the structure of the scalar field.
We characterize the amplitude of the wake  with the decay of the scalar field downstream of the sphere. This decay is represented on figure \ref{fig:decay_centerline} both for the scalar field and for the velocity deficit $\tilde{u}_d$. The latter is defined as 
{
\begin{eqnarray}
   \tilde{u}_d(\tilde{r}_\perp,\tilde{z}_\parallel)&=&-\frac{\left<u_{z_\parallel} \right>_{\theta t}(\tilde{r}_\perp,\tilde{z}_\parallel)}{\langle \lVert  \mathbf{u}_{p} \rVert \rangle_t}\;,
\end{eqnarray}
where $u_{z_\parallel}=\mathbf{u}\cdot\mathbf{e}_{\parallel}$ and $\left<\;.\;\right>_{\theta t}$ refers to the average over the azimuthal direction and in time.}
The evolution on the axis {$\tilde{r}_\perp=0$} of the velocity deficit as well as the mean temperature are represented on figure \ref{fig:decay_centerline}.
We observe a negligible influence of the wake regime in the rear of the sphere as both fields seem to be dominated by the influence of the recirculation: the velocity defect first increases because of the reverse fluid motion on the wake axis in the recirculation zone and then it further decreases. The corresponding temperature evolution is shown on figure \ref{fig:decay_centerline}(b). It features a rapid decrease within a distance of $0.2D$ to attain a value approximately equal to $0.4$ as already observed by \cite{Bagchi_PoF2008}.
After this zone, both velocity deficit and temperature decrease, with decay rates that differ from the expected trend ($z^{-1}$ for a laminar wake and $z^{-2/3}$ for a turbulent wake), most probably because this asymptotic state would be expected at larger distances from the sphere.
%
%
\begin{figure*}[t]
   \begin{minipage}{2ex}
      \rotatebox{90}{\centerline{$\left<T\right>_{\theta t}(\tilde{r},\tilde{z}_{{\parallel}i})/\left<T\right>_{\theta t}(0,\tilde{z}_{{\parallel}i})$}}
   \end{minipage}
   \begin{minipage}{0.23\linewidth}
      \centerline{(a)}
      \includegraphics[width=1.0\linewidth]
         {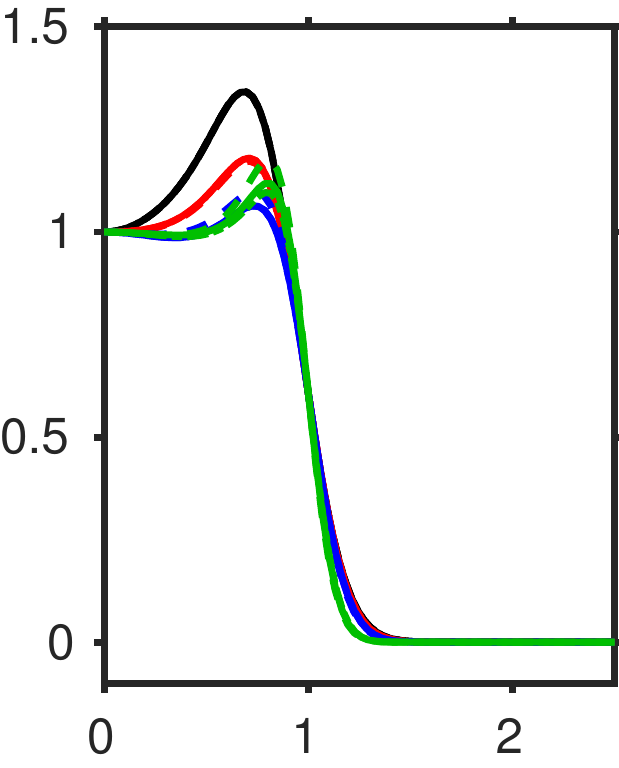}
      \centerline{{$\tilde{r}_\perp/L_{hw}^T(\tilde{z}_{{\parallel}i})$}}
   \end{minipage}
   \begin{minipage}{0.23\linewidth}
      \centerline{(b)}
      \includegraphics[width=1.0\linewidth]
         {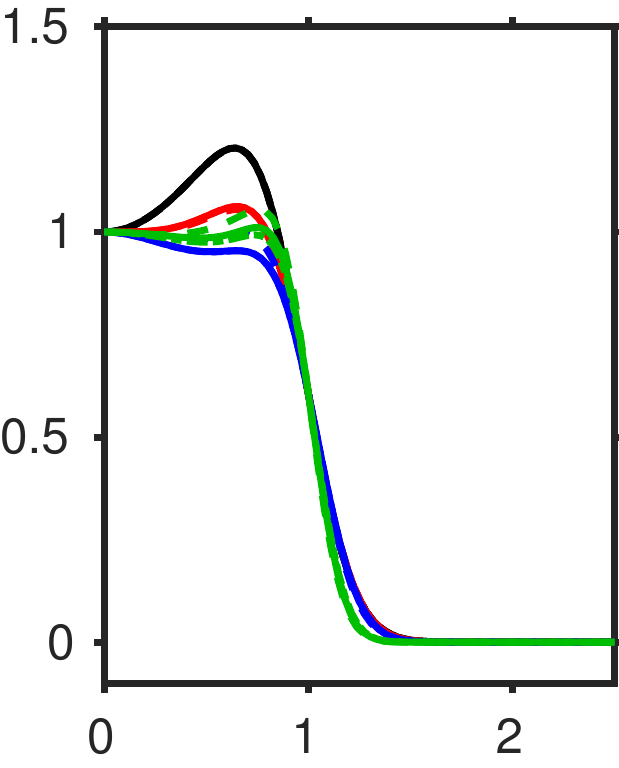}
      \centerline{{$\tilde{r}_\perp/L_{hw}^T(\tilde{z}_{{\parallel}i})$}}
   \end{minipage}
   \begin{minipage}{0.23\linewidth}
      \centerline{(c)}
      \includegraphics[width=1.0\linewidth]
         {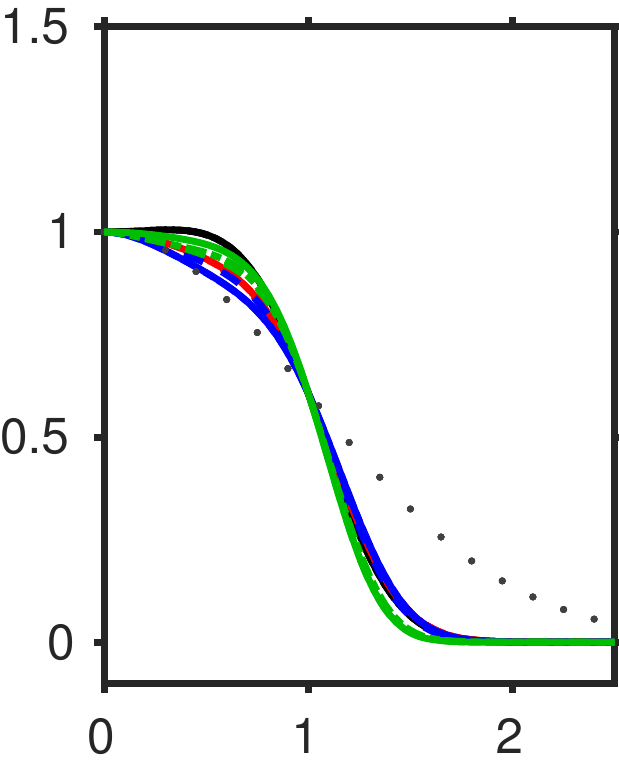}
      \centerline{{$\tilde{r}_\perp/L_{hw}^T(\tilde{z}_{{\parallel}i})$}}
   \end{minipage}
   \begin{minipage}{0.23\linewidth}
      \centerline{(d)}
      \includegraphics[width=1.0\linewidth]
         {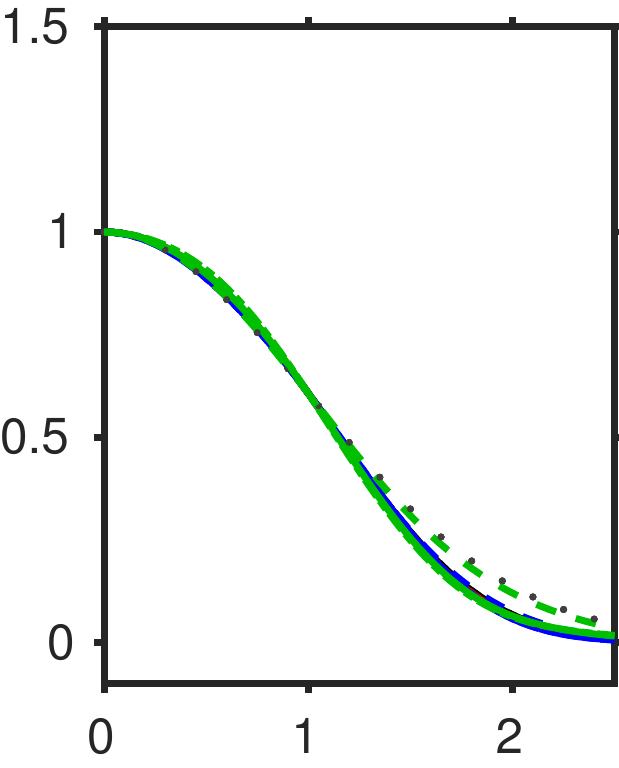}
      \centerline{{$\tilde{r}_\perp/L_{hw}^T(\tilde{z}_{{\parallel}i})$}}
   \end{minipage}
   \caption{{Transverse profiles} of the mean temperature at different locations downstream of the sphere $\tilde{z}_{{\parallel}i}$ {normalized by the profile half width $L_{hw}^T$} for $Ri_T=0$, $Pr=0.72$: (a) $\tilde{z}_\parallel=0.75$, (b) $\tilde{z}_\parallel=1$, (c) $\tilde{z}_\parallel=1.75$, (d) $\tilde{z}_\parallel=5$. 
   The dotted line on (c) and (d) indicates the Gaussian profile.
   Linestyle: ${\solidthick}$ $Ga=150$, $\color{red}{\solidthick}$ $Ga=170$, $\color{blue}{\solidthick}$ $Ga=200$, $\color{green}{\solidthick}$ $Ga=300$, dashed lines: $\rho_p/{{\rho_\infty}}=1.5$, solid lines: $\rho_p/{{\rho_\infty}}=10$ and dash-dotted lines: fixed particle}
   \label{fig:self_sim_near_wake}
\end{figure*}

The influence of the recirculation zone is also visible downstream of the sphere with the cross-stream evolution of the temperature at different positions as shown in figure \ref{fig:self_sim_near_wake}. 
Due to the recirculation, cold fluid is transported towards the downstream stagnation point on the sphere which will induce the large gradient observed on the temperature decay (fig. \ref{fig:decay_centerline}(b)), and from there hot fluid is transported to the shear region creating a local peak of temperature, in accordance with the temperature contours observed by \cite{Bagchi_ASME2001}, \cite{Bhattacharyya_IJHMT2008} and \cite{Bagchi_PoF2008}. 
The amplitude of this peak decreases for {for $Ga=150,170$ and $200$} and then slightly increases for the last Galileo number.
{This first decrease is consistent with the increase of the recirculation length for $Ga=\lbrace 150,\;170,\;200 \rbrace$ as visual inspections of the temperature field have shown that recirculation vortices with large streamwise extensions lead to colder fluid in the vortex core.
The small increase of the peak observed for $Ga=300$ compared to $Ga=200$, which is in contradiction with the increase of $L_r$, can be attributed to the decrease of the temperature in the inter-vortex region observed from $Ga=200$ to $Ga=300$ for the streamwise evolution of the temperature (fig. \ref{fig:decay_centerline}): fluid is getting colder in the intra-vortex region due to the increase of $L_r$, but it also gets colder in the inter-vortex region, decreasing with this the normalized amplitude of the temperature peak.}
After the recirculation zone, the temperature becomes a monotonic function of $\tilde{r}_\bot$ and attains a self-similar Gaussian profile for $\tilde{z}_\parallel > 5$ as shown in figure \ref{fig:self_sim_near_wake}, which is consistent with the observations made by \cite{Bagchi_PoF2008} who found Gaussian profiles in the thermal wake at streamwise distances of $6.5D$ and $13D$.
\begin{figure}[t]
   %
   %
   \begin{minipage}{4ex}
      \rotatebox{90}{\centerline{$L_{hw}^{u_d}(\tilde{z}_\parallel)$}}
   \end{minipage}
   \begin{minipage}{0.45\linewidth}
      \centerline{(a)}
      \begin{overpic}[width=1\linewidth]
         {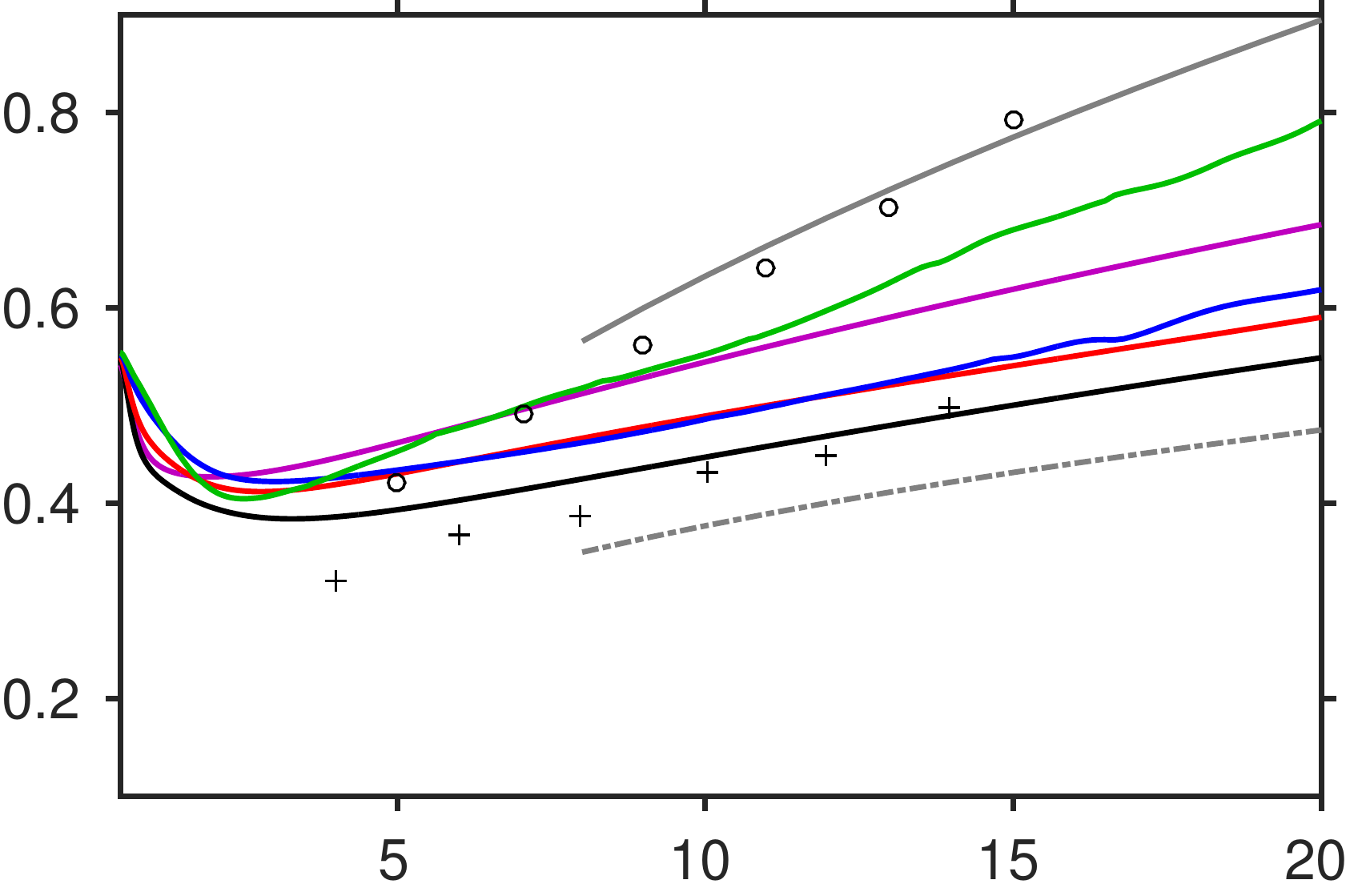}\\[-1.5ex]
         \put(60,53) {$z^{1/2}$}
         \put(70,25) {$z^{1/3}$}
      \end{overpic}
     \centerline{$\tilde{z}_\parallel$}
   \end{minipage}
   \begin{minipage}{2ex}
      \mbox{}
   \end{minipage}
   \begin{minipage}{4ex}
      \rotatebox{90}{\centerline{$L_{hw}^T(\tilde{z}_\parallel)$}}
   \end{minipage}
   \begin{minipage}{0.45\linewidth}
      \centerline{(b)}
      \begin{overpic}[width=1\linewidth]
         {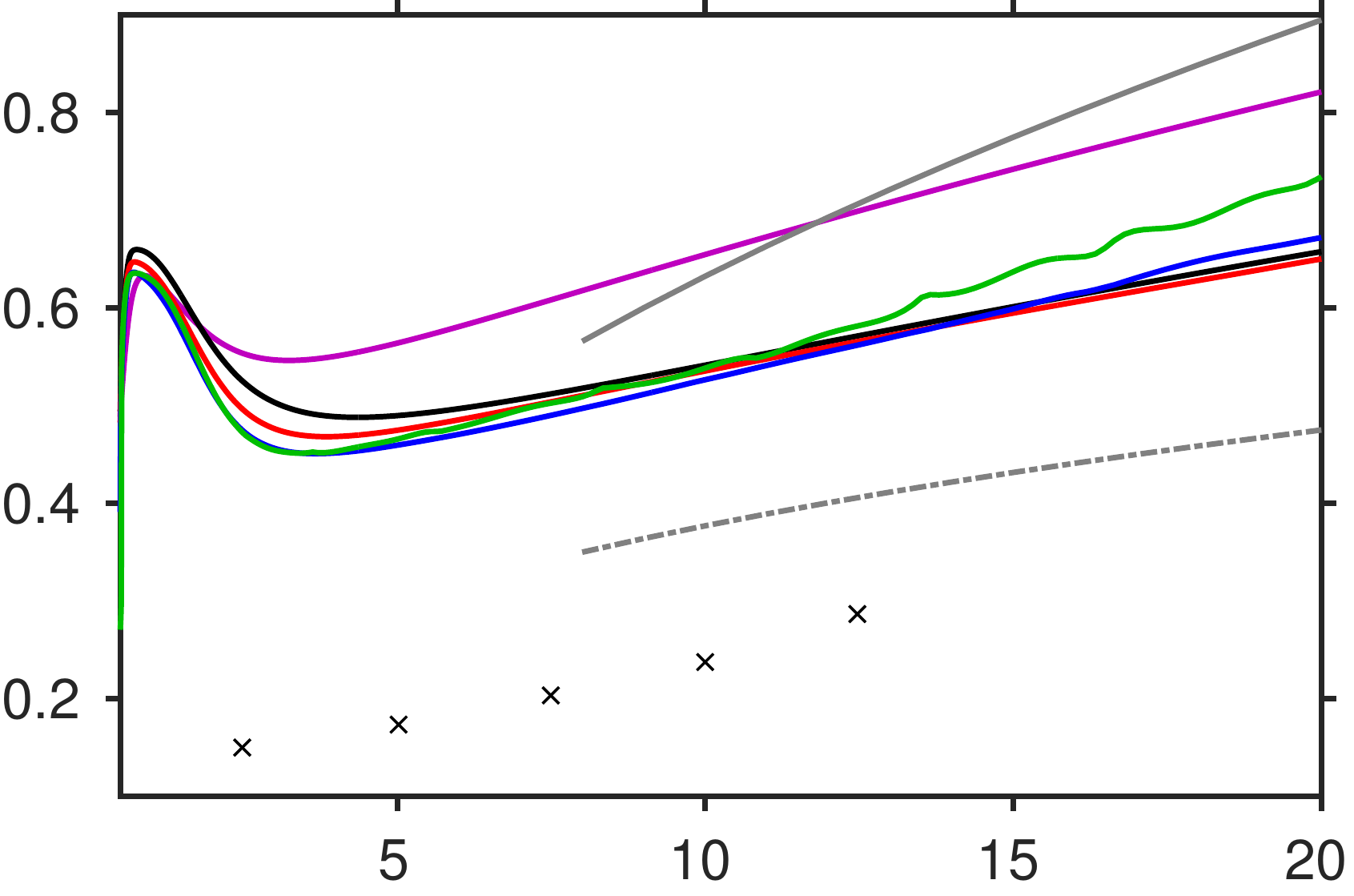}\\[-1.5ex]
         \put(60,53) {$z^{1/2}$}
         \put(70,25) {$z^{1/3}$}
      \end{overpic}
      \centerline{$\tilde{z}_\parallel$}
   \end{minipage}
   \caption{Evolution the half width of the velocity deficit $L_{hw}^{u_d}$ (a), and temperature $L_{hw}^{T}$ (b), for different Galileo numbers, $Ri_T=0$, $Pr=0.72$ and $\rho_p/{{\rho_\infty}}=10$, with comparison with the numerical work of \cite{Bagchi_JFM2004} and \cite{Bagchi_PoF2008}.
   \newline
   Linestyle:  $\color{magenta}{\solidthick}$ $Ga=100$, ${\solidthick}$ $Ga=150$, $\color{red}{\solidthick}$ $Ga=170$, $\color{blue}{\solidthick}$ $Ga=200$, $\color{green}{\solidthick}$ $Ga=300$. 
   Grey lines indicate the trends for a laminar (solid lines) and a turbulent wake (dash-dotted lines).
   \newline
   Data for the velocity field taken from \cite{Bagchi_JFM2004} for a fixed sphere in uniform flow at 
   $+$ $\left<Re\right>_t=610$, $\circ$ $Re=107$, and data from the temperature field taken from \cite{Bagchi_PoF2008} for a fixed sphere in uniform flow at $\left<Re\right>_t=250$, and $Pr=1.0$ $\times$. 
   }
   \label{fig:half_width_passive}
\end{figure}
The influence of the regime is more visible away from the centerline as can be clearly seen on the streamwise evolution of the half width of the wake (fig. \ref{fig:half_width_passive}). We introduce this quantity for the velocity deficit and the temperature, respectively $L_{hw}^{u_d}(\tilde{z}_\parallel)$ and $L_{hw}^T(\tilde{z}_\parallel)$. We define it as the radial coordinate $\tilde{r}_\bot$, for a given location $\tilde{z}_\parallel$ on the axis where the corresponding quantity is equal to $e^{-1/2}$ times the value on the centerline \citep[this definition has also been used by][]{Bagchi_JFM2004,Legendre_PoF2006,Bagchi_PoF2008}. The evolution of $L_{hw}$ is represented on figure \ref{fig:half_width_passive} and shows a similar trend for the scalar quantities and the velocity deficit: the width first decreases with $\tilde{z}_\parallel$ which is consistent with \cite{Legendre_PoF2006} and then increases almost linearly as observed with most free shear flows. 
The width of the hydrodynamic and thermal wakes follow different trends with $Ga$ as $L_{hw}^T$ shows similar size for the first three Galileo numbers while $L_{hw}^{u_d}$ is lower at $Ga=150$ than $Ga=170$ and $Ga=200$. 
We have also included a case featuring $Ga=100$ to emphasize the effects of both diffusion and wake regime, since diffusion should tend to increase the growth of the half width with the distance $\tilde{z}_\parallel$ and the structure of the wake should affect this width at the end of the recirculation region.
The evolution of {$L_{hw}^{u_d}$ confirms}  the intuition that the half width of the velocity deficit should increase much faster for $Ga=100$ than $Ga=150$ due to diffusion effects, which is in accordance with the observations of \cite{Bagchi_JFM2004} who observed a larger width at $Re=107$ than at $Re=241$ and $Re=261$.
The half width is surprisingly larger at $Ga=170$ than $Ga=150$ although diffusive effects should be expected to be slightly larger at $Ga=150$.
The difference in the viscosities are indeed not large enough to induce a significant difference in the slope of $L_{hw}^{u_d}$ while the obliqueness of the wake at $Ga=170$ tends to increase its width in the recirculation region once averaging in the azimuthal direction. In the steady oblique regime, the wake projected onto the oblique coordinate system is not axisymmetric \citep[see for example the visualizations in][]{Uhlmann_IJMF2014}.
The unsteadiness of the wake will also act as diffusive effects and tend to increase the width of the average wake, as we observe here.
This last point is in accordance with the simulations of \cite{Legendre_PoF2006} in the case of a bubble for which a thicker wake is observed at $Re=500$ compared to $Re=200$.
Considering now the half width of the temperature, this trend is less pronounced (fig. \ref{fig:half_width_passive}(b)). The influence of the viscosity is clearly visible with a larger thermal wake obtained at $Ga=100$ compared to the other Galileo numbers. But the configurations with $Ga=150,\;170$ {and }$200$ feature similar wake width. This can be attributed to the low difference in diffusivities between those cases. The enlargement of the wake attributed to the unsteadiness of the regime is mostly visible on the chaotic regime.

%
\begin{figure}[t]
\centering
   \begin{minipage}{3ex}
      \rotatebox{90}{\centerline{$L_{hw}^{u_d}(\tilde{z}_\parallel)$}}
   \end{minipage}
   \begin{minipage}{0.40\linewidth}
      \centerline{(a)}
      \begin{overpic}[width=1\linewidth]
         {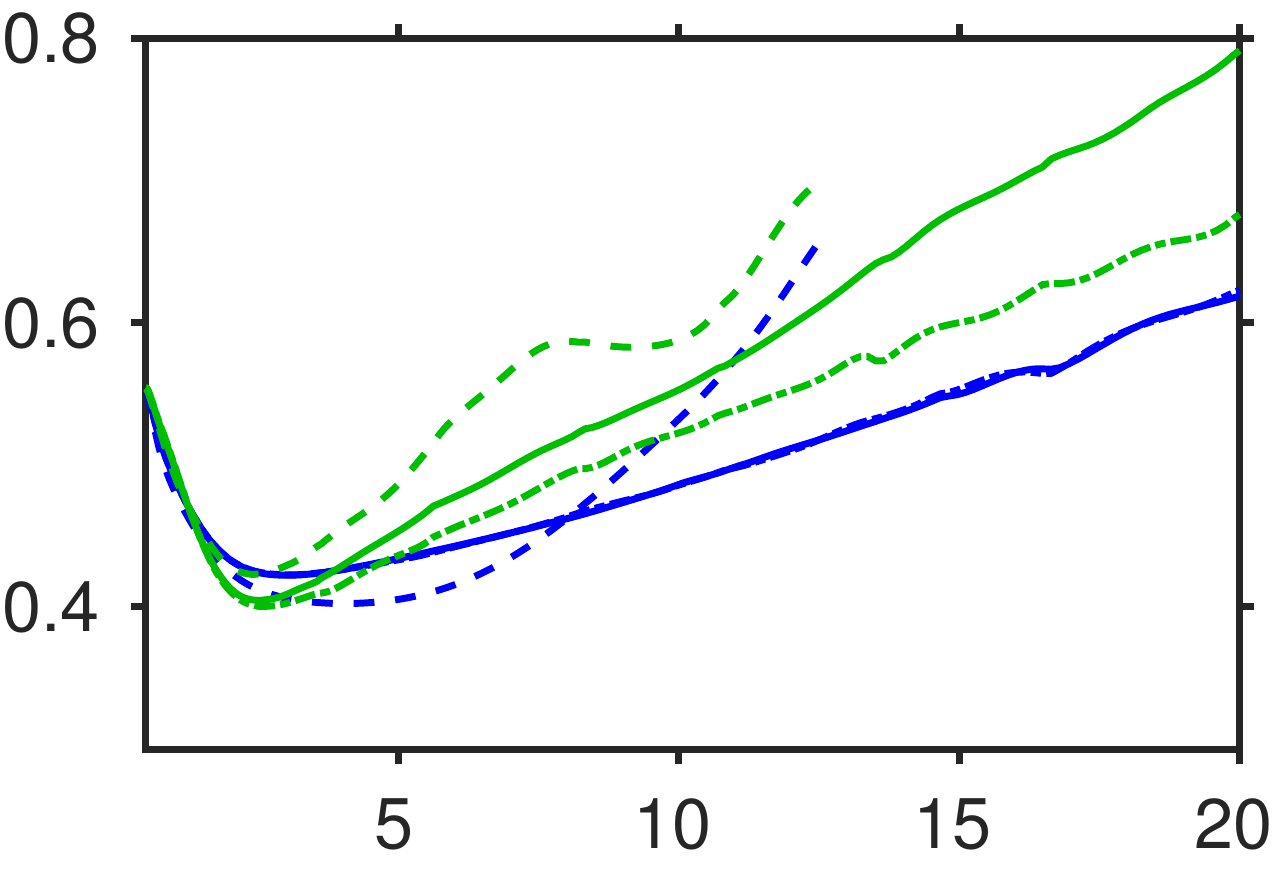}\\[-2.0ex]
      \end{overpic}
     \centerline{$\tilde{z}_\parallel$}
   \end{minipage}
   \begin{minipage}{2ex}
      \mbox{}
   \end{minipage}
   \begin{minipage}{3ex}
      \rotatebox{90}{\centerline{$L_{hw}^T(\tilde{z}_\parallel)$}}
   \end{minipage}
   \begin{minipage}{0.40\linewidth}
      \centerline{(b)}
      \begin{overpic}[width=1\linewidth]
         {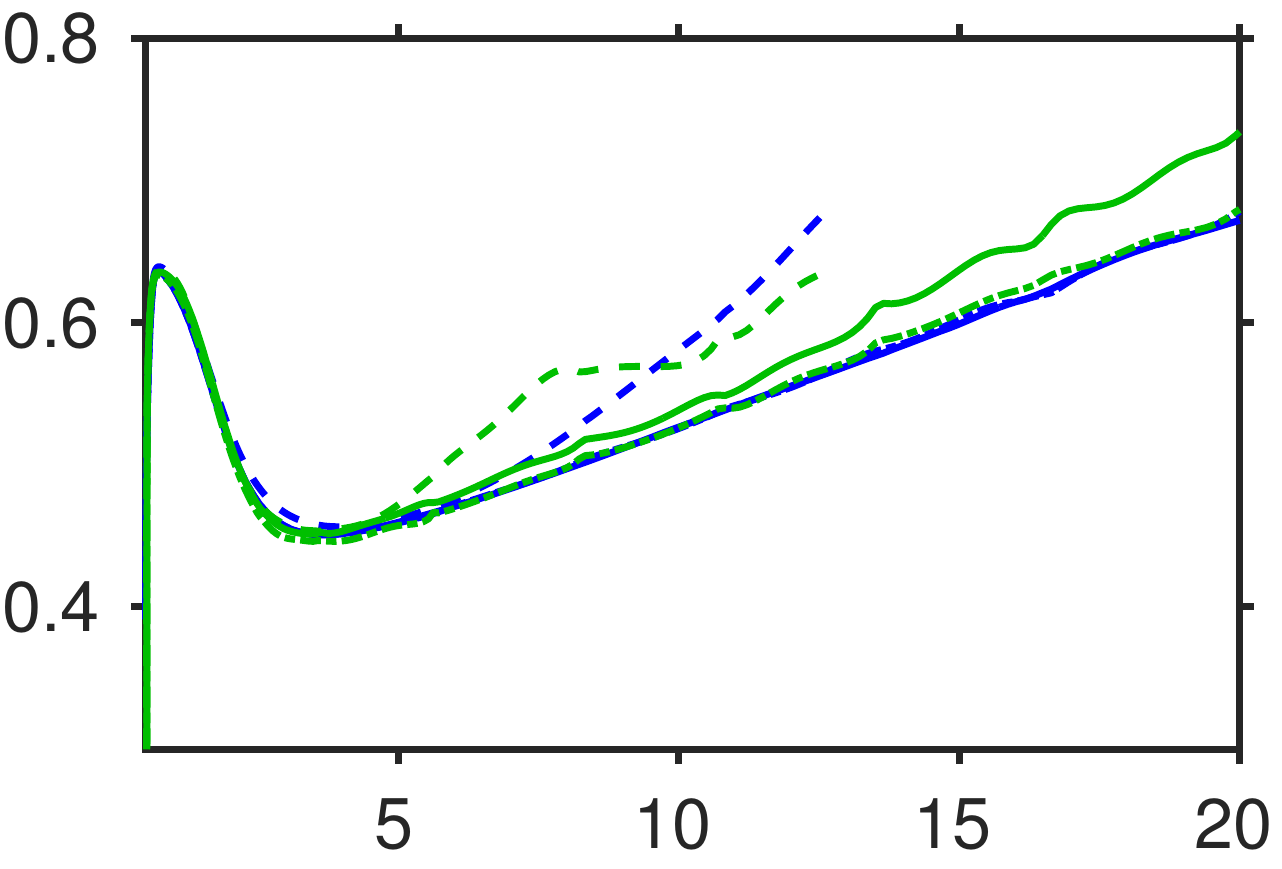}\\[-2.0ex]
      \end{overpic}
     \centerline{$\tilde{z}_\parallel$}
   \end{minipage}
   \caption{Evolution of the corresponding half width $L_{hw}^{(i)}$ of the velocity defect (a) and temperature (b) for $Pr=0.72$, $Ri_T=0$ and two Galileo numbers corresponding to unsteady regimes and different mobilities: $\rho_p/{{\rho_\infty}}=1.5$ (dashed lines), $\rho_p/{{\rho_\infty}}=10$ (solid lines) and fixed particle (dash-dotted lines).
   Colorstyle: $\color{blue}{\solidthick}$ $Ga=200$, $\color{green}{\solidthick}$ $Ga=300$.  
   }
   \label{fig:half_width_density}
\end{figure}
The density ratio does not have an important influence on streamwise evolution of the velocity deficit and on the temperature (figure omitted). 
The main difference can be attributed to the slow decrease of the recirculation length to be observed for $\rho_p/\rho_\infty=1.5$ compared to $\rho_p/\rho_\infty=10$, but which is not observed when comparing the larger density ratio to the fixed sphere configuration.
The half width shows a larger sensitivity to the sphere mobility as represented on figure \ref{fig:half_width_density} which displays an increase of the half width as the density ratio decreases. This can be attributed to the tendency for the wake to align differently than the particle trajectory when the density ratio decreases. With this the angle between the wake and the trajectory might increase which causes an enlargement of the wake once averaging.
{Another difference at the lowest density ratio is the convergence of the statistics that has been more difficult to reach for the half widths, for similar observation time, as can be seen in figure \ref{fig:half_width_density}. Therefore here the comparison should remain mostly qualitative.}
The scaled evolution of the temperature as a function of the radial distance shows little influence of the density ratio (fig \ref{fig:self_sim_near_wake}), apart from slightly larger peak of temperature in the recirculation region obtained for $\rho_p/\rho_\infty=1.5$, as a consequence of the small decrease of the extent of this zone.
{The evolution of $L_{hw}^T$ confirms that density has less influence on the thermal wake at short distances than on the velocity deficit.}

It appears therefore that in a meteorological context it is important to investigate configurations featuring a chaotic regime as it affects most of the statistics of the scalar field, starting from the characteristics of the recirculation zone. The mobility of the sphere can be neglected, or density ratio has to be taken as large.
%

%
\section{The case of active scalar transport}  \label{sec:active}
\begin{figure}[h]
\begin{minipage}{\linewidth}
\centering
\includegraphics[width=0.9\linewidth]
   {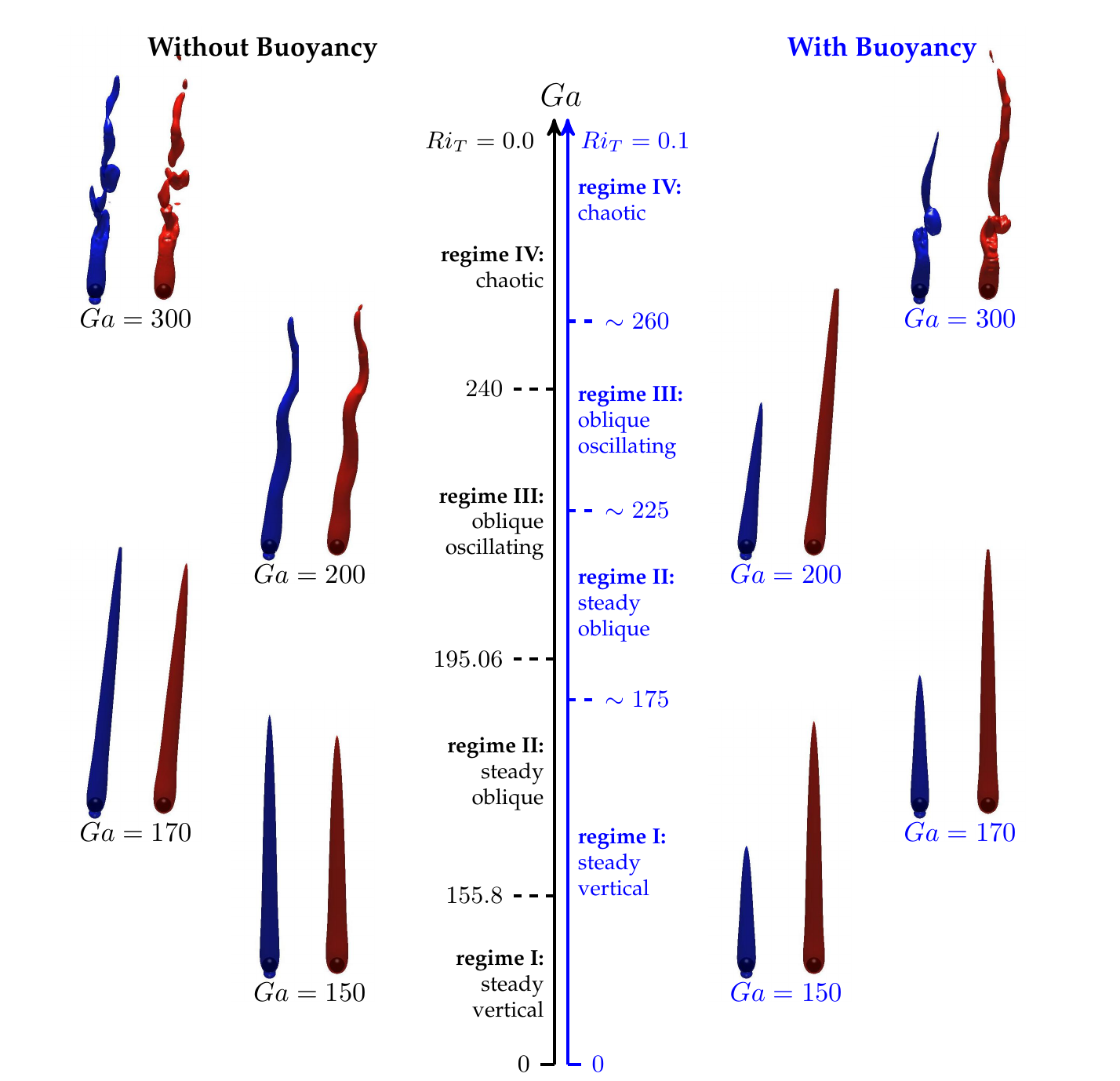}
\end{minipage}
\caption{{Settling regimes observed for a density ratio $\rho_p/\rho_\infty=10$, for different Galileo numbers and in the absence of buoyancy effects (left part) and with buoyancy effects (right part, for $Ri_T=0.1$), with illustration of the wake structure for the same four Galileo numbers as used in figure \ref{fig:sketch_regimes} (i.e. from left to right $Ga=\lbrace 150,\;170,\;200,\;300 \rbrace$). 
Blue surfaces represent isocontour of the velocity $\left|\mathbf{\tilde{u}-\tilde{u}_p}\right|=0.80\left|\mathbf{\tilde{u}_\infty}-\mathbf{\tilde{u}_p}\right|$ and red surfaces isocontour of the temperature $\tilde{T}=0.1$.}}
\label{fig:sketch_regimes_with_buoyancy}
\end{figure}
\begin{figure}[ht]
\centering
   \begin{minipage}{2ex}
      \rotatebox{90}{\centerline{$Ri_T$}}
   \end{minipage}
   \begin{minipage}{0.65\linewidth}
      \includegraphics[width=0.95\linewidth]
         {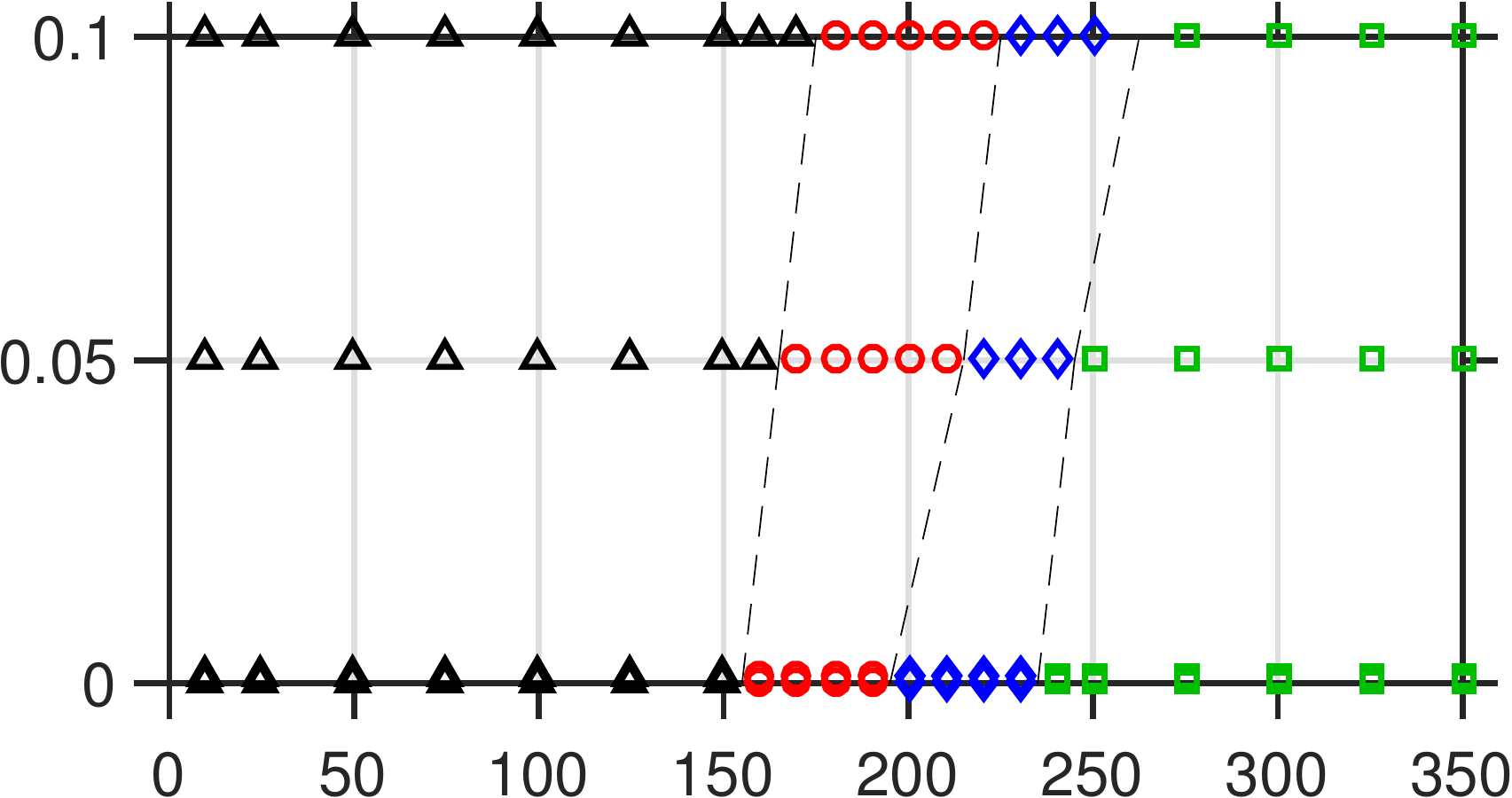}\\
      \centerline{$Ga$}
   \end{minipage}
   \caption{Map of the regimes observed as a function of $Ga$ and $Ri_T$ for $\rho_p/{{\rho_\infty}}=10$ and $Pr=0.72$ ($\mathlarger{\mathlarger{\vartriangle}}$ steady axisymmetric, $\mathlarger{\mathlarger{\mathlarger{\circ}}}$ steady oblique, $\lozenge$ oscillating oblique, $\mathsmaller{\square}$ chaotic)}
   \label{fig:regime_map}
\end{figure}

Figure \ref{fig:sketch_regimes_with_buoyancy} gives a visual impression of the three-dimensional structure of the velocity and temperature fields for the same Galileo numbers as described in the previous section, here for $Ri_T=0.1$. It first appears that buoyancy effects tend to elongate the {thermal} wake in the direction of gravity, due to the transport of {warmer} fluid in the upward direction.
Another major point is the modification of the regime of the wake for $Ga=170$ and $Ga=200$. We indeed observe that axisymmetry is preserved at $Ga=170$ while this case is steady oblique for $Ri_T=0$. In the same way at $Ga=200$ the wake does not feature space oscillations anymore and lost its unsteadiness.
Buoyancy, in the assisting flow configuration, therefore tends to stabilize the wake, in accordance with the previous observations of \cite{Kotouc_IJHMT2008} and \cite{Kotouc_JFM2009}.
Figure \ref{fig:regime_map} gives the map of the different regimes observed for $\rho_p/\rho_\infty=10$ and for $Ri_T$ ranging from 0 to 0.1. The stabilizing influence of buoyancy is clearly visible as the different transition thresholds are pushed towards larger values of $Ga$.\\
It is known from the literature that a buoyant plume will be induced at larger Richardson numbers, and the competition between the upward plume and the original downward motion tends to decrease the extent of the recirculation region. 
The evolution of this recirculation as a function of the Galileo number for those new Richardson numbers is included in figure \ref{fig:recirculation_length}.
The trend of the evolution of $L_r$ with $Ga$ is conserved: it will first increase with $Ga$ as long as the regime remains steady, then unsteadiness will be associated with a decay of the recirculation length until the wake becomes chaotic and then $L_r$ increases with $Ga$. 
The recirculation length decreases with $Ri_T$, in accordance with the observations of \cite{Bhattacharyya_IJHMT2008}.
{This decrease can be seen as the signature of the superposition of two effects: the initial recirculation induced in the rear of the sphere, and the plume with positive streamwise velocity induced by the buoyancy effect.}
It influences the heat and mass transfer by a decrease of the local Nusselt
number in the region of the rear stagnation point. This is consistent with the
observations of \cite{Bhattacharyya_IJHMT2008} and \cite{Kotouc_IJHMT2008}
(figure omitted). 
{The buoyancy induced plume reduces the transport of cold fluid from the shear region towards the rear stagnation point, and with this the local temperature gradient and therefore the local Nusselt number decrease.}
But this modification is relatively small at the Richardson numbers investigated here and no significant difference is observed on the mean transfer coefficient (figure omitted).
This is consistent with the simulations of \cite{Bhattacharyya_IJHMT2008} for which only small variations of the Nusselt number are observed in the range $0.0<Ri_T<0.1$ and $1<Re<200$.
\begin{figure}[h]
   \begin{minipage}{4ex}
      \rotatebox{90}{\centerline{$u_d(\tilde{r}_\bot=0,\tilde{z}_\parallel)$}}
   \end{minipage}
   \begin{minipage}{0.45\linewidth}
      \centerline{(a)}
      \begin{overpic}[width=1\linewidth]
         {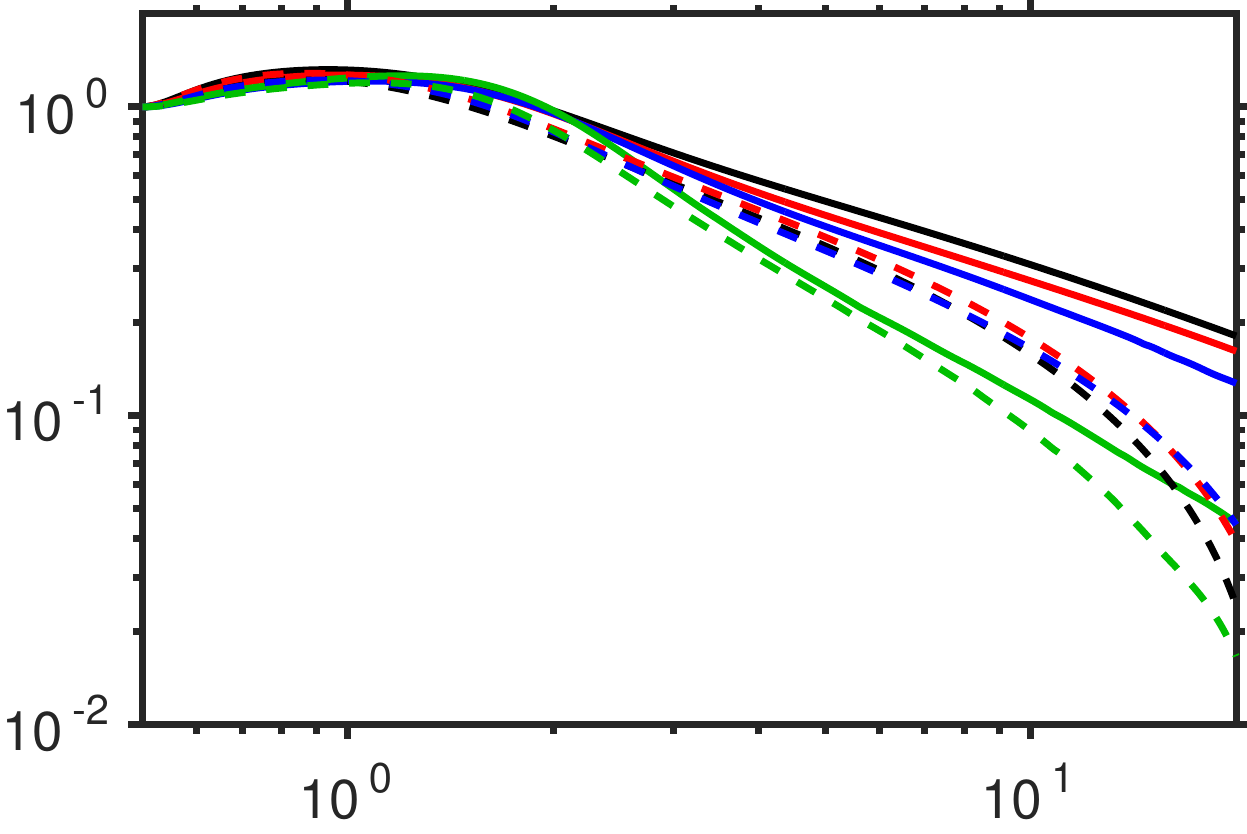}\\[-1.5ex]
      \end{overpic}
     \centerline{$\tilde{z}_\parallel$}\\
   \end{minipage}
   \begin{minipage}{2ex}
      \mbox{}
   \end{minipage}
   \begin{minipage}{4ex}
      \rotatebox{90}{\centerline{$\left<\tilde{T}\right>_t(\tilde{r}_\bot=0,\tilde{z}_\parallel)$}}
   \end{minipage}
   \begin{minipage}{0.45\linewidth}
      \centerline{(b)}
      \begin{overpic}[width=1\linewidth]
         {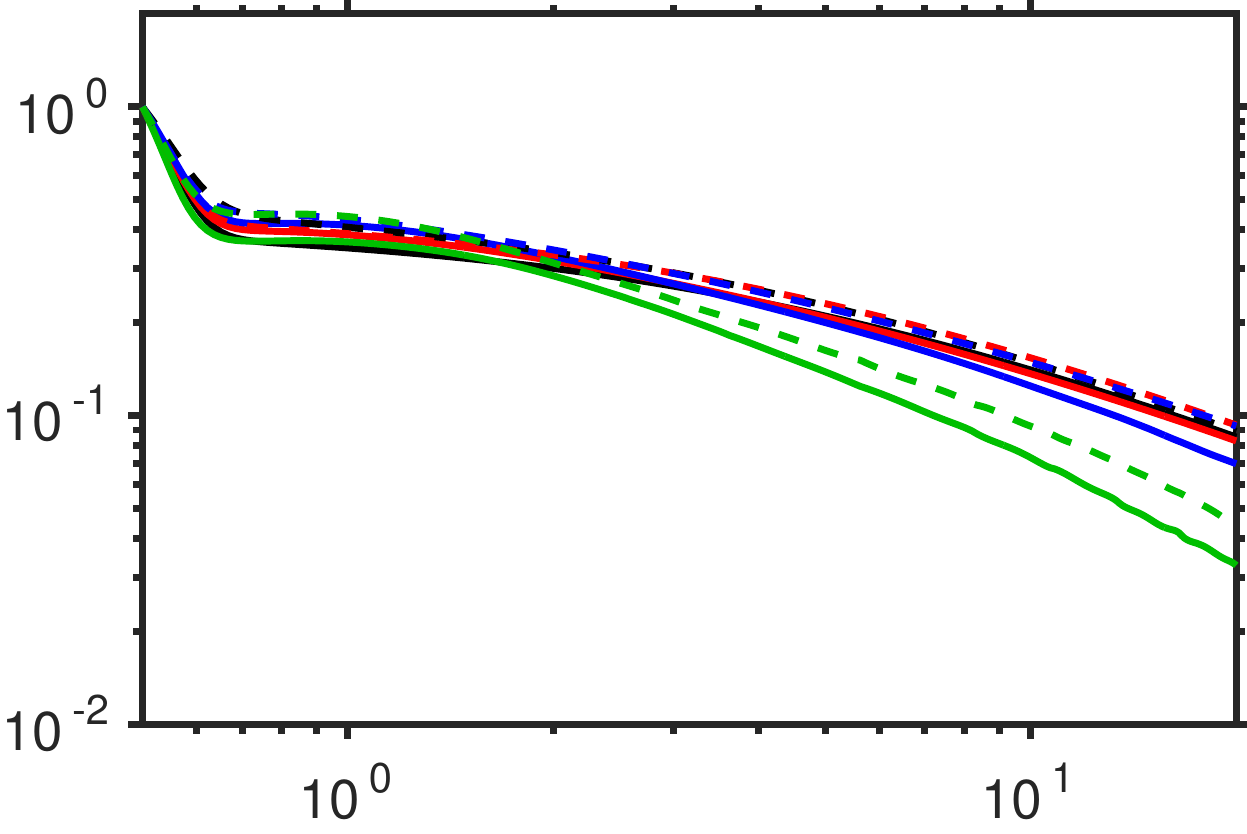}\\[-1.5ex]
      \end{overpic}
      \centerline{$\tilde{z}_\parallel$}\\
   \end{minipage}
      \begin{minipage}{4ex}
      \rotatebox{90}{\centerline{$L_{hw}^{u_d}(\tilde{z}_\parallel)$}}
   \end{minipage}
   \begin{minipage}{0.45\linewidth}
      \centerline{(c)}
      \begin{overpic}[width=1\linewidth]
         {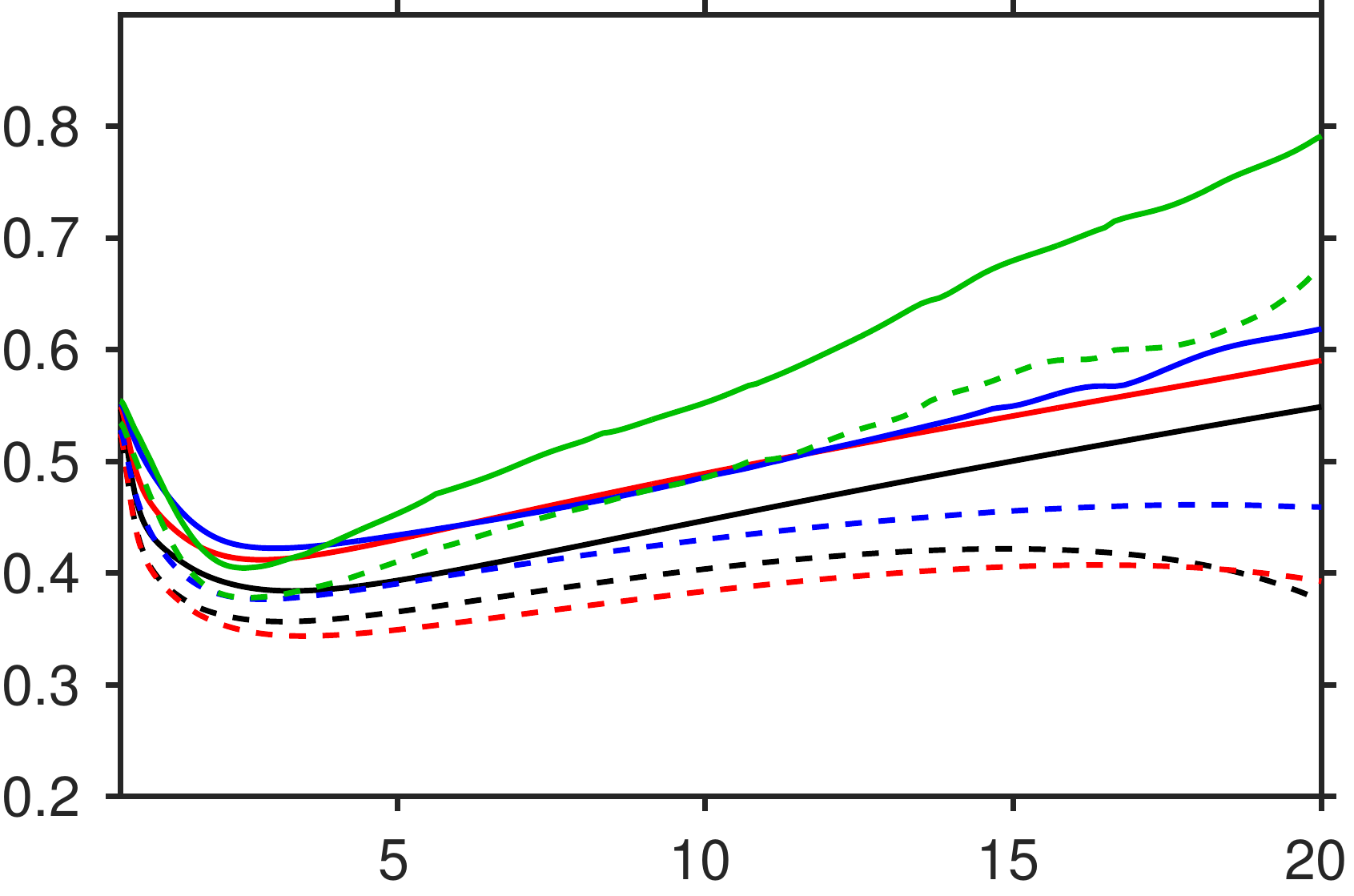}\\[-1.5ex]
      \end{overpic}
     \centerline{$\tilde{z}_\parallel$}\\
   \end{minipage}
   \begin{minipage}{2ex}
      \mbox{}
   \end{minipage}
   \begin{minipage}{4ex}
      \rotatebox{90}{\centerline{$L_{hw}^T(\tilde{z}_\parallel)$}}
   \end{minipage}
   \begin{minipage}{0.45\linewidth}
      \centerline{(d)}
      \begin{overpic}[width=1\linewidth]
         {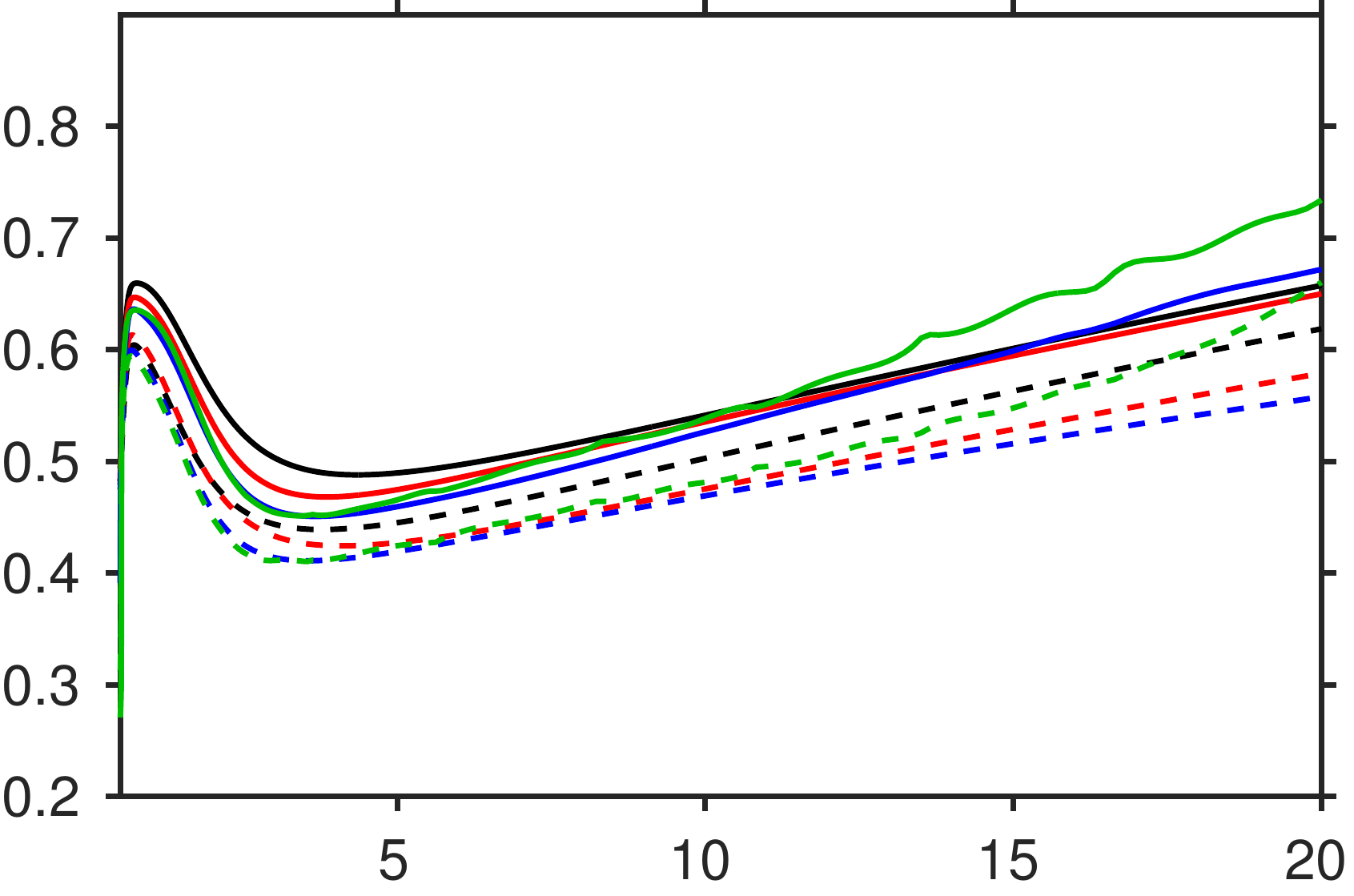}\\[-1.5ex]
      \end{overpic}
      \centerline{$\tilde{z}_\parallel$}\\
   \end{minipage}
   \caption{Evolution of the velocity deficit (a) and mean temperature (b) on the centerline of the wake as well as the corresponding half width $L_{hw}^{u_d}$ (c) and $L_{hw}^{T}$ (d), for a density ratio $\rho_p/{{\rho_\infty}}=10$ and for different Galileo numbers, at $Ri_T=0$ ($\solidthick$) or $Ri_T=0.1$ ($\dashed$).
   \newline
   Colorstyle:  ${\solidthick}$ $Ga=150$, $\color{red}{\solidthick}$ $Ga=170$, $\color{blue}{\solidthick}$ $Ga=200$, $\color{green}{\solidthick}$ $Ga=300$.    
   }
   \label{fig:decay_centerline_RiT}
\end{figure}

We consider now the influence of buoyancy on the evolution of the velocity and temperature field. 
We first {consider the evolution on the centerline (fig \ref{fig:decay_centerline_RiT}) and} observe that buoyancy not only decreases the size of the recirculation zone but also the magnitude of the velocity in this zone as a consequence of the upward flow induced by density variations. 
The influence of buoyancy on the velocity deficit on the centerline seems to be confined in this region as little influence of $Ri_T$ is to be observed at $\tilde{r}_\bot=0$ for larger streamwise positions.
The temperature is more affected on the centerline and it increases with $Ri_T$. This can be attributed to the modification of the recirculation: as this zone is more confined and less intense, the fluid transported to the back of the sphere is hotter than with $Ri_T=0$. This tends to decrease the cooling to be observed in this region of the flow, leading to larger temperatures.
{If we now consider the evolution of the half width, we see that the influence of buoyancy} is more pronounced as represented on figure \ref{fig:decay_centerline_RiT}, where it can be seen that buoyancy tends to decrease the width of both thermal and velocity wakes, meaning that the wake is more confined. 
A possible interpretation of this is that buoyancy tends to favor the alignment
of the wake in the vertical direction and through this the alignment with the sphere's trajectory leading to a thinner wake.

%
\section{Saturation profiles in the wake}  \label{sec:discussion}

\begin{figure}[ht]
\centering
\begin{minipage}{2ex}
      \rotatebox{90}{\centerline{$\tilde{r}_\bot$}}
\end{minipage}
\begin{minipage}{0.65\linewidth}
   \begin{overpic}[width=1.0\linewidth]
         {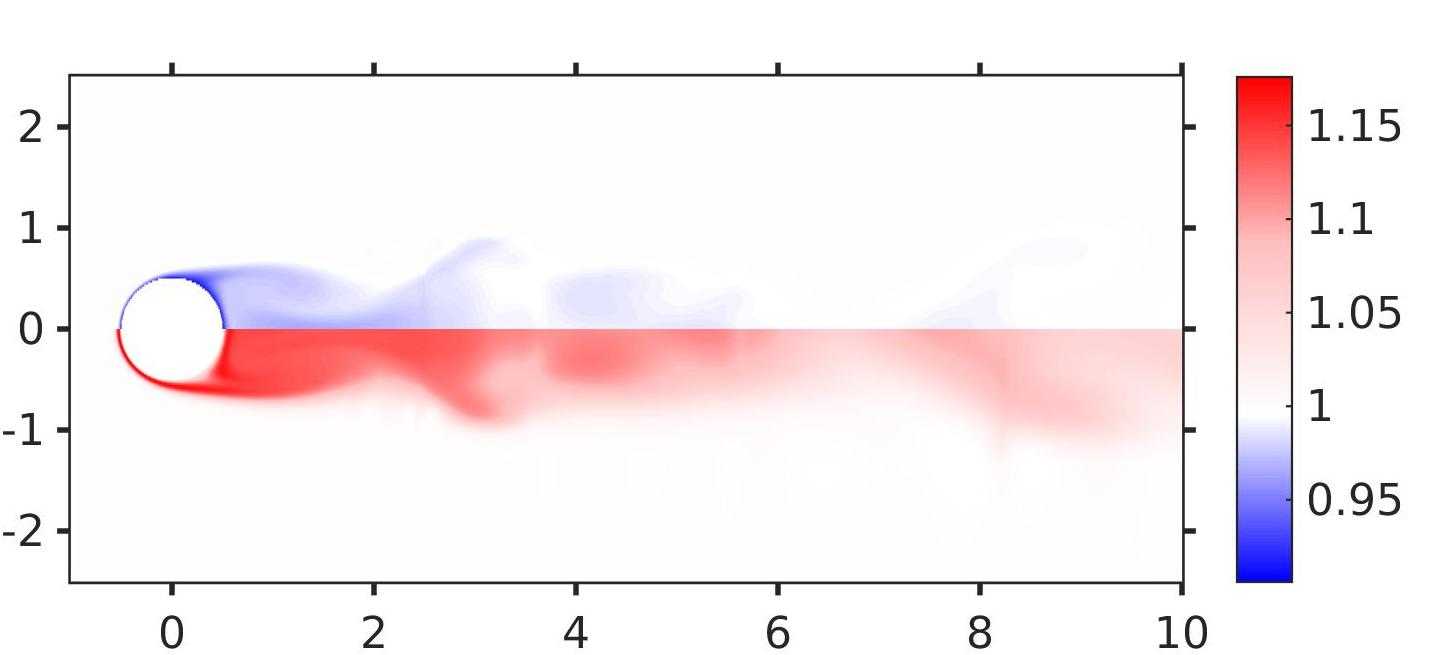}
      \put(32,35) {\textcolor{black}{$T_p=-10\degree C$, $T_\infty=-15\degree C$}}
      \put(32,8) {\textcolor{black}{$T_p=0\degree C$, $T_\infty=-15\degree C$}}
   \end{overpic}     
   \centerline{$\tilde{z}_\parallel$}    
\end{minipage}
\caption{{Instantaneous saturation with respect to liquid water shown for one half of an axial/radial plane, comparing two sets of boundary conditions, for $Ga=300$, a fixed particle, $Ri_T=0$, $Pr=0.72$ and $Sc=0.63$ (top: $T_p=-10\degree C$, $T_\infty=-15\degree C$, bottom: $T_p=0\degree C$, $T_\infty=-15\degree C$)}}
\label{fig:2D_RH}
\end{figure}

We now examine the evolution of the saturation in the wake of the sphere for physical conditions that are the most relevant in a meteorological context. 
For this reason the case with $Ga=300$ 
{is chosen, which features a chaotic behavior, with no
contribution from buoyancy ($Ri_T=0$). 
Since the scalar fields seem to be little affected by the particle mobility at high density ratios, we present the case of a fixed sphere, which also ensures consistency with former work
\citep{wang:13,cheng:14}.}
We will investigate the following set of temperatures {of the sphere:}
\begin{eqnarray}
   T_p=\lbrace -10,\; -5,\; 0\rbrace \;\degree C \;,
       \label{eq:Tp_range}
\end{eqnarray}
{and} for the temperatures of the incoming fluid: 
\begin{eqnarray}
   T_\infty=\lbrace -15,\; -10,\; -5\rbrace\;\degree C \;,
      \label{eq:Tinf_range}
\end{eqnarray}
with the condition that the particle will always be warmer than the fluid.
%
In the current section we will briefly describe the structure of the saturation field with respect to ice defined as the ratio between the local partial pressure $e$ and the saturation water pressure with respect to {liquid water $e_{sat,w}$}:
{\begin{eqnarray}
   S_w(\mathbf{x},t) &=& e/e_{sat,w} \\
   &=& \frac{k_b n_v(\mathbf{x},t) T(\mathbf{x},t)}{e_{sat,w}(T(\mathbf{x},t))}
   \label{eq:S_i_definition}\;.
\end{eqnarray}}
{The aim in the analysis is twofold. First we will test whether the local saturation can increase up to values that might trigger further ice nucleation \citep{hoose:12}.
Second, a methodological issue will be addressed with a discussion on the necessity to separate the transport equations of heat and water vapor.
The difference between Schmidt and Prandtl number is in the present case relatively small, and the boundary conditions in non-dimensional form are the same for the temperature and the vapor concentration, which could suggest to use the approximation $\tilde{n}_v=\tilde{T}$. In the current study, our goal is to have access to the detail of the saturation field in the wake of the particle in order to address these two questions, namely what is the range of saturation reached? And does the approximation $\tilde{n}_v=\tilde{T}$ have a negligible influence?}%
%
{As previously explained in section \ref{sec:methodology} we set the saturation to be equal to unity with respect to the liquid phase at infinity and equal to unity with respect to ice at the particle surface such as to model the presence of micro-droplets at the inflow. It could also represent a wet growth regime at $T_p=0\degree C$ since we would have then $e_{sat,w}=e_{sat,i}$.}
The saturation field with respect to {water} will depend on the set of temperatures taken at the boundaries and figure \ref{fig:2D_RH} gives a visual impression of the saturation {$S_w$} for the smallest and largest temperature differences between the particle and the updraft.
It shows that the largest temperature difference (i.e. a very warm graupel) will induce the largest {saturation}, which can be larger than {1.1} in a large portion of the wake. For the lowest temperature difference (i.e.  configurations with the coldest hailstone) the level of saturation reached is much lower. This last configuration does therefore not appear to be favorable for secondary ice nucleation.
%

From the definitions \ref{eq:nv_definition}, \ref{eq:S_i_definition} and \ref{eq:murphy_ice} one can show that the incremental change of {$S_w$} is linked to the incremental change of $\tilde{n}_v$ and $\tilde{T}$ according to
{\begin{eqnarray}
   \frac{\mathrm{d}S_w}{S_w}
      &=&
      \left(
      \frac{1}{T}-\frac{1}{e_{sat,w}}\frac{\partial e_{sat,w}}{\partial T}
      \right)(T_p-T_\infty)\mathrm{d}\tilde{T}     
      +\frac{n_{v,p}-n_{v,\infty}}{n_v}\mathrm{d}\tilde{n}_v\;.
      \label{eq:coeff_proportionality}
\end{eqnarray}
}
This shows that very close to the particle, the variations of $\tilde{n}_v$ and $\tilde{T}$ will act in different ways, since the coefficient in front of $\mathrm{d}\tilde{n}_v$ is always positive and the coefficient in front of $\mathrm{d}\tilde{T}$ always negative, due to the evolution of {$e_{sat,w}$} in the range of temperature considered here.
It explains the small saturation observed in the vicinity of the particle for very small $\tilde{z}_\parallel$ on figure \ref{fig:2D_RH}.
Indeed, in this region we will have a decrease of both $\tilde{n}_v$ and $\tilde{T}$. It leads on the one hand to a decrease of the local partial pressure, but on the other hand to the decrease of temperature $T$ and inducing a decrease of the threshold {$e_{sat,w}(T)$} at which we attain supersaturation.

If we neglect the differences in the diffusivity of mass and heat (i.e.\ $\tilde{n}_v \approx \tilde{T}$), then {equation \ref{eq:S_i_definition} can be written in the form $S_w=f(\tilde{T},T_p,T_\infty)$.}
%
\begin{figure}[t]
\centering
\begin{minipage}{2ex}
      \rotatebox{90}{\centerline{$\left< \mathscr{S}_w \right>(T_c)$, $S_w^{T}(T_c)$}}
\end{minipage}
\begin{minipage}{0.45\linewidth}
   \begin{overpic}[width=1.0\linewidth]
         {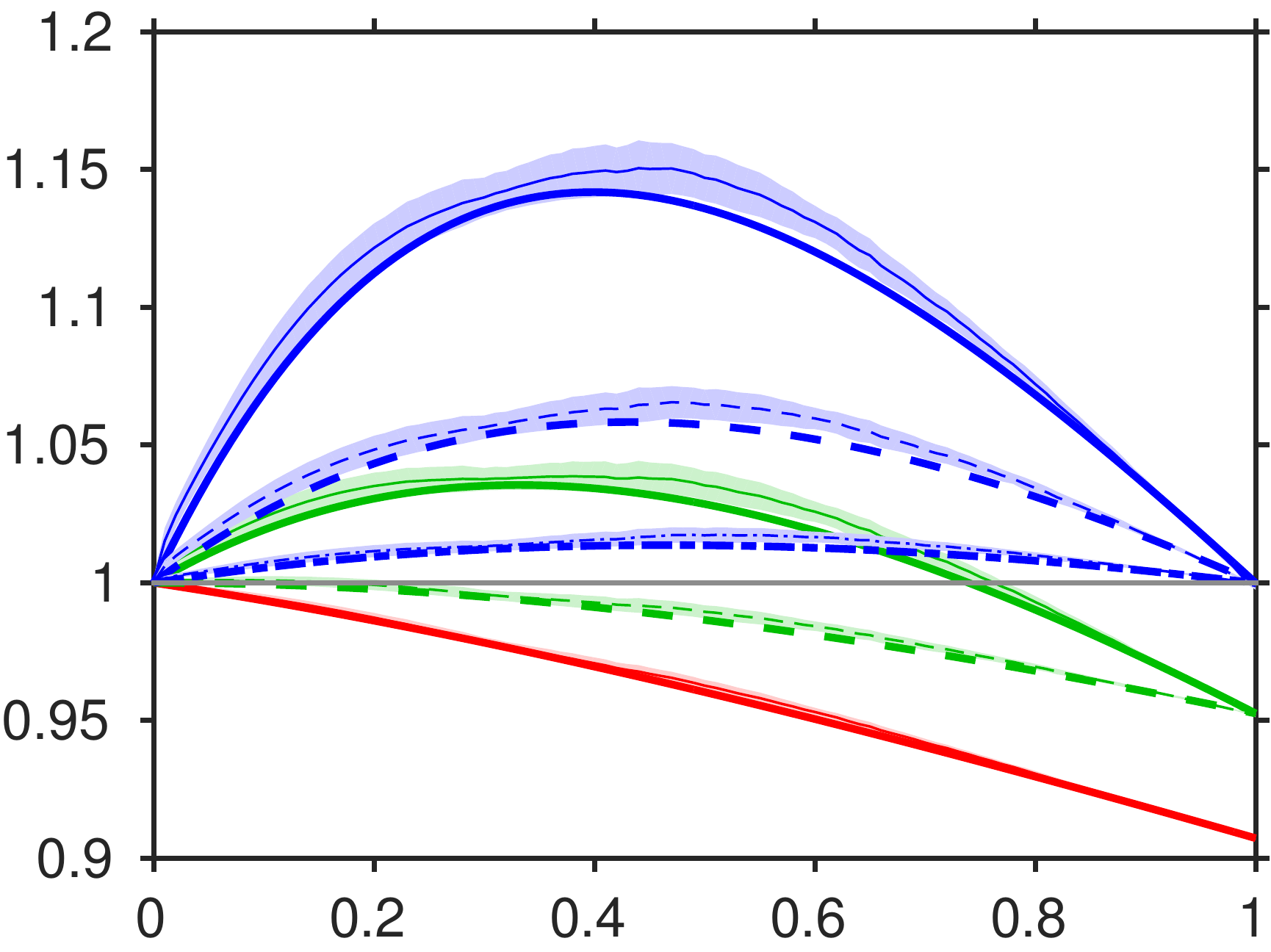}
   \end{overpic}     
   \centerline{$\tilde{T}_c$}    
\end{minipage}
\begin{minipage}{2ex}
   \mbox{}
\end{minipage}
\begin{minipage}{2ex}
      \rotatebox{90}{\centerline{$\left< \mathscr{S}_w \right>(T_c)$, $S_w^{T}(T_c)$}}
\end{minipage}
\begin{minipage}{0.45\linewidth}
   \begin{overpic}[width=1.0\linewidth]
         {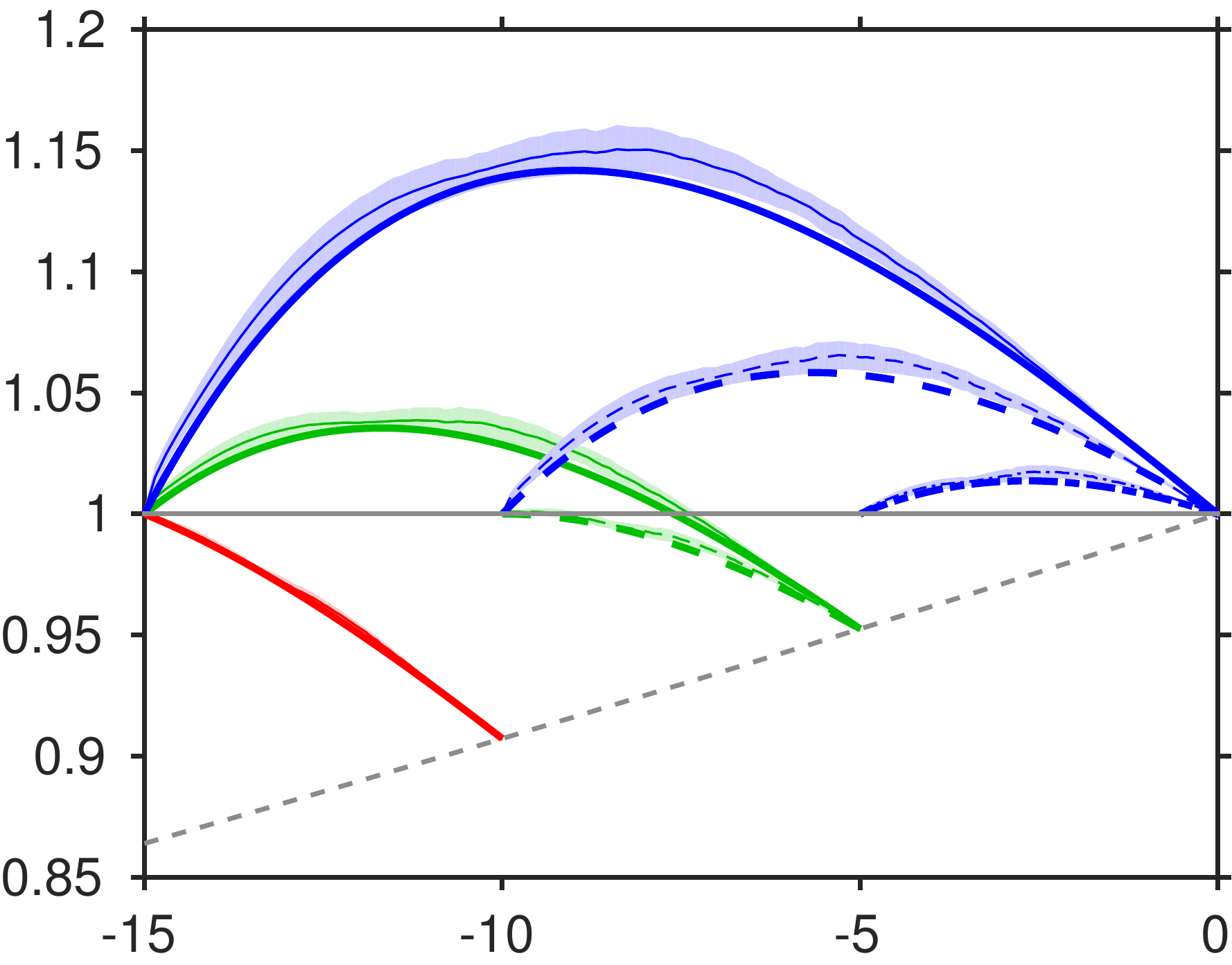}
      \put(37,12) {\textcolor{grey!120}{\rotatebox{17}{$S_w=e_{sat,i}/e_{sat,w}$}}}
   \end{overpic}     
   \centerline{$T_c\;(\degree C)$}    
\end{minipage}
\caption{{Evolution of the mean real ($Sc \neq Pr$, thin lines) and ideal ($Sc=Pr$, heavy lines) saturation with respect to liquid water as a function of the conditional temperature in non-dimensional form $\tilde{T}_c$ (a) or dimensional form (b), for a given set of temperatures $(T_p,T_\infty)$.  The notation $\langle \cdot \rangle$ refers here to the ensemble and time averaging operator. The colored areas indicate the amplitude of the standard deviation for each case, the solid grey lines indicate a saturation of unity with respect to water and the dashed grey line on (b) indicates the saturation of unity with respect to ice (i.e. \ $S_w=e_{sat,i}(T)/e_{sat,w}(T)$).
Linestyle: 
$\color{red}{\solidthick}$ ${T_p=-10\degree C}$, 
$\color{green}{\solidthick}$ ${T_p=-5\degree C}$, 
$\color{blue}{\solidthick}$ ${T_p=0\degree C}$,
solid lines ${T_\infty=-15\degree C}$,
dashed lines ${T_\infty=-10\degree C}$,
dash-dotted lines  ${T_\infty=-5\degree C}$
}}
\label{fig:ideal_vs_real_saturation}
\end{figure}
%
We can introduce with this the surrogate saturation {$S_w^T$} that would be obtained by computing $n_v$ from the non-dimensional temperature, hence providing information on the expected saturation.
{It is computed as a function of $\tilde{T}$ only and the boundary conditions according to:
\begin{eqnarray}
   S_w^T=\frac{k_b\times\left(n_{v,\infty}+\tilde{T}\times(n_{v,p}-n_{v,\infty})\right)\times
   \left(T_\infty+\tilde{T}\times(T_p-T_\infty)\right)}{e_{sat,w}(T_\infty+\tilde{T}\times(T_p-T_\infty))}.
\end{eqnarray}}
The evolution of this "ideal" saturation {$S_w^T$} is represented as a function of the temperature and for different sets of boundary conditions on figure \ref{fig:ideal_vs_real_saturation}. 
%

\begin{table}[h] 
   \centering
   \renewcommand{\arraystretch}{1.1}
   \begin{tabular}{ccc}
      \noalign{\smallskip}\hline\noalign{\smallskip} 
      $T_p$ ($\degree C$) & $T_\infty$ ($\degree C$)  & Max relative error \\
      \noalign{\smallskip}\hline\noalign{\smallskip} 
      $0$ & $-15$   &  $ 7.02 \times 10^{-2}$ \\
      $0$ & $-10$   &  $ 3.53 \times 10^{-2}$ \\
      $0$ & $-5$    &  $ 1.38 \times 10^{-2}$ \\
      $-5$ & $-15$  & $ 3.41 \times 10^{-2}$ \\
      $-5$ & $-10$  & $ 1.23 \times 10^{-2}$ \\
      $-10$ & $-15$ & $ 1.08 \times 10^{-2}$ \\
      \noalign{\smallskip}\hline\noalign{\smallskip} 
   \end{tabular}
   \caption{{Maximal relative error between $S_w$ and $S_w^T$ for different sets of boundary conditions.}}
   \label{table:relative_error_saturation_definitions}
\end{table}

\begin{figure}
\centering
\begin{minipage}{2ex}
      \rotatebox{90}{\centerline{$\tilde{r}_\bot$}}
\end{minipage}
\begin{minipage}{0.65\linewidth}
   \begin{overpic}[width=1.0\linewidth]
         {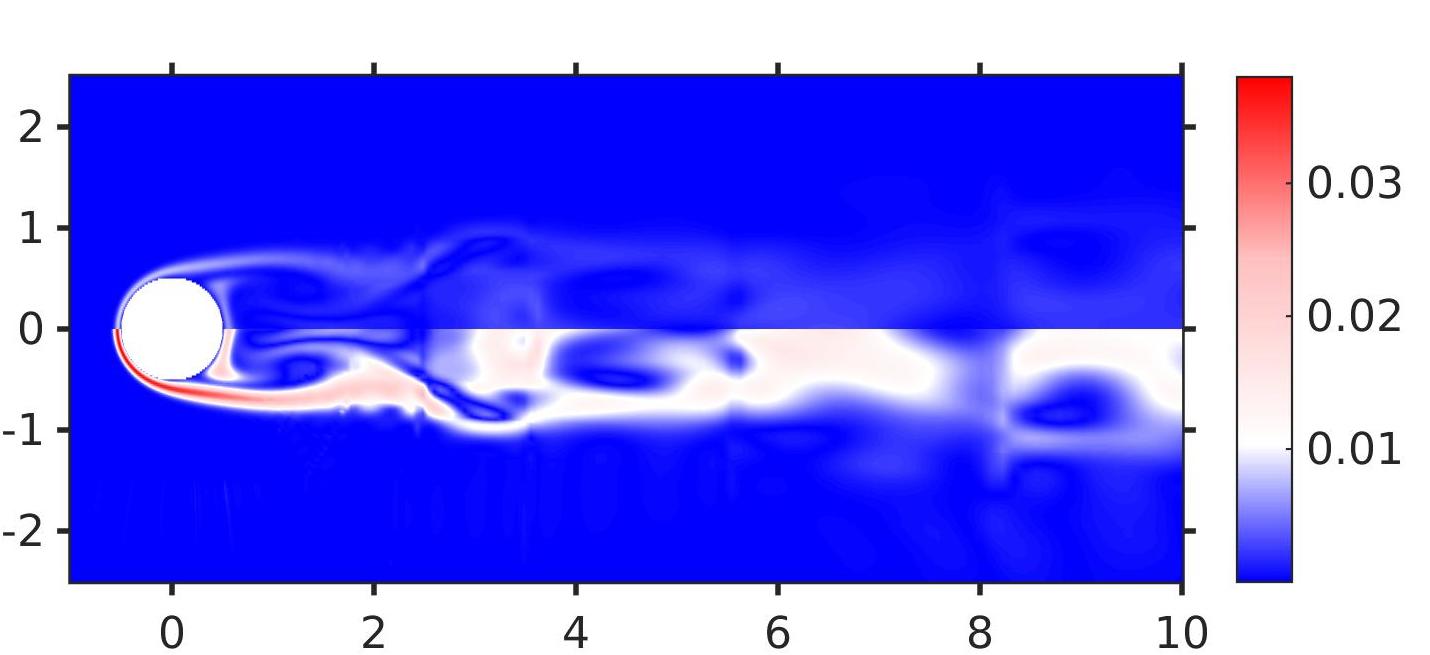}
      \put(32,35) {\textcolor{white}{$T_p=-10\degree C$, $T_\infty=-15\degree C$}}
      \put(32,8) {\textcolor{white}{$T_p=0\degree C$, $T_\infty=-15\degree C$}}
   \end{overpic}   
   \centerline{$\tilde{z}_\parallel$}      
\end{minipage}
\caption{{Instantaneous relative error between the definitions of the saturation with respect to liquid water shown for one half of an axial/radial plane, comparing two sets of boundary conditions, for $Ga=300$, fixed particle, $Pr=0.72$ and $Sc=0.63$ (top: $T_p=-10\degree C$, $T_\infty=-15\degree C$, bottom: $T_p=0\degree C$, $T_\infty=-15\degree C$).}}
\label{fig:2D_RHError}
\end{figure}

Equation \ref{eq:coeff_proportionality} shows {not only that partial pressure and temperature have opposite contributions but also the necessity to account for both variations.}
We now investigate to which extent the approximation {$\tilde{n}_v = \tilde{T}$} would affect the results obtained here.
{For this we introduce the relative error, $\varepsilon(\textbf{x},t)=\left| S_w(\mathbf{x},t)/S_w^T(\mathbf{x},t)-1\right|$ to quantify the differences reached.}
Table \ref{table:relative_error_saturation_definitions} summarizes the maximal relative error between {$S_w^T$} and {$S_w$} obtained for different sets of boundary conditions. 
It shows that the error can attain up to roughly {7 }percent for the largest temperature difference.
Figure \ref{fig:2D_RHError} depicts the evolution in a $(\tilde{r}_\bot,\tilde{z}_\parallel)$ plane of the relative error obtained on an instantaneous field. It shows that the largest error is reached in the mixing region and in the recirculation region next to the sphere, indicating that the local temperature might have an influence on this error.

{Let us take a closer look at the typical saturation $S_w$ that is encountered for a given temperature. For this we proceed as follows: we first divide the temperature domain into bins of temperature $T_c$ and we denote by $\mathscr{S}_w(T_c,t)$ the ensemble of saturation values obtained on the subvolume $\Omega(T_c,t)$, itself defined by the condition $T(\mathbf{x} \in \Omega(T_c,t),t)=T_c$. We then average, for each temperature $T_c$, over the samples collected in time by the ensemble of saturation $\mathscr{S}_w(T_c,t)$.
We denote the corresponding mean by $\left< \mathscr{S}_w \right>(T_c)$.
The evolution of this conditional-averaged saturation as a function of the conditional temperature $T_c$ is represented on figure \ref{fig:ideal_vs_real_saturation}, where the standard deviation is also indicated.}
It confirms the trend given by %
{the evolution of $S_w^T$, namely that for the smallest values of $\tilde{T}$ (i.e. far away from the sphere)
$S_w$ increases with $\tilde{T}$ until a non-dimensional temperature $\tilde{T}_c$ ranging between 0.2 and 0.4, 
after which the saturation $S_w$ decreases with $\tilde{T}$.
It also explains the increase of the saturation observed on figure \ref{fig:2D_RH} in the vicinity of the particle, as $\tilde{T}$ will decrease when going from the particle towards the fluid (negative components of $\mathrm{d}\tilde{T}$).
The evolution of $S_w$ emphasizes also the influence of the boundary conditions since it appears that a large temperature difference tends to increase $S_w$ in the regions $\tilde{T} < \tilde{T}_c$, leading to larger increase of the saturation for a given temperature difference.
This last point explains the difference observed for two sets of boundary condition represented on figure \ref{fig:2D_RH}. 
The case  $(T_p=0\degree C,T_\infty=-15\degree C)$ appears indeed to be more favorable to the development of supersaturated regions than the configuration $(T_p=-10\degree C,T_\infty=-15\degree C)$.
The comparison between the evolutions of $S_w^T$ and $\mathscr{S}_w$ shows that
the "ideal" saturation $S_w^T$ provides a good approximation of the effective
saturation, with the tendency to underestimate it.}
%

Let us now finish with a brief discussion on the evolution of the saturation as a function of the temperature represented on fig. \ref{fig:ideal_vs_real_saturation}(b). It indeed shows that transport of both heat and mass increases the saturation above the level of saturation equal to unity with respect to the liquid phase, which was already known to be a value at which a marked increase in ice-nucleating activity is achieved \citep{beard:92}.
This indicates that, as proposed in the literature \citep{pruppacher:10,dye:68,rosinski:91}, a falling graupel could be at the origin of a local increase of the saturation that might trigger further secondary ice nucleation.
{%
A deeper analysis would be required to quantify more precisely the potential of secondary ice nucleation and will be the object of future analysis. At this state of the work several conclusions can be formulated:
\begin{itemize}[leftmargin=*]%
\setlength\itemsep{0em}
\item The presence of the graupel can induce an increase of the saturation compared to the ambient. This may have several implications: it could "trigger" the formation of secondary ice nuclei by increasing $S_w$ up to the threshold of ice nucleating particle (INP) activation (which depends on the type of INP). If this threshold is already reached in the ambient then the increase of $S_w$ would also increase the activation rate. Furthermore, it could finally favor the growth of an ice crystal that was already formed.
\item The difference between the saturations $S_w(\mathbf{x},t)$ and $S_w^T(\mathbf{x},t)$ is more important in the zones of high saturations, with error up to 7 percent and the tendency to underestimate the saturation if only one transport equation is considered. This means that both INP activation and activation rate might be underestimated, as well as the growth rate of ice crystals or micro-droplets already present in the flow.
\item It is difficult to generally state whether this underestimation could lead or not to substantial error as it requires to know the typical regions that could be occupied by the micro-particles (ice nucleating particles, micro-droplets or ice crystals).
To our knowledge, the concentration of micro-particles in the wake of a sphere has been only studied in the work of \cite{homann15}. A key parameter is the Stokes number of the micro-particles $St_{mp}$ which is defined as the ratio between the relaxation time scale of the micro-particles $\tau_{mp}=\rho_{mp}D_{mp}^2/(18\nu\rho_\infty)$ (with $\rho_{mp}$ and $D_{mp}$ respectively the density and diameter of the micro-particles) and the time-scale of the flow $\tau_f=D/\langle \lvert u_{p,z}\rvert\rangle_t$.
Based on this the Stokes number will be a function of the properties of the micro-particle and the graupel according to:
\begin{eqnarray}
   St_{mp} 
      &=&
      \frac{1}{18}\frac{\rho_{mp}}{\rho_\infty}\left(\frac{D_{mp}}{D}\right)^2
      \frac{\langle \lvert u_{p,z}\rvert\rangle_t}{u_g}Ga
\end{eqnarray}
A large range of Stokes numbers can be reached depending on the type of micro-particles (featuring a small or very small diameter ratio $D_{mp}/D$) and the type of graupel (featuring a large or very large Galileo number).
The largest Stokes will be reached by micro-droplets since they feature large diameter ratios (ranging from 0.001 to 0.5 depending on the type of graupel). 
The density ratio is expected to be of the order of $\mathcal{O}(10^3)$ for micro-droplets, and for aerosols similar to the ones investigated by \cite{kanji11}. The size ratio can vary between $\mathcal{O}(10^{-6})$ and $\mathcal{O}(10^{-3})$, leading either to very small Stokes number ($St_{mp}=\mathcal{O}(10^{-6})$) or intermediate Stokes numbers ($St_{mp}\approx0.1$). For the smallest Stokes number, $\tau_{mp}$ is so small that the micro-particles react instantaneously to the fluid and can sample the entire wake. But for larger Stokes number and at the Reynolds numbers of interest \cite{homann15} have shown the formation of a cylinder of high concentration around the sphere, with a low-concentration shadow behind the sphere, followed by a zone of high concentration for larger streamwise distances.
The zone with the largest saturation and largest error due to the approximation $\tilde{n}_v=\tilde{T}$ in the shear region coincides with the shadow region, such that the largest error would not have a significant influence at this Stokes number.
For the largest temperature difference the error features blobs with errors of the order of one percent for larger streamwise distances in the high concentration cone (fig. \ref{fig:2D_RHError}). It is difficult to estimate whether or not this error could have an influence on the activation fraction as the corresponding saturation is close to the onset saturation \citep{hoose:12}. Indications on the history of the saturation seen by micro-particles would be necessary to properly determine the importance of the error.
Simulations accounting for the transport of aerosols should therefore be
considered to clearly state whether this underestimation could lead to substantial error.
\end{itemize}
}

%
\section{Conclusion} \label{sec:conclusion}

Direct numerical simulations have been conducted to study the case of a falling sphere made out of ice in moist air with the aim of exploring the saturation field in the wake of this sphere and test whether it might be  large enough to be at the origin of secondary ice nucleation.
For this we used a spectral/spectral-element solver to simulate the Navier-Stokes equations coupled to two transport equations to account for variation of heat and vapor content in the flow, under the Boussinesq assumption. The sphere is assumed to be at constant temperature, warmer than the surroundings, and at saturation of unity with respect to {ice}. The surrounding fluid is assumed to have a constant velocity with incoming constant temperature and constant saturation of unity with respect to {liquid} water.

The goal of the present study is to address several methodological questions to build a numerical framework that can be used in the future to investigate the ability of a graupel, for a given set of conditions, to trigger secondary ice nucleation. 

The first question concerns the influence of the settling regime on the structure of the scalar fields, in the absence of buoyancy effects. 
The analysis focused first on the temperature field. Since heat and mass in this system have similar diffusivities the conclusions hold also for the vapor partial pressure. 
As expected it shows that the regime plays an important role on the structure of the temperature field, and that a configuration with a chaotic wake should be chosen for the application aimed here. 
The recirculation region appears to play a significant role on the temperature field as it drives the transport of cold fluid from the wake toward the back of the sphere and conversely.
The influence of the mobility is considered here either by varying the density ratio between the particle and the upcoming fluid or by fixing the sphere, and it appears {necessary to represent the graupel by an object featuring large density or by a fixed object}. 

{The relatively small temperature differences and large density ratio together imply in the current context to have small Richardson numbers (less than $0.001$).}
Simulations accounting for buoyancy effects, for larger but still small Richardson number (i.e.\ up to $0.1$) have been performed in order to test whether such small values might affect the results. It appears that light buoyancy effects at $Ri_T=0.1$ are sufficient to stabilize the wake and trajectory and the sphere by pushing the thresholds of appearance of each regime towards larger Galileo numbers, {with qualitative accordance} with the observation from the literature for fixed spheres \citep{Kotouc_IJHMT2008, Kotouc_JFM2009}.
{Simulations involving $Ri_T=0.05$ and $Ri_T=0.1$ confirmed that the recirculation zone is also affected by the buoyancy with a decrease of the extent of this region,
at much smaller Richardson numbers than that of \cite{Bhattacharyya_IJHMT2008}
and in non-axisymmetric and unsteady chaotic regimes.}
The actual Richardson number of $Ri_T=10^{-4}$ appears to barely affect the flow, meaning that buoyancy effects do not need to be accounted for in the meteorological context.

Finally, the saturation in the wake of the sphere is explored for different sets of boundary conditions and it shows that substantial supersaturation can indeed be reached in the wake of the sphere, and that large temperature differences between the particle and the surrounding are more favorable to the development of this zone. Here as well the recirculation region has an important impact on the level of the saturation.
{This is, to our knowledge, the first time that this increase of saturation has been observed for chaotic regimes with the aid of direct numerical simulation. }
The necessity to separate both heat and mass transport equations is discussed, and small differences in the saturation field are obtained, with the tendency to underestimate the saturation if mass diffusivity is approximated with the heat diffusivity.
{No general recommendation can be formulated on the necessity to separate both transport equations since the error introduced would depend on the explored system. Configurations with small ice nucleating particles featuring small Stokes numbers appear to be the most sensitive ones since the micro-particles might sample the regions of the flow with the largest error (up to 7$\%$) at saturation close to the activation threshold. Additional simulations involving the transport of micro-particles would therefore be required to discuss the necessity to simulate separately the transport of heat and water vapor.}
The main prospects of this work concern the more systematic analysis of the saturation field, for different sets of boundary conditions in order to discuss in detail the expectation to observe secondary ice nucleation in the wake of a graupel.

\section*{Acknowledgements}
The simulations were partially performed at SCC Karlsruhe. The computer resources, technical expertise and assistance provided by {this center} are thankfully acknowledged.
{We acknowledge support by Deutsche Forschungsgemeinschaft and open access publishing fund of Karlsruhe Institute of Technology.}
\begin{appendices}
{%
\section{Governing equations} \label{sec:governing_equations}
The dimensional form of the Navier-Stokes equations under the Boussinesq approximation and the advection-diffusion equations for heat and mass,
both formulated in an inertial frame of reference with independent variables $(x^*,y^*,z^*,t^*)$, read
\begin{eqnarray}
   \boldsymbol{\nabla}^*\cdot\mathbf{u} &=& 0, \\
   \frac{\partial \mathbf{u}}{\partial t^*} + \mathbf{u}\cdot\boldsymbol{\nabla}^*\mathbf{u}
      &=&
      -\frac{1}{\rho_\infty}\boldsymbol{\nabla}^*P + \nu \boldsymbol{\nabla}^{*2}\mathbf{u} +\frac{\rho}{\rho_\infty}\mathbf{g}, \\
    \frac{\partial T}{\partial t^*} + \mathbf{u}\cdot\boldsymbol{\nabla}^*T
       &=&
       \mathcal{D}_T\boldsymbol{\nabla}^{*2} T, \\
    \frac{\partial n_v}{\partial t^*} + \mathbf{u}\cdot\boldsymbol{\nabla}^*n_v
       &=&
       \mathcal{D}_m\boldsymbol{\nabla}^{*2} n_v,
\end{eqnarray}
where $\mathbf{u}$ is the fluid velocity, $P$ the pressure and $T$, $n_v$ the two scalars.
The Boussinesq approximation states that the variation in density due to variations in temperature and vapor concentration 
affects only the buoyancy term $\rho/\rho_\infty \mathbf{g}$.}

{%
We subtract the contribution of the hydrostatic pressure from the pressure gradient term by introducing a modified pressure field $p$ such that $\mathbf{\nabla}^*P=\mathbf{\nabla}^*p+\rho_\infty\mathbf{g}$. With this the momentum equation becomes
\begin{eqnarray}
\frac{\partial \mathbf{u}}{\partial t} + \mathbf{u}\cdot\boldsymbol{\nabla}^*\mathbf{u}
      &=&
      -\frac{1}{\rho_\infty}\boldsymbol{\nabla}^*p + \nu \boldsymbol{\nabla}^{*2}\mathbf{u} +\mathbf{B},
\end{eqnarray}
where the buoyancy term $\mathbf{B}$ is defined as $\mathbf{B}=\left((\rho-\rho_\infty)/\rho_\infty\right)\mathbf{g}$.}

{%
In order to evaluate our governing equations on a mesh which is translating with the particle motion, we introduce the following independent variables of the moving mesh
$(x,y,z,t) = (x^*-x^*_p(t^*),y^*-y^*_p(t^*),y^*-z^*_p(t^*),t^*)$. The origin is located at the center of the sphere and 
the axes are aligned with the frame of reference. 
Each field $A(\mathbf{x},t)$ is then given in the mesh by
\begin{eqnarray}
  \label{eq:appendix-cstrans}
   A(\mathbf{x}(\mathbf{x}^*,t^*),t) &\equiv& A(\mathbf{x}^*,t^*).
\end{eqnarray}
Please note, that
the components of vector fields remain expressed in the inertial frame, and hence merely the indexing has changed.
The transformation is constant in space, and thus
\begin{eqnarray}
  \boldsymbol{\nabla}^*A(\mathbf{x}^*,t^*)
     &=&
     \boldsymbol{\nabla} A(\mathbf{x},t).
\end{eqnarray}}

{%
The temporal derivative of $A$ on the moving mesh can be obtained by applying the multivariable chain rule to 
\eqref{eq:appendix-cstrans}, which results in
\begin{eqnarray}
   \frac{\partial A (\mathbf{x}^*,t^*)}{\partial t^*}
      &=&
        \frac{\partial A (\mathbf{x},t)}{\partial t} 
        + \frac{\partial \mathbf{x}(\mathbf{x}^*,t^*)}{\partial t^*} \cdot \boldsymbol{\nabla A}(\mathbf{x},t) \nonumber\\
      &=&
        \frac{\partial A (\mathbf{x},t)}{\partial t} 
        - \mathbf{u}_p \cdot \boldsymbol{\nabla A}(\mathbf{x},t),
\end{eqnarray}
where $\mathbf{u}_p \equiv \frac{\partial \mathbf{x}^*_p}{\partial t^*}$ is the particle velocity in the inertial
frame. The substantial derivative of quantity $A$ can therefore be expressed on the translating mesh by
\begin{eqnarray}
   \frac{\partial A(\mathbf{x}^*,t^*)}{\partial t^*}+\mathbf{u}(\mathbf{x}^*,t^*) \cdot \boldsymbol{\nabla}^* A(\mathbf{x}^*,t^*)
      &=&
      \frac{\partial A(\mathbf{x},t)}{\partial t} + (\mathbf{u}(\mathbf{x},t) -\mathbf{u}_p) \cdot \boldsymbol{\nabla} A(\mathbf{x},t).
\end{eqnarray}
Thus, the formulation of the Navier-Stokes equations and transport equations of the scalar quantities, on the particle-attached grid read
\begin{eqnarray}
   \boldsymbol{\nabla} \cdot \mathbf{u} &=& 0,\\
   \frac{\partial \mathbf{u}}{\partial t}
      +(\mathbf{u}-\mathbf{u}_p)\cdot\boldsymbol{\nabla}\mathbf{u}
      &=&
      -\frac{1}{\rho_\infty}\boldsymbol{\nabla}P
      +\nu\boldsymbol{\nabla}^2\mathbf{u}+\mathbf{B},\\
   \frac{\partial T}{\partial t}
      +(\mathbf{u}-\mathbf{u}_p)\cdot\boldsymbol{\nabla} T
      &=&
      \mathcal{D}_T\boldsymbol{\nabla}^2 T,\\
   \frac{\partial n_v}{\partial t}
      +(\mathbf{u}-\mathbf{u}_p)\cdot\boldsymbol{\nabla} n_v
      &=&
      \mathcal{D}_m\boldsymbol{\nabla}^2 n_v.
\end{eqnarray}
The system of equations can then be reformulated in non-dimensional form, after normalizing the spatial coordinates according to $\tilde{x}_i=x_i/D$, the time $\tilde{t}=t D/u_g$, velocity as $\tilde{\mathbf{u}}=\mathbf{u}/u_g$, pressure as $\tilde{p}=p/( \rho_\infty u_g^2)$, and temperature and concentration according to eq. \ref{eq-T_NonDim} and \ref{eq-nv_NonDim}, leading to the set of equations \ref{eq-continuity}-\ref{eq-mass}.}

{%
We decompose the buoyancy term into two contributions with help of a first-order Taylor series approximation around $\rho_\infty$.
The first contribution, $\mathbf{B}_T$, accounts for the modification of density due to variations in temperature and the second one, $\mathbf{B}_{n_v}$,
due to variations in vapor concentration. Their definitions are
\begin{eqnarray}
   \mathbf{B}_T 
      &=&
       \frac{1}{\rho_\infty}\frac{\partial \rho}{\partial T}\bigg\vert_\infty(T-T_\infty)\mathbf{g}, \\
   \mathbf{B}_{n_v}
      &=&
       \frac{1}{\rho_\infty}\frac{\partial \rho}{\partial n_v}\bigg\vert_\infty(n_v-n_{v,\infty})\mathbf{g}.
\end{eqnarray}
We assume that the fluid behaves as a perfect gas, implying that the partial derivatives of $\rho$ with respect to $T$ and $n_v$ yield
\begin{eqnarray}
  \frac{\partial \rho}{\partial T}\bigg\vert_{\infty} &=& -\frac{1}{T_\infty} \rho_\infty, \\
  \frac{\partial \rho}{\partial n_v}\bigg\vert_{\infty} &=& \frac{M_w - M_d}{N_A},
\end{eqnarray}
where $N_A$ is the Avogadro constant, $M_w$ the molar mass of water and $M_d$ the mixture molar mass of dry air.
Both buoyancy contributions can therefore be written in non-dimensional form as 
$\tilde{\mathbf{B}}_T=Ri_T\tilde{T}\mathbf{k}$ and $\tilde{\mathbf{B}}_{n_v}=Ri_{n_v}\tilde{n}_v\mathbf{k}$, 
with the thermal Richardson number defined as in equation \ref{eq:Ri_heat} and the vapor concentration Richardson number according to
\begin{eqnarray}
   Ri_{n_v}
      &=&
      -\frac{1}{\left(\frac{\rho_p}{\rho_\infty} -1\right)} \frac{M_w-M_d}{N_A \rho_\infty} (n_{v,p}-n_{v,\infty}),
\end{eqnarray}
where the term $(\rho_p/\rho_\infty-1)^{-1} = D\lvert  \mathbf{g}\rvert/u_g^2$ comes from the non-dimensionalization.
}%
\section{Computation of the saturation field} \label{sec:appendix_MurphyKoop}
%
\begin{figure}[h]
\centering
\begin{minipage}{3ex}
      \rotatebox{90}{\centerline{$e_{sat,w}(T)$, $e_{sat,i}(T)$ { (Pa)}} }
\end{minipage}
\begin{minipage}{0.45\linewidth}
   \begin{overpic}[width=1.0\linewidth]
         {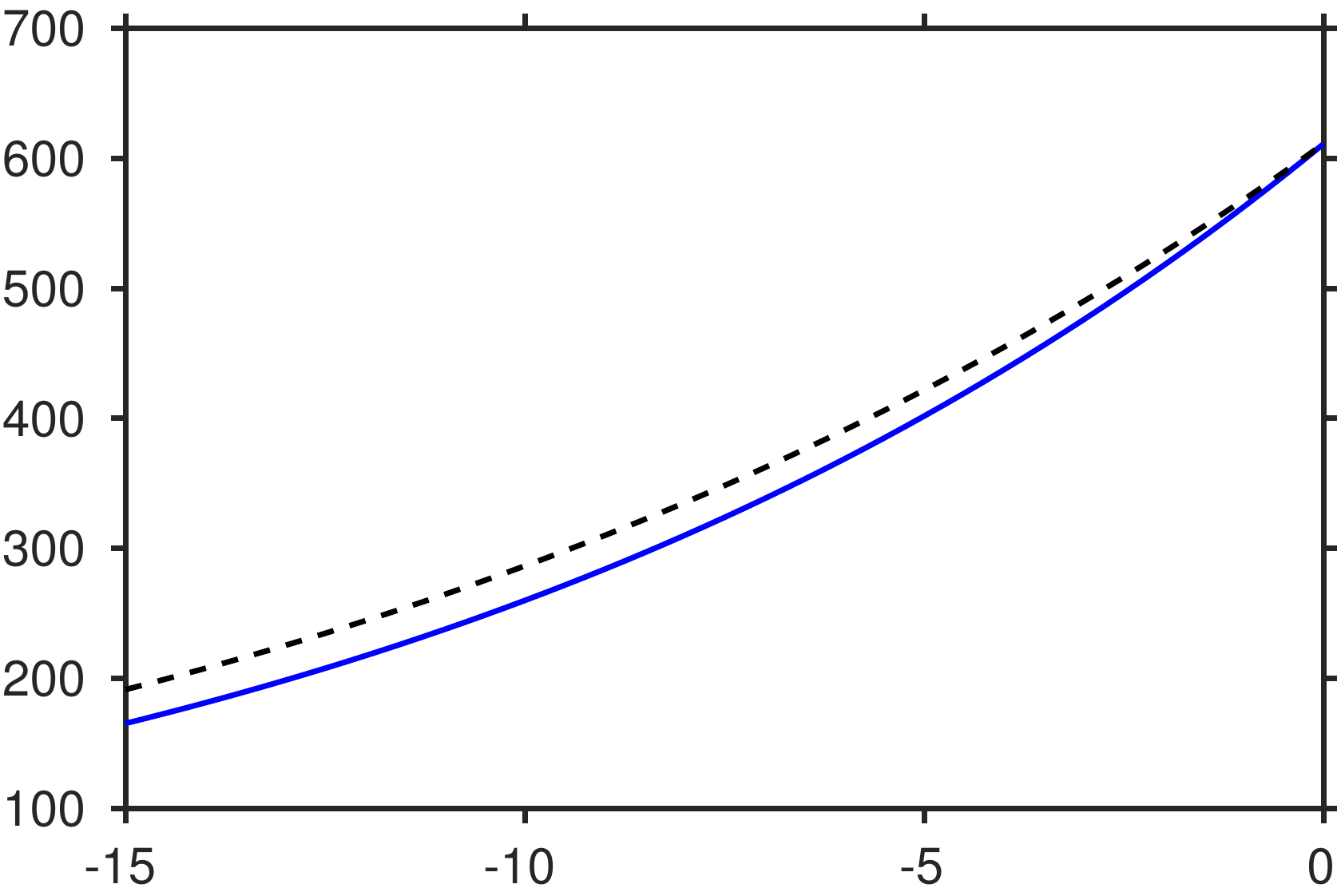}
   \end{overpic}     
   \centerline{$T\;(\degree C)$}    
\end{minipage}
\caption{ Evolution of the saturation vapor pressure with respect to liquid $e_{sat,w}$ and with respect to ice $e_{sat,i}$ as a function of the temperature, according to the correlations of \cite{murphy:05}.
Linestyle: $\color{blue}{\solidthick}$ $e_{sat,i}$, $\dashed$ $e_{sat,w}$.}
\label{fig:e_MurphyKoop}
\end{figure}

%
The saturation $S_j$ with respect to the phase $j$ is defined as the ratio between the vapor partial pressure $e$ and the saturation vapor pressure with respect to this phase $e_{sat,j}$.
We used the correlations of \cite{murphy:05} for the computation of the saturation vapor pressure with respect to ice and liquid water, respectively $e_{sat,i}$ and $e_{sat,w}$, namely
\begin{eqnarray}
   e_{sat,i} &=&
      \mathrm{exp}[
         9.550426 - 5723.265/T + 3.53068 \mathrm{ln}(T ) - 0.00728332T
      ]
      \label{eq:murphy_ice}
      \\
   e_{sat,w} &=&  
      \mathrm{exp} [
      54.842763 - 6763.22/T - 4.210 \mathrm{ln}(T ) + 0.000367T \nonumber  \\
    &+&   \mathrm{tanh}\left(0.0415(T - 218.8)\right) 
       (53.878 - 1331.22/T - 9.44523 \mathrm{ln}(T ) + 0.014025T )
      ]
      \label{eq:murphy_water}
      \;,
\end{eqnarray}
with temperatures expressed in degrees Kelvin and pressures expressed in Pascal. Figure \ref{fig:e_MurphyKoop} depicts the evolution of the saturation vapor pressure as a function of the dimensional temperature for both liquid and solid phases.

\end{appendices}
\bibliography{./references.bib}

\begin{thebibliography}{}

\bibitem[Bagchi and Balachandar, 2004]{Bagchi_JFM2004}
Bagchi, P. and Balachandar, S. (2004).
\newblock Response of the wake of an isolated particle to an isotropic
  turbulent flow.
\newblock {\em Journal of Fluid Mechanics}, 518:95--123.

\bibitem[Bagchi et~al., 2001]{Bagchi_ASME2001}
Bagchi, P., Ha, M.~Y., and Balachandar, S. (2001).
\newblock {Direct numerical simulation of flow and heat transfer from a sphere
  in a uniform cross-flow}.
\newblock {\em Trans. ASME. J. Fluids Eng.}, 123(2):347--358.

\bibitem[Bagchi and Kottam, 2008]{Bagchi_PoF2008}
Bagchi, P. and Kottam, K. (2008).
\newblock {Effect of freestream isotropic turbulence on heat transfer from a
  sphere}.
\newblock {\em Phys. Fluids}, 20(7).

\bibitem[Beard, 1992]{beard:92}
Beard, K. (1992).
\newblock Ice initiation in warm-base convective clouds: An assessment of
  microphysical mechanisms.
\newblock {\em Atmospheric Research}, 28(2):125--152.

\bibitem[Bhattacharyya and Singh, 2008]{Bhattacharyya_IJHMT2008}
Bhattacharyya, S. and Singh, A. (2008).
\newblock {Mixed convection from an isolated spherical particle}.
\newblock {\em Int. J. Heat Mass Transf.}, 51(5-6):1034--1048.

\bibitem[Bouchet et~al., 2006]{Bouchet_EJM2006}
Bouchet, G., Mebarek, M., and Du\v{s}ek, J. (2006).
\newblock Hydrodynamic forces acting on a rigid fixed sphere in early
  transitional regimes.
\newblock {\em European Journal of Mechanics - B/Fluids}, 25(3):321--336.

\bibitem[Cheng et~al., 2014]{cheng:14}
Cheng, K., Wang, P., and Wang, C. (2014).
\newblock A numerical study on the ventilation coefficients of falling
  hailstones.
\newblock {\em Journal of the Atmospheric Sciences}, 71(7):2625--2634.

\bibitem[Clift et~al., 1978]{Clift_1978}
Clift, R., Grace, J., and Weber, M. (1978).
\newblock {\em Bubbles, drops and particles}.
\newblock Academic Press.

\bibitem[Dan and Wachs, 2010]{Dan_IJHMT2010}
Dan, C. and Wachs, A. (2010).
\newblock {Direct Numerical Simulation of particulate flow with heat transfer}.
\newblock {\em Int. J. Heat Fluid Flow}, 31(6):1050--1057.

\bibitem[Dye and Hobbs, 1968]{dye:68}
Dye, J.~E. and Hobbs, P.~V. (1968).
\newblock The influence of environmental parameters on the freezing and
  fragmentation of suspended water drops.
\newblock {\em Journal of the Atmospheric Sciences}, 25(1):82--96.

\bibitem[Ern et~al., 2012]{Ern_AnnuRev2012}
Ern, P., Risso, F., Fabre, D., and Magnaudet, J. (2012).
\newblock {Wake-Induced Oscillatory Paths of Bodies Freely Rising or Falling in
  Fluids}.
\newblock {\em Annu. Rev. Fluid Mech.}, 44:97--121.

\bibitem[Fabre et~al., 2012]{fabre:12}
Fabre, D., Tchoufag, J., and Magnaudet, J. (2012).
\newblock The steady oblique path of bouyancy-driven disks and spheres.
\newblock 707:24--36.

\bibitem[Gan et~al., 2003]{Gan_JFM2003}
Gan, H., Chang, J., Feng, J.~j., and Hu, H.~H. (2003).
\newblock {Direct numerical simulation of the sedimentation of solid particles
  with thermal convection}.
\newblock {\em J. Fluid Mech.}, 481:385--411.

\bibitem[Ghidersa and Du{\v{s}}ek, 2000]{Ghidersa_JFM2000}
Ghidersa, B. and Du{\v{s}}ek, J. (2000).
\newblock {Breaking of axisymmetry and onset of unsteadiness in the wake of a
  sphere}.
\newblock {\em J. Fluid Mech.}, 423:33--69.

\bibitem[Homann and Bec, 2015]{homann15}
Homann, H. and Bec, J. (2015).
\newblock Concentrations of inertial particles in the turbulent wake of an
  immobile sphere.
\newblock {\em Physics of Fluids}, 27(5):053301.

\bibitem[Hoose and M\"{o}hler, 2012]{hoose:12}
Hoose, C. and M\"{o}hler, O. (2012).
\newblock Heterogeneous ice nucleation on atmospheric aerosols: a review of
  results from laboratory experiments.
\newblock {\em Atmospheric Chemistry and Physics}, 12(20):9817--9854.

\bibitem[Jenny and Du{\v{s}}ek, 2004]{Jenny_JCP2004}
Jenny, M. and Du{\v{s}}ek, J. (2004).
\newblock {Efficient numerical method for the direct numerical simulation of
  the flow past a single light moving spherical body in transitional regimes}.
\newblock {\em J. Comput. Phys.}, 194(1):215--232.

\bibitem[Jenny et~al., 2004]{Jenny_JFM2004}
Jenny, M., Du{\v{s}}ek, J., and Bouchet, G. (2004).
\newblock {Instabilities and transition of a sphere falling or ascending freely
  in a Newtonian fluid}.
\newblock {\em J. Fluid Mech.}, 508:201--239.

\bibitem[Johnson and Patel, 1999]{johnson:99}
Johnson, T. and Patel, V. (1999).
\newblock Flow past a sphere up to a reynolds number of 300.
\newblock {\em Journal of Fluid Mechanics}, 378:19--70.

\bibitem[Kanji et~al., 2011]{kanji11}
Kanji, Z.~A., DeMott, P.~J., M\"{o}hler, O., and Abbatt, J. P.~D. (2011).
\newblock Results from the university of toronto continuous flow diffusion
  chamber at {ICIS} 2007: instrument intercomparison and ice onsets for
  different aerosol types.
\newblock {\em Atmospheric Chemistry and Physics}, 11(1):31--41.

\bibitem[Kotou{\v{c}} et~al., 2008]{Kotouc_IJHMT2008}
Kotou{\v{c}}, M., Bouchet, G., and Du{\v{s}}ek, J. (2008).
\newblock {Loss of axisymmetry in the mixed convection, assisting flow past a
  heated sphere}.
\newblock {\em Int. J. Heat Mass Transf.}, 51(11-12):2686--2700.

\bibitem[Kotou{\v{c}} et~al., 2009]{Kotouc_JFM2009}
Kotou{\v{c}}, M., Bouchet, G., and Du{\v{s}}ek, J. (2009).
\newblock {Transition to turbulence in the wake of a fixed sphere in mixed
  convection}.
\newblock {\em J. Fluid Mech.}, 625:205.

\bibitem[Legendre et~al., 2006]{Legendre_PoF2006}
Legendre, D., Merle, A., and Magnaudet, J. (2006).
\newblock {Wake of a spherical bubble or a solid sphere set fixed in a
  turbulent environment}.
\newblock {\em Phys. Fluids}, 18(4):048102.

\bibitem[Magnaudet et~al., 1995]{magnaudet:95}
Magnaudet, J., Rivero, M., and Fabre, J. (1995).
\newblock Accelerated flows past a rigid sphere or a spherical bubble. part 1.
  steady straining flow.
\newblock {\em Journal of Fluid Mechanics}, 284(-1):97.

\bibitem[Murphy and Koop, 2005]{murphy:05}
Murphy, D.~M. and Koop, T. (2005).
\newblock Review of the vapour pressures of ice and supercooled water for
  atmospheric applications.
\newblock {\em Quarterly Journal of the Royal Meteorological Society},
  131(608):1539--1565.

\bibitem[Ormi{\`{e}}res and Provansal, 1999]{ormieres:99}
Ormi{\`{e}}res, D. and Provansal, M. (1999).
\newblock Transition to turbulence in the wake of a sphere.
\newblock {\em Physical Review Letters}, 83(1):80--83.

\bibitem[Prabhakaran et~al., 2017]{prabhakaran:17}
Prabhakaran, P., Weiss, S., Krekhov, A., Pumir, A., and Bodenschatz, E. (2017).
\newblock Can hail and rain nucleate cloud droplets?
\newblock {\em Physical Review Letters}, 119(12).

\bibitem[Pruppacher and Klett, 2010]{pruppacher:10}
Pruppacher, H. and Klett, J. (2010).
\newblock {\em Microphysics of Clouds and Precipitation}.
\newblock Springer Netherlands.

\bibitem[Ranz and Marshall, 1952]{ranz:52a}
Ranz, W. and Marshall, W. (1952).
\newblock Evaporation from drops.
\newblock {\em Chem. Eng. Prog}, 48(3):141--146.

\bibitem[Rosinski and Morgan, 1991]{rosinski:91}
Rosinski, J. and Morgan, G. (1991).
\newblock Cloud condensation nuclei as a source of ice-forming nuclei in
  clouds.
\newblock {\em Journal of Aerosol Science}, 22(2):123--133.

\bibitem[Schiller and Naumann, 1935]{Schiller_1935}
Schiller, L. and Naumann, A. (1935).
\newblock A drag coefficient correlation.
\newblock {\em VDI Zeitung}, 77(318):51.

\bibitem[Uhlmann and Du{\v{s}}ek, 2014]{Uhlmann_IJMF2014}
Uhlmann, M. and Du{\v{s}}ek, J. (2014).
\newblock {The motion of a single heavy sphere in ambient fluid: A benchmark
  for interface-resolved particulate flow simulations with significant relative
  velocities}.
\newblock {\em Int. J. Multiph. Flow}, 59:221--243.

\bibitem[Vali et~al., 2015]{vali15}
Vali, G., DeMott, P.~J., M\"{o}hler, O., and Whale, T.~F. (2015).
\newblock Technical note: A proposal for ice nucleation terminology.
\newblock {\em Atmospheric Chemistry and Physics}, 15(18):10263--10270.

\bibitem[Wang and Kubicek, 2013]{wang:13}
Wang, P. and Kubicek, A. (2013).
\newblock Flow fields of graupel falling in air.
\newblock {\em Atmospheric Research}, 124:158--169.

\bibitem[Young, 1993]{young:93}
Young, K. (1993).
\newblock {\em Microphysical Processes in Clouds}.
\newblock Oxford University Press.

\bibitem[Zhou and Du{\v{s}}ek, 2015]{Zhou_IJMF2015}
Zhou, W. and Du{\v{s}}ek, J. (2015).
\newblock {Chaotic states and order in the chaos of the paths of freely falling
  and ascending spheres}.
\newblock {\em Int. J. Multiph. Flow}, 75:205--223.

\end{thebibliography}

\end{document}